\documentclass{article}
\usepackage[top=1.5in, bottom=1.5in, left=1.2in, right=1.2in]{geometry}
\usepackage{authblk}
\usepackage{amssymb, amsmath,framed}
\usepackage{graphicx}
\usepackage[round]{natbib}

\usepackage{bm}
\usepackage[colorlinks=true,linkcolor=blue,urlcolor=blue,citecolor=blue,anchorcolor=blue]{hyperref}
\expandafter\ifx\csname package@font\endcsname\relax\else
 \expandafter\expandafter
 \expandafter\usepackage
 \expandafter\expandafter
 \expandafter{\csname package@font\endcsname}
\fi
\def\bq{\begin{equation}}
\def\eq{\end{equation}}
\def\bqy{\begin{eqnarray}}
\def\eqy{\end{eqnarray}}

\makeatother

\begin{document}

\title{Subsurface Exolife}

\author{Manasvi Lingam \thanks{Electronic address: \texttt{manasvi.lingam@cfa.harvard.edu}}}

\author{Abraham Loeb \thanks{Electronic address: \texttt{aloeb@cfa.harvard.edu}}}
\affil{Harvard-Smithsonian Center for Astrophysics, 60 Garden St, Cambridge, MA 02138, USA}
\affil{Institute for Theory and Computation, Harvard University, Cambridge MA 02138, USA}

\date{}

\maketitle

\begin{abstract}
We study the prospects for life on planets with subsurface oceans, and find that a wide range of planets can exist in diverse habitats with ice envelopes of moderate thickness. We quantify the energy sources available to these worlds, the rate of production of prebiotic compounds, and assess their potential for hosting biospheres. Life on these planets is likely to face challenges, which could be overcome through a combination of different mechanisms. We estimate the number of such worlds, and find that they may outnumber rocky planets in the habitable zone of stars by a few orders of magnitude.
\end{abstract}

\section{Introduction} \label{SecIntro}
The concept of the circumstellar habitable zone (HZ), i.e. the region around a host star where liquid water can exist on the surface of a planet with a given atmospheric composition, has a complex history \citep{Gon05}. Over the past two decades, since its first modern formulation \citep{KWR93}, there has been a tendency in some quarters to conflate the HZ with the broader notion of habitability as pointed out by the likes of \citet{SG16,TT17,ML17}.\footnote{As per NASA's Astrobiology Strategy: ``\emph{Habitability has been defined as the potential of an environment (past or present) to support life of any kind.}''} Hence, it is necessary to clearly distinguish between these two concepts and recognize the limitations (and strengths) of the HZ as a signpost for life. In an early treatise on the HZ, \citet{Sag96} emphasized the fact that a diverse range of planets (and moons) lying outside the HZ are not precluded from having water or life-as-we-know-it.\footnote{By ``life-as-we-know-it'', we will refer henceforth to organisms which involve carbon-based chemistry, and water constitutes the solvent.}

If one takes into account the possibility that potentially habitable worlds outside the HZ can exist, a wide range of habitats are feasible \citep{Lam09}. Planets and satellites with subsurface oceans are amongst the most commonly studied worlds in terms of their capacity to sustain biospheres. In our own Solar system, Europa and Enceladus fall distinctly under this category and have been widely considered as possible abodes for life \citep{Chy00,MFE03,PLY08,SBE09,VHP16,WG17} since they appear to host many of the necessary ingredients. In addition, some theoretical models appear to indicate that the outer planets of the TRAPPIST-1 system \citep{GTD17} also possess subsurface oceans \citep{BDK17}. The class of planets and moons with deep subterranean biospheres \citep{Go92,Sleep,MOM18} also merits consideration, since it widens the boundaries of the conventional HZ \citep{MOM13,Cock14}.

Hitherto, we have restricted our discussion only to objects (planets, moons and planetoids) around stars \citep{Dys03,AM11}. However, it was pointed out in \citet{S99} that free-floating planets with thick atmospheres may exist in interstellar space with the appropriate conditions for surface life. A related proposal was advocated in \citet{AS11}, who suggested that free-floating potentially habitable Earth-sized planets with subsurface oceans may exist. Looking even further beyond, several authors have discussed the possibility of life based on alternative biochemistry \citep{Bains,BRC04,SI08,SLC15}. Thus, it is evident that life in the Universe has a vast range of niches that it could occupy, and worlds with subsurface oceans under ice envelopes constitute an important category.

Hence, we shall concern ourselves with the likelihood of life-as-we-know-it existing within subsurface oceans henceforth in our analysis. In Sec. \ref{SecIce}, we present a simple model for the thickness of the ice layer and examine the range of objects that can exist in different environments.\footnote{The object, which can be either free-floating or gravitationally bound, could refer to a planet, moon or planetoid, but we shall label it a ``planet'' henceforth to simplify our notation.} Next, we examine the energy sources for prebiotic chemistry on these planets and the potential routes to the origin of life in Sec. \ref{SecAbio}. We discuss the biological potential of these worlds in Sec. \ref{SecEco}, and determine the rate of biomass production through different avenues. In Sec. \ref{SecNumb}, we determine the total number of subsurface planets that may exist and delineate some of the consequences for panspermia and detection. We conclude with a summary of our main results in Sec. \ref{SecConc}.

\section{Icy worlds: temperature profile and habitats}\label{SecIce}
First, we will formulate a simple model for the thickness of the ice envelope, and identify certain ``habitats'' where icy worlds can exist.

\subsection{Temperature profile of icy worlds}

\begin{figure}
\quad\quad\quad \includegraphics[width=7.2cm]{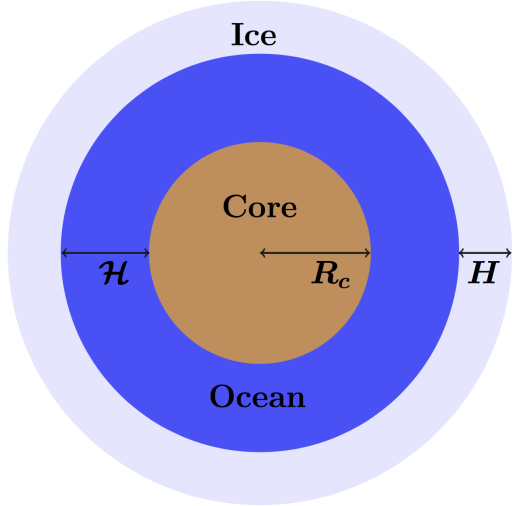} \\
\caption{Schematic illustration of the planet with an ice envelope, a liquid water ocean and an inner rocky/metallic mantle and core.}
\label{Fig0}
\end{figure}

In our analysis, we shall assume that the object under consideration comprises a surface ice layer with an subsurface ocean situated below. For the sake of simplicity, we do not consider worlds where the outer layers consist of both ice and rock, which are believed to exist on some Solar system satellites \citep{SASM,NiPa}.\footnote{One such example is Titan, which is believed to possess a subsurface ocean \citep{IJ12,MM14}, but we do not consider such moons because of their relatively complex interior structure. These worlds have been predicted to be fairly common around M-dwarfs at distances of $\sim 1$ AU \citep{Lun10}.} Our assumption of an ice envelope and a subsurface ocean implicitly assumes that the water content is sufficiently high to enable the existence of these two layers. However, it must be recognized that the water inventory of planets and satellites has been predicted to vary considerably \citep{RSM,MCMP,CMPA15,BBC15}, and hence the depth and existence of the subsurface ocean cannot be estimated \emph{a priori}.

The planet's heat flux is assumed to arise from a combination of radiogenic and primordial (gravitational contraction) heating; on Earth, it is known that the former contributes approximately $50\%$ of the total heat flux \citep{KC11}. The effects of tidal heating are assumed to be negligible in our model, unlike Europa and Enceladus \citep{BS09,HCL10,SN13,CTCB17} as well as some free-floating satellite-planet systems \citep{DS07}, where tidal dissipation is expected to play an important role. In addition, we do not explicitly consider the heating due to serpentinization reactions that is anticipated to be quite significant on small worlds like Enceladus and Mimas \citep{MP13}. A schematic figure of the planet has been depicted in Fig. \ref{Fig0}.

We shall determine the thermal profile for the outer ice layer by assuming that the heat is transported via conduction, and not through convection. The latter depends on a wide range of properties for rocky and icy planets, such as the mass, ice grain size, availability of water, and mantle rheology to name a few \citep{STO01,BMK07,BCFS} and the putative existence of bistable behavior, multiple steady states, mixed heating and temporal evolution only serves to complicate matters further \citep{LC12,Kor17}. In our own Solar system, it has been proposed that non-Newtonian creep mechanisms are responsible for convective shutdown on some dwarf planets and satellites with potential subsurface oceans \citep{McK06}, such as Callisto and Pluto. It has also been suggested that the inclusion of contaminants (e.g. ammonia and salts), which lower the melting point of ice, may influence the degree of convection \citep{AS11,TPS12}.\footnote{The presence of clathrate hydrates in the ice shell is predicted to reduce its thickness compared to the pure-ice case for a fixed value of the heat flux \citep{HCCC}.} Despite these complexities, provided that the shell's thickness or the heat flow is sufficiently high, thermal convection will occur on such objects, and the thickness of the ice layer can be estimated accordingly \citep{HSS06,FOCS}.

Given the above assumptions, the temperature profile is determined via Fourier's law
\begin{equation} \label{Four}
    \mathcal{Q} +  \kappa \frac{d T}{d r} = 0, 
\end{equation}
where $\mathcal{Q}(r)$ is the geothermal heat flux at radius $r$ (in units of W/m$^2$), $T \equiv T(r)$ denotes the temperature at radius $r$, and $\kappa$ is the thermal conductivity of ice. Note that $\kappa = \mathcal{C}/T$ with $\mathcal{C} \approx 651$ W/m based on Eq. (3.11) of \citet{PW99}. The heat flux $\mathcal{Q}(r)$ within the ice envelope can be expressed as
\begin{equation} \label{Hflux}
    \mathcal{Q} = \frac{Q}{4\pi r^2} \times \frac{r^3}{R^3} = \frac{Q\, r}{4\pi R^3},
\end{equation}
where $R$ is the radius of the planet, and $Q$ represents the total internal heat flow from the planet to space (in units of W). Thus, the heat flux is given by the ratio of the heat flow within the enclosed region $Q_\mathrm{enc}$ and the area of this region ($4\pi r^2$). Note that $Q_\mathrm{enc}$ is approximately equal to $Q \times \left(\frac{4\pi}{3}r^3/\frac{4\pi}{3} R^3\right)$ and the second factor follows from the assumption that the heating sources are uniformly distributed throughout the planet's volume. Furthermore, we will make use of the ansatz
\begin{equation} \label{Qans}
    Q = \Gamma Q_\oplus \left(\frac{M}{M_\oplus}\right)^\alpha,
\end{equation}
where $Q_\oplus$ is the internal heat flow of the Earth \citep{KC11} and $M$ is the mass of the planet, while $\Gamma$ and $\alpha$ are free parameters. Note that both $Q$ and $Q_\oplus$ are implicitly time-dependent, as radiogenic heating declines exponentially with time. However, if we restrict the discussion to planets that were formed a few Gyr ago, it can be verified that $Q$ changes merely by an order unity factor \citep{TS02} when only long-lived isotopes are taken into consideration. Our analysis has excluded short-lived isotopes \citep{NDCR}, but these elements can play a potentially important role in the thermal evolution of icy worlds.

Equation (\ref{Qans}) is a generalization of the standard convention wherein $\Gamma = 1$ and $\alpha = 1$. This choice amounts to stating that $Q/M$, namely the heating rate per unit mass (in W/kg) is constant \citep{VC09}. For radiogenic heating, $Q$ depends upon the mass of the rocky mantle - usually assumed to have a chondritic composition that contains the radioactive elements \citep{SS03} - which is not necessarily linearly proportional to the mass of the planet. Hence, there is no \emph{a priori} reason for assuming $\alpha = 1$. Similarly, since planets with metallicities different than that of Earth exist \citep{JA09,BB14}, it is therefore conceivable that the abundance of radionuclides may also vary accordingly; this variability is encapsulated in our model by means of the parameter $\Gamma$.\footnote{In referring to ``metallicity'' in this paper, we will work with the astrophysical definition, namely the mass fraction of elements other than hydrogen or helium. The quantity [Fe/H] is often used as a measure of metallicity, and it quantifies the logarithm of the ratio of Fe and H \emph{relative} to the Sun's ratio of Fe and H.} Alternatively, $\Gamma$ can also be used to encapsulate the degree of heating from other sources (e.g. tidal dissipation).

We turn our attention to the planetary radius which can be determined through a mass-radius relationship. It is not possible to identify a single scaling since it depends on the composition, and is not always a power-law \citep{SK07}. Nevertheless, for the sake of simplicity, we assume
\begin{equation}
    M = \lambda M_\oplus \left(\frac{R}{R_\oplus}\right)^\beta,
\end{equation}
where $R$ is the planet's radius, with $\lambda$ and $\beta$ representing free parameters. The value of $\beta$ is dependent both on the H$_2$O content and the mass, but $\lambda \approx 1$ in most cases. We shall use $\beta \approx 3.3$ for $M \lesssim M_\oplus$ \citep{SGM07} and $\beta \approx 3.8$ for $M \gtrsim M_\oplus$ \citep{VS07,FOCS}.

\begin{figure*}
$$
\begin{array}{cc}
  \includegraphics[width=7.2cm]{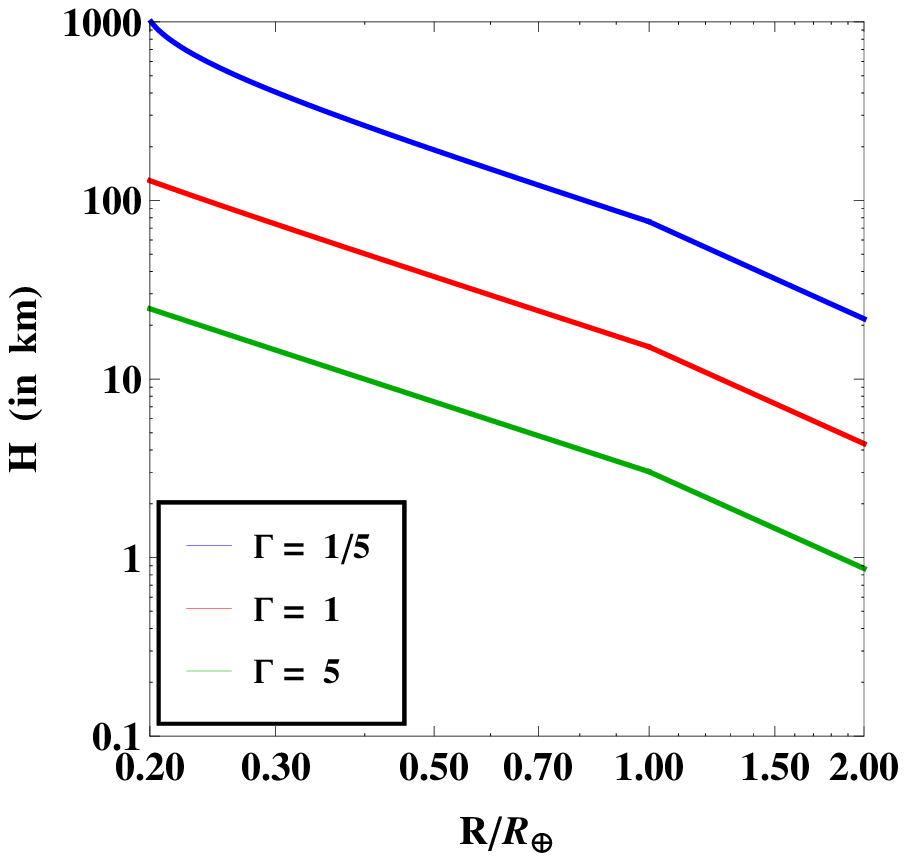} &  \includegraphics[width=7.2cm]{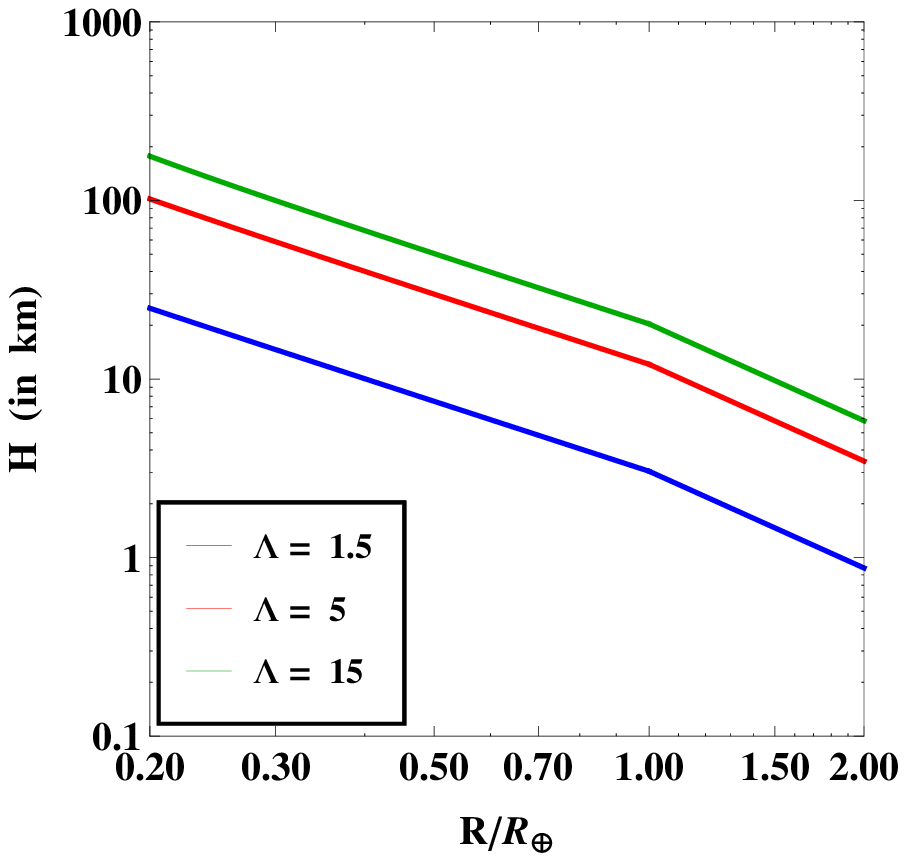}\\
\end{array}
$$
\caption{Left panel: thickness of the ice shell as a function of the planet's radius for differing radionuclide concentrations (represented by $\Gamma$) and a fixed surface temperature arising from Earth's radiogenic heating (corresponding to $\Lambda = 7.5$). Right panel: thickness of the ice envelope for differing surface temperatures (encapsulated by $\Lambda$) and a fixed radionuclide concentration ($\Gamma = 1$).}
\label{Fig1}
\end{figure*}

Next, we solve (\ref{Four}) by imposing the boundary condition $T(r=R) = T_s$, where $T_s$ denotes the temperature at the surface. The temperature profile is given by
\begin{equation} \label{Tsol}
    \ln \left(\frac{T}{T_s}\right) = \frac{Q \left(R^2 - r^2\right)}{8\pi \mathcal{C} R^3}.
\end{equation}
We are now in a position to determine the thickness of the ice layer. The phase-diagram of H$_2$O implies that the melting point of ice ranges between $\approx 250-270$ K provided that the pressure $P$ is lower than $620$ MPa \citep{CG07}. As the temperature-dependence is logarithmic in nature, we can assume a melting point of $T_m = 260$ K for pure ice. In contrast, the presence of ammonia (as a contaminant) can lower the melting point of ice to the peritectic temperature of $176$ K; this corresponds to a $33\%$ NH$_3$ concentration in the ocean \citep{LMN}. The value of $r$ at which $T = T_m$, denoted by $R_m$, is
\begin{equation}
    R_m = R \left[1 - \frac{8\pi \mathcal{C} R}{Q} \ln \left(\frac{T_m}{T_s}\right)\right]^{1/2},
\end{equation}
and the thickness $H$ of the ice layer is determined via $H = R - R_m$. Thus, we obtain
\begin{equation} \label{Height}
    \frac{H}{R_\oplus} =  \left(\frac{R}{R_\oplus}\right) \left(1 - \left[1 - 2.4 \times 10^{-3}\,\frac{\ln \Lambda}{\Gamma}\left(\frac{R}{R_\oplus}\right)^{1-\gamma}\right]^{1/2}\right),
\end{equation}
where we have introduced the auxiliary notation $\gamma = \alpha \beta$ and $\Lambda = T_m/T_s$. By inspecting this formula, it is evident that $H$ will decrease whenever $\Lambda$ is reduced or $\Gamma$ is increased for fixed values of $R/R_\oplus$ and $\gamma$. Hence, provided that $D > H$, where $D$ denotes the average water depth, the planet can host sub-surface oceans. As noted earlier, it is not possible to quantify $D$ beforehand since the water content of the planet can vary widely.

As an example, let us suppose that the Earth were to be ejected into space. We obtain the surface temperature upon solving $\sigma T_s^4 = \mathcal{F}_\oplus$, where $\mathcal{F}_\oplus = 0.087$ W/m$^2$, and find that $T_s \approx 35$ K. Hence, it follows that $\ln \Lambda \approx 2$ if we use $T_m \approx 260$ K. Using this value in (\ref{Height}) along with $\Gamma = 1$ and $R = R_\oplus$, we find $H \approx 15$ km. The Earth's average water depth $D_\oplus$ is $\approx 3.7$ km \citep{CS10}. Hence, it seems plausible that the Earth's oceans would be entirely frozen as per this model;\footnote{However, even if Earth's oceans were wholly frozen, the possibility of chemolithoautotrophic life in the deep biosphere (e.g. hydrated regions of subduction zones) ought not be discounted \citep{PKG17}.} see also \citet{LA00} for a related discussion of this question. If the Earth was ejected shortly after its formation, the geothermal heat flux could have been higher by a factor of order unity due to short-lived radionuclides and primordial heat. Hence, provided that the criterion $\Gamma \gtrsim 4$ was valid, the primordial Earth might have retained a global subsurface ocean if it had become a free-floating planet. For this value of $\Gamma$, we obtain $H \approx 3.75$ km for the Earth, and the condition $D_\oplus > H$ is satisfied since $D_\oplus$ for the early Earth was approximately twice its present-day value \citep{Kor08}. 

In Fig. \ref{Fig1}, we have plotted $H$ as a function of the radius for different values of the free parameters $\Gamma$ and $\Lambda$. If we consider the case where the second term inside the square brackets of (\ref{Height}) is much smaller than unity, we can use the binomial theorem to obtain the following expression:
\begin{equation}\label{SimpH}
    H \approx 7.6\,\mathrm{km}\,\frac{\ln \Lambda}{\Gamma}\left(\frac{R}{R_\oplus}\right)^{2-\gamma},
\end{equation}
and, for $\alpha = 1$ in conjunction with the values of $\beta$ discussed earlier, the condition $\gamma > 3$ is always satisfied. Hence, the thickness will be a monotonically decreasing function of the radius. Before proceeding further, it must be recognized that (\ref{Height}) is not valid for all values of $R$. This limitation arises since the quantity inside the square brackets must be positive (to ensure that that its square root is real). Hence, we find that the model has a lower bound (denoted by $R_c$) given by
\begin{equation}
    \frac{R_c}{R_\oplus} = \left(2.4 \times 10^{-3}\,\frac{\ln \Lambda}{\Gamma}\right)^{1/(\gamma-1)},
\end{equation}
and for the choices $\ln \Lambda \approx 2$ and $\Gamma = 1$, we obtain $R_c \approx 0.1\,R_\oplus$. The value of $R_c$ increases monotonically with $\Lambda$ and decreases monotonically with $\Gamma$. On the other hand, we can assume that $\mathcal{Q}$ is constant in (\ref{Four}) along the lines of \citet{AS11}. We find that the solution resembles (\ref{SimpH}) except for the additional factor of $\Gamma$ and a different power-law exponent. 

Our analysis is not expected to yield accurate results for Europa and Enceladus since the effects of tidal dissipation are not included in our model. For instance, if we employ (\ref{Height}) with $\Gamma = 1$ and use $\Lambda \approx 2.5$ for Europa, the resultant value of $H$ is about a factor of $4$ higher compared to more sophisticated predictions \citep{Lun17}.

\subsection{Habitats for potential subsurface oceans}\label{SSecHab}
We briefly discuss some of the cases where subsurface oceans might exist.\\

{\bf Type B} (involving objects {\bf bound} to a star): When planets lie beyond the HZ of their host star, they can host sub-surface oceans \citep{EC07,FOCS}. Examples within our Solar system include Europa and Enceladus, and it has been posited that water-rich extrasolar planets \citep{LS04} such as OGLE 2005-BLG-390Lb may also fall under this category \citep{EL06}. Two important points are worth noting at this stage concerning this class of planets.
\begin{itemize}
    \item The value of $\Lambda$ can vary by an order of magnitude because $\Lambda \approx 1-10$. For instance, the average (morning/evening) surface temperature of Ceres is $168$ K \citep{KOR14}, which leads to $\Lambda \approx 1.5$ upon choosing $T_m \approx 260$ K; a minimum of $\Lambda \approx 1.1$ can be attained for Ceres at the subsolar point. Recent observations of carbonate deposits in the Occator crater on Ceres \citep{DS16} and the detection of volatiles \citep{RRA16,CMT16} lend further credence to the possibility that this dwarf planet has/had a subsurface ocean on account of its relatively high surface temperature \citep{CRM10}. On the other hand, for trans-Neptunian objects (TNOs) like Eris, it can be shown that $\Lambda \approx 10$.
    \item The HZ need not necessarily correspond only to a main-sequence star. Exoplanets with subsurface oceans (lying outside the HZ) could also exist around white dwarfs and brown dwarfs \citep{Ag11,BH13,LM13}, as well as post-main-sequence \citep{LL97,LSD05,RK16} and pre-main-sequence \citep{RK14,LB15} stars.
\end{itemize}
If the Type B planet is not very far from its host star, the surface temperature can be estimated via
\begin{equation}
    T_s \approx 213\,\mathrm{K}\,\left(\frac{1-A}{0.36}\right)^{1/4}\left(\frac{L_\star}{L_\odot}\right)^{1/4} \left(\frac{a}{1\,\mathrm{AU}}\right)^{-1/2},
\end{equation}
where $A$ is the albedo of the planet, $L_\star$ denotes the luminosity of the host star, and $a$ is the average star-planet distance. We have normalized $1-A$ by $0.36$ since this happens to be the value for Europa. For main-sequence stars, note that $L_\star/L_\odot \propto \left(M_\star/M_\odot\right)^{\xi}$ is commonly used, where $M_\star$ represents the mass of the host star. The value of $\xi$ is dependent on the stellar mass range and is approximately equal to $3.1$ for $0.08 M_\odot < M_\star < 0.25 M_\odot$, $4.5$ for $0.25 M_\odot < M_\star < 0.75 M_\odot$ and $3.5$ for $0.75 M_\odot < M_\star < 3 M_\odot$ \citep{LBS16}.\\

{\bf Type U} (involving {\bf unbound} objects): When objects are ejected from the planetary system by means of gravitational interactions with giant planets \citep{RF96,PT06}, they end up as ``rogue'' (i.e. free-floating) planets. Most studies concerning the habitability of these planets have focused on surface-based life-as-we-know-it \citep{S99,DS07,Bad11}, but it is also possible that they can host subsurface oceans with ice/rock envelopes \citep{LA00,AS11}.

At first glimpse, it is tempting to conclude that $\Lambda \approx 10$ must hold true in most instances because of the low surface temperature. However, there might exist certain environments where $\Lambda \sim 1$ becomes feasible. We consider one such example in detail, namely galactic nuclei. During their quasar phase, the luminosity of supermassive black holes $L_\mathrm{BH}$ approximately equals the Eddington luminosity ($L_\mathrm{Edd}$) \citep{KL14}, and we have
\begin{equation}
    L_\mathrm{BH} \approx L_\mathrm{Edd} = 1.3 \times 10^{37}\,\mathrm{W}\,\left(\frac{M_\mathrm{BH}}{10^{6}\,M_\odot}\right),
\end{equation}
where $M_\mathrm{BH}$ is the mass of the supermassive black hole. We will choose $T_s = 100$ K as a fiducial value, which leads to $\Lambda \approx 2$ when the ice layer has a few percent ammonia.
If we denote the distance from the supermassive black hole by $\mathcal{R}$,
\begin{equation}
    \sigma T_s^4 \approx \frac{L_\mathrm{BH}}{4\pi \mathcal{R}^2},
\end{equation}
thereby yielding,
\begin{equation}
    \mathcal{R} \approx 12\,\mathrm{pc}\, \left(\frac{L}{10^{37}\,\mathrm{W}}\right)^{1/2} \left(\frac{T_s}{100\,\mathrm{K}}\right)^{-2}.
\end{equation}
Although this distance is $2$-$3$ orders of magnitude smaller than the inner edge of the conventional Galactic Habitable Zone (GHZ) which is a few kpc from the Galactic centre \citep{LFG04,FDCL} - see, however, \citet{Pran08} and \citet{MG15} - it must be recognized that the GHZ presupposes the existence of host stars and surficial life. The presence of a sufficiently thick ice layer could potentially shield the planet from ionizing radiation from supernovae, Gamma Ray Bursts (GRBs) as well as other astrophysical catastrophes \citep{Dart11,MT11}, and effects like atmospheric erosion due to hydrodynamic escape \citep{FL17,BT17} would also be rendered irrelevant.\footnote{The erosion rate of Europa's ice layer due to sputtering by energetic ions is $\sim 20$ m/Gyr \citep{CJM01}. Hence, unless the sputtering rate near the Galactic centre is $2$-$3$ orders of magnitude higher, its effects on the ice layer are relatively insignificant over Gyr timescales.} However, at such close distances, it is possible that gravitational interactions play a disruptive role \citep{GEG10}. The characteristic quasar lifetime is $\sim 10^7-10^8$ years \citep{Mar04}, after which the surface temperature would drop and the thickness of the ice envelope would increase by a factor of a few. Nevertheless, even a relatively short interval of time ($\sim 10^7$ yrs) might have sufficed for the origin of life on Earth \citep{OF89,LM94,LD02} and some worlds may be characterized by similar timescales for the origin of microbial life.\footnote{Yet, it is equally important to recognize that a great deal remains unknown about the pathways and timescales for abiogenesis on Earth and other habitable planets \citep{Org98,ST12}.}

Giant molecular clouds at the Galactic centre (e.g. Sagittarius B2) possess a wide range of organic molecules at relatively high concentrations \citep{HVD13}, whose putative significance in prebiotic chemistry has been extensively investigated. Examples include hydrogen cyanide \citep{JBC}, aldehydes \citep{HLJ00,RT08}, and nitriles \citep{BM08,BM13}.\footnote{\citet{KC03} claimed to have detected glycine in the Central Molecular Zone, but their evidence does not appear to be supported by subsequent studies \citep{SL05,CJ07}.} Hence, any Type B or Type U planets existing in these regions may be characterized by $\Lambda \sim 1$ and also have access to these prebiotic molecules, although they will subsequently have to be transported across the ice shell into the ocean. 

A second avenue for achieving $\Lambda \sim 1$ is through the cosmic microwave background (CMB). The CMB energy density is redshift dependent, and we determine the surface temperature by equating the CMB energy flux to $\sigma T_s^4$. Upon doing so, we arrive at
\begin{equation} \label{TCMB}
    T_s \approx 82\,\mathrm{K}\,\left(\frac{1+z}{30}\right)
\end{equation}
where $z$ is the redshift \citep{Wein08}. For $z \approx 30$, we obtain $\Lambda \approx 2$ when the ice has contaminants. Hence, planets that formed during this epoch are expected to have relatively thinner ice envelopes and the likelihood of subsurface life originating at this juncture should not be ruled out; our proposal is very akin to the idea that surface life was possible at $z \sim 100$ \citep{Loeb14} when $T_s \sim 273$ K. A few of the challenges for abiogenesis in the early Universe are described below.
\begin{itemize}
    \item The first stars must have formed, and seeded the Galaxy with metals through supernovae. The first stars are expected to have formed at $z \lesssim 30$ \citep{LF13,Bro13}, and hence this condition could have been satisfied.
    \item Although exoplanets have been observed around stars with a wide range of metallicities \citep{BL12}, the formation of planetesimals has been predicted to depend on the metallicity \citep{JYM}. Hence, the typically low-metallicity environment of the early Universe may have posed difficulties for planet formation \citep{Lin01,JL12}, although some theoretical models suggest that the first terrestrial planets were capable of forming $\approx 13$ Gyr ago \citep{BP15,ZCG16}.
    \item C, H, N, O, P and S are necessary for life-as-we-know-it, implying that they must be available in sufficient quantities. The putative existence of carbon-enhanced metal-poor (CEMP) planets at high redshifts indicates the potential availability of C \citep{MaLo}, but the abundance of P and S on high-redshift planets remains poorly constrained.
    \item The formation of water in molecular clouds, and its subsequent delivery to protoplanetary disks and planets, is an essential requirement \citep{VDB14}. Since water vapour is believed to have been abundant even at low metallicity ($\sim 10^{-3}$ of the solar value) in these clouds \citep{BSL15}, it suggests that life at high redshifts, only insofar the availability of water is concerned, ought not be ruled out.
    \end{itemize}
If we consider the current epoch, the energy densities of Galactic interstellar radiation, cosmic rays and the CMB are similar \citep{Ferr01}, and collectively yield a surface temperature of a few K; the value of $T_s$ cannot drop below this value. Lastly, for most Type U planets, the surface temperature is set by the geothermal heat flux, and is given by
\begin{equation}
    T_s \approx 35\,\mathrm{K}\,\,\Gamma^{1/4}\left(\frac{R}{R_\oplus}\right)^{(\gamma-2)/4}.
\end{equation}

\section{Energy sources and paths for abiogenesis}\label{SecAbio}
Previously, we have seen that free-floating (and bound) planets with subsurface oceans and ice envelopes can exist. However, the availability of liquid water is evidently only a necessary, but \emph{not} sufficient, condition for the planet to be habitable (or inhabited). First, it must be noted that there exist several other factors that must be taken into account in analyses of habitability from a biological standpoint \citep{CB16}. Second, even when liquid water is present, there exist additional stringent constraints set by water activity, chaotropicity, ionic strength, temperature and pressure \citep{PD13,BH15,FHCC,LiLo17} since it ought not be regarded merely as a passive background \citep{Ball08,BHH16}. These reasons collectively serve to explain why the \emph{majority} of Earth's aquasphere (about $88\%$) is ``not known to host life'' \citep{JL10}.

Hence, it is more instructive to adopt alternative approaches, such as the ``follow the energy'' strategy \citep{HAS07,SH07}, instead of the ``follow the water'' \emph{modus operandi}; the former may also have the advantage of addressing, to some degree, the prospects for life based on non-standard biochemistries.\footnote{In our terminology, life that is not based on carbon and does not involve water as the solvent is taken to be composed of ``non-standard biochemistry''.} In this section, we will explore energy considerations and possible routes available for the origin of life (abiogenesis) that must be taken into consideration for assessing the likelihood of these planets to support simple or complex biospheres. 

\subsection{Energy sources for prebiotic synthesis}\label{SSecEnS}
There remains a great deal that is unknown about the processes that led to abiogenesis on Earth \citep{Fry,RBD14,Lu16} although it is likely that no single microenvironment or physicochemical process was responsible for the emergence of life on Earth \citep{DL05,Spit17}.\footnote{Note that life need not have originated on Earth \emph{in situ}, and could have been transported from elsewhere by means of panspermia, as discussed further in Sec. \ref{SSecPanS}.} Despite these inherent uncertainties, the availability of free energy \citep{Sch44,Deam97,Dys99,PPS13,SW17} is widely regarded as a necessary requirement for abiogenesis to take place. Other basic requirements include: (i) raw materials, (ii) suitable solvent, and (iii) appropriate environmental conditions \citep{Ho04,Ho07}. As we focus only on planets where subsurface oceans can exist, (ii) is automatically satisfied. Although (i) and (iii) are undoubtedly important, they are also harder to quantify, and we shall assume these criteria are fulfilled in our subsequent analysis.

On Earth, ultraviolet (UV) radiation has been identified as one of the most dominant energy sources for enabling prebiotic reactions \citep{DW10}. There are several lines of evidence indicating that UV radiation played an important role in prebiotic chemistry on Earth \citep{Pat15,RV16,Suth17}. By considering the UV radiation with wavelength $< 200$ nm emitted by the host star, we can heuristically evaluate the far-UV flux received by the planet in this range (denoted by $\Phi_\mathrm{UV}$). Upon assuming that the Ly$\alpha$ emission serves as an approximate proxy for FUV radiation, we obtain for Type B planets,
\begin{equation} \label{PhUV}
    \Phi_\mathrm{UV} \sim 10^{6}\,\mathrm{J\,m^{-2}\,yr^{-1}} \left(\frac{a}{1\,\mathrm{AU}}\right)^{-2} \left(\frac{M_\star}{M_\odot}\right)^\nu,
\end{equation}
where $\nu \approx 1.2$ for $M_\star \lesssim M_\odot$ and $\nu \approx 6.8$ for $M_\odot < M_\star \lesssim 2\,M_\odot$ following the scaling relations in \citet{LL17}. We note that the power-law exponent is not constant since a significant portion of the FUV emission from low-mass stars (especially M-dwarfs) is from the chromosphere region \citep{FF13} in contrast to solar-type stars. The above formula implies that larger planets at closer distances around higher-mass stars (but outside the HZ) may be more conducive to prebiotic synthesis. In reality, it must be recognized that $\Phi_\mathrm{UV}$ also depends on other stellar parameters such as the rotation rate \citep{LFA13}.

Several studies have focused on irradiating interstellar ice analogs at low temperatures with UV radiation \citep{Ob16} - often through a flowing-hydrogen discharge lamp, whose output is divided between the Ly$\alpha$ line and a $20$ nm band centered around $160$ nm - suggesting that our use of the Ly$\alpha$ proxy in (\ref{PhUV}) could be a reasonable assumption since a significant fraction of the FUV flux in laboratory experiments comprises of Ly$\alpha$ photons. Some of the organic molecules thus synthesized include amino acids (e.g. alanine, glycine, serine), and the RNA/DNA nucleobases \citep{MC02,ED07,TT07,NAB08,NM12,MNS17}. Type U planets, which are free-floating, can traverse through interstellar regions close to O/B-type stars and receive high (but transient) doses of UV radiation leading to the rapid formation of biologically relevant molecules \citep{Roop}. Using (\ref{PhUV}) and the quantum yield of $\approx 10^{-4}$ for the pathway studied in \citet{MC02}, the mass of amino acids ($\mathcal{M}_\mathrm{UV}$) produced per unit time is
\begin{equation}
    \mathcal{M}_\mathrm{UV} \sim 10^{10}\,\mathrm{kg/yr}\,\left(\frac{R}{R_\oplus}\right)^2\left(\frac{a}{1\,\mathrm{AU}}\right)^{-2} \left(\frac{M_\star}{M_\odot}\right)^\nu.
\end{equation}
The production rates of organics determined in the paper ought not be extrapolated \emph{ad infinitum} since they are clearly contingent on the availability of the appropriate reactants,\footnote{Insofar interstellar ices are concerned, it has been concluded recently that they are abundantly available to most young planetary systems \citep{CS12,CBA14}.} and will also be subject to decomposition by ionizing radiation and other processes. It should also be noted that the above value should be taken with due caution since the quantum yield used is pathway-dependent. Finally, laboratory experiments are undertaken in controlled environments endowed with a plentiful supply of the requisite ingredients, implying that our estimates are likely to be upper bounds. 

A few caveats regarding UV radiation are in order here. We begin by noting that many of laboratory experiments already presuppose the existence of ``feedstock'' organic molecules and typically operate at temperatures that are a few tens of K. There is no guarantee that either these molecules are present, or that the same reactions can function at higher temperatures; for methanol-rich ices, these two factors are likely to be more important than the UV flux and the thickness of the ice monolayers \citep{OG09}. It must also be recognized that UV photolysis can not only facilitate the formation of these compounds but also aid in their decomposition; for instance, at $\sim 100$ K it has been shown that the concentration of amino acids (within the uppermost meter of a pure ice layer) will be halved in a span of $\sim 10$ yrs \citep{OGJ07}. Hence, any organic molecules deposited or formed on the icy surface must be eventually transported to the subsurface ocean.

In addition, some of the pathways studied in the laboratory (mimicking Earth-like conditions) driven by UV irradiation presuppose the existence of liquid water \citep{SCCD,Ba13,Mill13,Pat15}, and the latter does not exist on the surface of either Type B or Type U worlds on a long-term basis. However, transient water could exist as a result of tectonic processes, cryovolcanism, and impact cratering events \citep{KK00}. With regards to the latter, the total energy required to melt the ice shell completely ($E_m$) is
\begin{equation}\label{LatHeat}
    E_m \approx 1.6 \times 10^{26}\,\mathrm{J}\, \left(\frac{R}{R_\oplus}\right)^2 \left(\frac{H}{1\,\mathrm{km}}\right),
\end{equation}
which is determined from multiplying the latent heat of fusion for ice with the total mass of the ice shell. By equating (\ref{LatHeat}) with the kinetic energy of the asteroid, we determine find its mass as follows:
\begin{equation}
    E_m = \frac{1}{2} M_a \left(v_e^2 + v_\infty^2\right),
\end{equation}
where $M_a$ is the asteroid mass, $v_e = \sqrt{2 G M/R}$ is the escape velocity of the planet, while $v_\infty$ is determined from {\"O}pik's theory of gravitational encounters, and has typical values of a few km/s \citep{SAL}. For an Earth-sized planet with an ice envelope that is a few kms thick, the required asteroid mass is $\mathcal{O}\left(10^{18}\right)$ kg. In our Solar system, only a few asteroids (e.g. Vesta and Pallas) exceed this mass. Even if the ice envelope is completely obliterated, organisms in the deep (subsurface) ocean may still be able to avoid extinction, and can undergo rapid evolutionary diversification afterwards in some instances \citep{Al08,GGH17}.

Next, the energy flux from the CMB, which is a function of the redshift, is found to be
\begin{equation}\label{PhiCMB}
\Phi_\mathrm{CMB} \sim 4.6 \times 10^2\,\left(1+z\right)^4\,\mathrm{J\,m^{-2}\,yr^{-1}},
\end{equation}
where $\Phi_\mathrm{CMB}$ signifies the CMB energy flux. It is evident that the dependence on the redshift is quite strong; for $z \sim 30$, we find that $\Phi_\mathrm{CMB}$ increases by almost six orders of magnitude compared to the present-day value. However, even at such high redshifts, the peak wavelength of the CMB radiation lies in the far-infrared (far-IR). At such low energies, it appears unlikely that this energy could be effectively utilized for enabling prebiotic chemistry. On the other hand, the surface temperature (\ref{TCMB}) is governed by the CMB radiation, and can therefore facilitate the existence of subsurface oceans at sufficiently high redshifts since $H$ will decrease when the value of $z$, and therefore $T_s$, is increased.

Other energy sources for prebiotic chemistry include electrical discharges, shock waves from impacts \citep{MPG13,FNS15} and volcanism. We need not consider the first two sources because the corresponding prebiotic synthesis has been shown to occur in the atmosphere \citep{RBD14}, and we have assumed that these planets either lack an atmosphere altogether or possess a very tenuous one. Similarly, estimating the energy flux of cryovolcanism is not straightforward since it remains poorly constrained for icy worlds like Europa \citep{Fa03,SH16}. The next source that we consider is radioactivity, whose surficial flux is denoted by $\Phi_\mathrm{rad}$. Upon utilizing (\ref{Hflux}) and (\ref{Qans}), we end up with
\begin{equation} \label{PhiR}
    \Phi_\mathrm{rad} \sim 2.7 \times 10^6\,\mathrm{J\,m^{-2}\,yr^{-1}}\,\Gamma\,\left(\frac{R}{R_\oplus}\right)^{\gamma-2}.
\end{equation}
Several theoretical and experimental analyses suggest that radioactivity, especially in the form of naturally occurring surficial nuclear reactors \citep{DDA,Ad07,Ada16}, played a potentially important role in the origin of life on Earth \citep{ACC87,GG01,Zag03,Par04,AHC18}. An interesting point worth mentioning in this context is the putative existence of organisms on Type B and U planets, such as \emph{Desulforudis audaxviator} on Earth \citep{CB08}, that derive their energy from radioactive decay.

Since we are not aware of any \emph{in situ} experiments that have yielded the G-values for the prebiotic synthesis of amino acids through natural radioactivity,\footnote{The G-value represents the number of molecules formed as products of the chemical reaction when $100$ eV of energy has been supplied.} we must resort to an indirect strategy for estimating the yields of biomolecules. As the alpha and beta particle decay mechanisms are akin to irradiation by energetic protons and electrons respectively, we will posit that their efficiencies are similar. Several studies have been undertaken in connection with the synthesis of organic compounds by means of energetic particles with energies of KeV-MeV. A wide range of organic molecules such as hydrogen cyanide, aldehydes, formamide, amino acids and nucleosides have been synthesized either directly or after acid hydrolysis \citep{GMH04,BK07,HM08,CC10,KK11,SC15}; here, we note that the possible importance of formamide in abiogenesis has been extensively studied \citep{SC12}.

\citet{KK95} irradiated cometary ice analogs at $77$ K with $3$ MeV protons, and obtained a G-value of $\sim 10^{-4}-10^{-5}$ for their \emph{specific} setup that yielded amino acids upon hydrolysis. We will work with the smaller value since we anticipate the efficiency of natural radioactivity-mediated synthesis to be lower. Hence, we find that the mass rate of amino acids synthesized ($\mathcal{M}_\mathrm{rad}$) is
\begin{equation}
    \mathcal{M}_\mathrm{rad} \sim 6.8 \times 10^4\,\mathrm{kg/yr}\,\Gamma\,\left(\frac{R}{R_\oplus}\right)^{\gamma-1}\left(\frac{H}{1\,\mathrm{km}}\right)
\end{equation}
An important point worth bearing in mind concerning the prior discussion is that the G-values, which are dependent on the radiation dose, are non-constant \citep{BLP02}. Furthermore, as with UV radiation, high-energy particles contribute to both the formation and destruction of organic molecules. Lastly, $\mathcal{M}_\mathrm{rad}$ represents an upper bound since we have assumed that all of the energy from radioactive heating in the ice layer, roughly approximated by $\sim Q \times \left(4\pi R^2 H\right)/\left(4\pi R^3/3\right) \sim Q \times \left(3H/R\right)$, is available for prebiotic synthesis. In actuality, amino acids and other organics are likely to be produced through radiolysis only in local environments where radionuclides occur in high concentrations (e.g. natural fission reactors).

Next, we turn our attention to another energy source: energetic particles. As noted previously, they produce a wide variety of biologically relevant compounds. It must be noted that there exist three sources of energetic particles for Type B planets, but just one for Type U planets. Stellar Energetic Particles (SEPs) and energetic particles emanating from (giant) planetary magnetospheres are unique to Type B, while Galactic Cosmic Rays (GCRs) are common to both Type B and Type U. In order for particles from planetary magnetospheres to be a significant energy source, which is expected to be true for Europa \citep{BB15}, it follows that our Type B ``planet'' must be a moon. 

We will start by determining the flux $\Phi_\mathrm{GP}$ received by a moon orbiting a Jupiter-analog; if we consider a Saturn-analog instead, the value of $\Phi_\mathrm{GP}$ becomes much lower and is dependent on complex magnetospheric physics that will not be studied here \citep{Crav04}. From the data provided in Table II of \citet{CJM01}, we obtain
\begin{equation} \label{PhiGP}
    \Phi_\mathrm{GP} \sim 4 \times 10^6\,\mathrm{J\,m^{-2}\,yr^{-1}}\,\left(\frac{a_m}{4.5 \times 10^{-3}\,\mathrm{AU}}\right)^{-2},
\end{equation}
where $a_m$ is the distance from the moon to the giant planet, the normalization factor $4.5\times 10^{-3}$ AU represents the Europa-Jupiter distance, and we have assumed that the energetic particle flux obeys an inverse square-law behaviour \citep{FSWG}; note that this assumption works well for the ratio of particle fluxes at Ganymede and Callisto \citep{CJM01}. Using the G-values from \citet{KK95}, we find
\begin{equation}
    \mathcal{M}_\mathrm{GP} \sim 1.9 \times 10^8\,\mathrm{kg/yr}\,\left(\frac{R}{R_\oplus}\right)^{2}\left(\frac{a_m}{4.5 \times 10^{-3}\,\mathrm{AU}}\right)^{-2},
\end{equation}
where $\mathcal{M}_\mathrm{GP}$ is the mass rate of amino acids synthesized through bombardment of the icy surface by energetic particles from the giant planet's magnetosphere. 

It is difficult to evaluate the SEP energy flux for two reasons. First, the physics behind SEPs is complex and the integrated fluence is dependent on the particle acceleration mechanisms and the sites of origin \citep{Ream13}. Second, the SEP flux is highly variable since it depends on the stellar age, mass and rotation; it is expected to be significant for low-mass stars with high activity and close-in planets \citep{LDF18,YF17}, but it may not prove to be a dominant player for middle aged G-type stars like the Sun. We turn our attention to the GCR flux $\Phi_\mathrm{CR}$ near Earth, which is estimated to be
\begin{equation}
    \Phi_\mathrm{CR} \sim 4.6 \times 10^2\,\mathrm{J\,m^{-2}\,yr^{-1}},
\end{equation}
based on the value provided in \citet{KK98}. It can be seen from (\ref{PhiCMB}) that the GCR energy flux is approximately equal to the CMB energy flux at $z=0$. Although this value may appear to be small, it is worth recalling that the cosmic-ray flux increases towards the Galactic centre and, more importantly, constitutes one of the few sources that is universally accessible to Type B and U planets. The corresponding mass $\mathcal{M}_\mathrm{CR}$ of amino acids produced per unit time is
\begin{equation}
     \mathcal{M}_\mathrm{CR} \sim 1.7 \times 10^4\,\mathrm{kg/yr}\,\left(\frac{R}{R_\oplus}\right)^{2}.
\end{equation}

Another important source of prebiotic compounds is the exogenous delivery of organic molecules via interplanetary dust particles (IDPs), comets and meteorites \citep{Pizz06,Mum11}. From Fig. 1 of \citet{CS92}, it is evident that the delivery rates of organics for IDPs are $\sim 3$ orders of magnitude greater than comets and $\sim 5$ orders of magnitude higher than meteorites, although the latter (carbonaceous chondrites in particular) tend to be very rich in organics \citep{Sep02,CSC11,PS17} and can facilitate selective catalysis \citep{RTR16}. For Europa, the average organic delivery rate of $\sim 10^3-10^4$ kg/yr \citep{PC02} through cometary impacts is approximately consistent with the corresponding rate of $\sim 10^3-10^6$ kg/yr for Earth \citep{CS92}. It is also evident that exogenous delivery of prebiotic compounds via comets and meteorites does not apply to Type U planets.

Hence, we shall restrict our attention to considering exogenous delivery of organics via IDPs. One crucial point worth mentioning here is that the `soft landings' of IDPs on the planetary surface has been predicted to require a sufficiently thick atmosphere \citep{CP02}, and it is therefore unclear as to whether IDPs could accumulate on the surface when the atmosphere is either absent or rarefied. However, for the sake of completeness, it is still instructive to estimate the mass of organic molecules delivered by IDPs. Clearly, a universal mass accretion rate for all Type B and U planets is not feasible. However, we suggest that the following expression for the mass accretion rate $\dot{M}$ constitutes a reasonable approximation:
\begin{equation}
    \dot{M} \approx 4 \pi R_\mathrm{max}^2\, \rho_d\, \sigma,
\end{equation}
where $\sigma = \sqrt{V^2 + c_s^2}$, $\rho_d$ is the density of the ambient dust particles, $c_s$ denotes the sound speed, and $V$ is the relative velocity between the object and the dust. Here, $R_\mathrm{max} = \mathrm{max}\{R,R_B\}$, where $R_B$ is the modified Bondi radius \citep{Bond52} defined as $R_B = G M/\sigma^2$. It must be noted here that the accretion of IDPs can occur even when bolides have been decoupled from the gas after the dispersion of the solar nebula. If we consider the scenario where gas is not present, $\sigma$ should be replaced with $V$, and in this regime $R_B$ becomes the Hoyle-Lyttleton radius \citep{HL39}.

For $R_\mathrm{max} = R$, we obtain the geometric mass accretion rate. In contrast, for the case $R_\mathrm{max} = R_B$, gravitational focusing leads to the Bondi-Hoyle-Lyttleton accretion rate used in many fields of astronomy \citep{Ed04}. We assume that the accretion rate of organics from IDPs, represented by $\mathcal{M}_\mathrm{DP}$, is proportional to $\dot{M}$. For Earth, we will choose an organic deposition rate of $\sim 5.7 \times 10^{7}$ kg/yr; this value is based on the geometric mean of the present-day estimate and the rate at $4.4$ Gya that was $\sim 3$ orders of magnitude higher \citep{CS92}. Consequently, we can express $\mathcal{M}_\mathrm{DP}$ as
\begin{eqnarray}
    &&\mathcal{M}_\mathrm{DP} \sim 5.7 \times 10^7\,\mathrm{kg/yr}\,\left(\frac{R}{R_\oplus}\right)^{2\beta} \left(\frac{\sigma}{26\,\mathrm{km/s}}\right)^{-3} \nonumber \\
    && \hspace{0.5 in} \times \left(\frac{\rho_d}{2 \times 10^{-24}\,\mathrm{kg/m^3}}\right)
\end{eqnarray}
when $R_\mathrm{max} = R_B$ holds true. The dust density near the heliosphere is $\sim 2 \times 10^{-24}$ kg/m$^3$ and the relative inflow velocity (which dominates over the sound speed) of the dust is $\sim 26$ km/s \citep{GG94,KS15}. Although the values for $\sigma$ and $\rho_d$ in the ISM \citep{MK00} can be quite different, we anticipate that the planetary radius will play the most dominant role in governing the magnitude of $\mathcal{M}_\mathrm{DP}$. In contrast, for the case $R_\mathrm{max} = R$,  we find that $\mathcal{M}_\mathrm{DP}$ is given by
\begin{eqnarray}
    &&\mathcal{M}_\mathrm{DP} \sim 5.7 \times 10^7\,\mathrm{kg/yr}\,\left(\frac{R}{R_\oplus}\right)^{2} \left(\frac{\sigma}{26\,\mathrm{km/s}}\right) \nonumber \\
    && \hspace{0.5 in} \times \left(\frac{\rho_d}{2 \times 10^{-24}\,\mathrm{kg/m^3}}\right).
\end{eqnarray}
Before moving on, we note that $\mathcal{M}_\mathrm{DP}$ quantifies the total amount of organics delivered (not just amino acids), and should be viewed as an upper bound.

Lastly, we turn our attention to the abiotic production of amino acids from hydrothermal vents that have attracted much attention as one of the potential sites for abiogenesis \citep{BH85,MB08,SHW16} through the abiotic synthesis and polymerization of prebiotic compounds \citep{MCS07,BWD07,BBS09,RHM,STL13,HOM15}. In order for hydrothermal vents to exist, a rock-ocean interface is required - in reality, this is a non-trivial condition because sufficiently high pressures at the bottom of the ocean may result in the formation of high-pressure ices \citep{SCK10}. As a result, the subsurface ocean would become trapped between two ice layers, which is conventionally expected to have important, but probably negative, ramifications for the habitability of such worlds \citep{Lam09,NH16};\footnote{On the other hand, despite the existence of high-pressure ice, a combination of convection and melting could enable the slow transport of salts and nutrients. These mechanisms, which have been predicted to operate on Ganymede \citep{CTS17,KSC18}, may collectively offset the challenges posed to long-term habitability.} we shall not tackle the pressure requirements herein, and will assume henceforth that an ocean-bare rock interface does exist. 

Although there are multiple variables involved,\footnote{One such example is the role of water activity (often governed by water-rock interactions) in regulating the rate of serpentinization of olivine; the latter has been observed to decrease in laboratory micro-reactors when the salinity is increased \citep{LR17}.} we will attempt to quantify the rate of abiotic amino acids produced from alkaline, relatively low-temperature hydrothermal vents \citep{MR07,STL13}. Recent observations by the \emph{Cassini-Huygens} mission suggest that this microenvironment is ostensibly present on Enceladus \citep{HP15,WG17}, whose high pH has been interpreted as a consequence of serpentinization through the alteration of ultramafic rocks \citep{GBW15}. The importance of this process stems from the fact that it serves as the ``mother engine'' responsible for the origin of life as per some authors \citep{RNB13}. In contrast, if the rock-ocean interface is acidic or characterized by a higher temperature, the rate of serpentinization will be significantly altered; furthermore, RNA nucleobases and amino acids have short half-lives at high temperatures and pressures \citep{LM98,ACB09,KL11,LWP18}. In this regard, we note that it remains controversial as to whether the first lifeforms on Earth were thermophilic \citep{ANY13,Weiss16} or mesophilic \citep{ML95,BL02,CF17}.

\citet{VHK07} used a detailed thermal cracking model for small planets/satellites with oceans and olivine-dominated lithospheres. From Table 2 of that paper, it can be seen that the flux of molecular hydrogen production (H$_2$) is nearly constant, and ranges between $\sim 10^{13}$ to $\sim 10^{14}$ molecules m$^{-2}$ s$^{-1}$. Choosing the lower bound, we obtain
\begin{equation} \label{NH2}
    \mathcal{N}_{H_2} \sim 2.7 \times 10^{11}\,\mathrm{mol/yr}\,\left(\frac{R}{R_\oplus}\right)^2,
\end{equation}
and for Enceladus, we obtain $\mathcal{N}_{H_2} \sim 13$ mol/s, which is nearly equal to the value of $\sim 11$ mol/s obtained in Table 3 of \citet{SDM17}.  Equivalently, we obtain $\mathcal{N}_{H_2} \sim 4 \times 10^8$ mol/yr, and it agrees fairly well with the estimate of $\sim 10^9$ mol/yr determined from \emph{Cassini} observations of the Enceladus plume \citep{WG17}. For Europa, we arrive at $\mathcal{N}_{H_2} \sim 1.6 \times 10^{10}$ mol/yr, which in very good agreement with the value of $\sim 10^{10}$ mol/yr obtained in \citet{VHP16}. For Earth, $\mathcal{N}_{H_2}$ computed from (\ref{NH2}) for water-rock interactions is comparable to the production rates of H$_2$ from the Precambrian continental lithosphere \citep{LO14}. Theoretically, (\ref{NH2}) can be understood by adopting the following approach \citep{VHP16,SDM17}:
\begin{equation}
     \mathcal{N}_{H_2} = \epsilon\,V,
\end{equation}
where $V$ is the volume of the region subject to alterations by serpentinization and $\epsilon$ is the conversion factor. $V$ can be further expressed as
\begin{equation}
    V = \frac{4\pi}{3} \left[R_c^3 - \left(R_c - \langle{z}\rangle\right)^3\right] \approx 4\pi R_c^2 \langle{z}\rangle,
\end{equation}
for $\langle{z}\rangle < R_c$, where $R_c$ denotes the radius of the `core' region (which comprises of both silicates and metals) and $\langle{z}\rangle$ denotes the width of the serpentinization front. It might be feasible to approximate it via the root mean square diffusion distance, i.e. by using
\begin{equation}
    \langle{z}\rangle \approx \sqrt{2Dt},
\end{equation}
with $D$ denoting an effective diffusion constant for the advancement of the serpentinization reaction front, and $t$ is the elapsed time. Moreover, we will use the ansatz $R_c \propto R$ that is known to be valid for terrestrial planets \citep{VOCS06}. Combining these equations, it can be seen that the scaling $\mathcal{N}_{H_2} \propto R^2$ follows as a result, and is consistent with the estimate provided in (\ref{NH2}). However, we wish to caution that our analysis may not be valid for planets larger than the Earth \citep{VHK07}. Next, we wish to calculate the mass rate of \emph{abiotic} amino acids produced ($\mathcal{M}_\mathrm{HV}$) from hydrothermal vents. Using the fact that $\mathcal{M}_\mathrm{HV}$ is proportional to $\mathcal{N}_{H_2}$ and the data from Sec 2.3 of \citet{SDM17}, we arrive at
\begin{equation}
     \mathcal{M}_\mathrm{HV} \sim 6.7 \times 10^8\,\mathrm{kg/yr}\,\left(\frac{R}{R_\oplus}\right)^{2}.
\end{equation}
The abiotic and biotic production rates of amino acids have been claimed to be comparable on Enceladus \citep{SDM17}, but the latter estimate is contingent on factors such as the existence of methanogens and using adenosine triphosphate (ATP) as a measure of the biomass. 

\subsection{The routes to abiogenesis}
Hitherto, our analysis has mostly focused on the \emph{surficial} production of prebiotic compounds, with the exception of radiolysis and hydrothermal vents where the organic molecules would be released into the ocean. In the rest of this section, we shall briefly examine how organics synthesized close to the surface may lead to abiogenesis; the various steps that could have lead to the origin of life from hydrothermal vents have already been documented in detail elsewhere \citep{RBB14,BBT15,KCT17}.

As noted previously, it is necessary for these molecules to penetrate deeper into the surface before they are subject to total decomposition by sputtering, electromagnetic radiation and charged particles. In this context, we observe that gardening \citep{CP02} in conjunction with tectonics and volcanism (if present) can lead to vertical mixing,\footnote{Note that there exists intriguing evidence favouring the presence of both cryovolcanism \citep{SSM17} and subduction \citep{KP14} on Europa.} and thereby transport the organics to lower regions where they are protected from ionizing radiation \citep{Dart11}. However, in the case of Type U planets, we anticipate that gardening, a nonlinear process facilitated due to surface bombardment by micrometeorites, is likely to be absent or minimal. 

It is believed that one of the significant challenges faced by prebiotic chemistry is that the appropriate organic compounds must be present in sufficiently high concentrations to undergo chemical reactions \citep{BS10}. Second, even at sufficiently high concentrations, these molecules must undergo polymerization to eventually yield peptides and nucleic acids without forming ``tar'' \citep{Sha84,BKKR,BKC12}. It has been shown that wet-dry cycles and thermal gradients can play an important role in facilitating these processes \citep{KKLB,RD16}. On Earth, a wide range of environments have been identified that fulfill the requisite criteria, such as coastal regions and intermountain valleys \citep{Ling17}. As these environments are not likely to exist on planets with subsurface oceans, it raises a potentially important difficulty. 

However, when the flexibility of ice as a medium is taken into account, many of these concerns are alleviated. Several studies have concluded that eutectic freezing serves as an effective mechanism for concentrating prebiotic compounds \citep{LM00,MJCM,MiCMi,Bad04,Pri07}. Laboratory experiments have shown that freeze-thaw cycles, in the presence of suitable catalysts, play a beneficial role in the formation of RNA polymerase ribozymes \citep{MS08,AWPC,AWH13,MWH15} owing to the stabilizing properties of ice \citep{BBC}. However, it is unclear as to whether these cycles are sufficiently important on a global scale since they tend to operate over geologically slow timescales, and do not alter fractional concentrations significantly (for pure ice). Nonetheless, taken collectively, a reasonable case could be built for ice as one of the possible sites for abiogenesis to take place \citep{TSB05}. If this hypothesis were indeed valid, the possibility that life may have originated on Type B and Type U planets ought not be ruled out. We have implicitly assumed that the requisite minerals \citep{Lam08,HS10} and other raw materials are available in sufficient quantities, but this supposition is not always valid \citep{GDD05}.

With regards to the above discussion, it is helpful to further quantify the requirements for polymerization. This can be done by evaluating the conditions under which these reactions become exergonic, i.e. the condition $\Delta G < 0$ must be satisfied, where $\Delta G$ denotes the Gibbs free energy of formation \citep{ALM13}. The corresponding values of $\Delta G$ for aqueous and crystalline organic compounds at different temperatures and pressures are well documented \citep{SH88,LH06}. A similar study was undertaken in \citet{KK15}, and it was shown (see Fig. 1 of that work) that $\Delta G$ is weakly dependent on the pressure, and becomes negative for $T_c \sim 50-110$ K based on the choice of polymerization reaction. Assuming that $T_s < 50$ K, we can solve for the depth $H_c$ at which these temperatures are attained by employing (\ref{Tsol}). Upon solving for $H_c$, we find that the solution is identical to (\ref{Height}) except for the fact that $\Lambda$ must be replaced by $\Lambda_c = T_c/T_s$. For planets not much smaller than Earth, the following approximation is valid:
\begin{equation}
   \frac{H_c}{H} \approx \frac{\ln \Lambda_c}{\ln \Lambda}.
\end{equation}
As an illustrative example, let us suppose that we consider a free-floating planet with a geothermal heat flux and radius similar to Earth. For these parameters, we obtain $\ln \Lambda_c \approx 0.36-1.15$ and $H_c \approx 2.7-8.6$ km. Hence, at a depth of a few km, the formation of peptides would become favourable on thermodynamic grounds. Note that $H_c$ will be lowered as one moves towards the central regions of the Galaxy, where $T_s$ can be higher. Although these reactions are exergonic at $H_c$, it does not necessarily imply that these reactions take place because their rates are proportional to the Boltzmann factor $\exp\left(-E_a/k_BT\right)$, where $E_a$ is the activation energy and $T$ is the ambient temperature. Hence, given that $T_c$ is much lower than the room temperature, polymerization may occur at very low rates. 

Thus, through a combination of the above mechanisms, prebiotic compounds could be concentrated, polymerized and delivered to the subsurface ocean underneath the ice envelope where they can undergo subsequent prebiotic evolution and possibly lead to abiogenesis. An important point worth bearing in mind is that the dependency of the probability of life on the concentration of prebiotic compounds (e.g. amino acids) and nutrients is not well understood \citep{SM87}. Based on laboratory experiments, it has been suggested that a monomer concentration of $\sim 0.1-1$ mM (mmol/L) would be necessary for initiating prebiotic self-assembly processes \citep{SFO67,BS10}. If we assume that the volume of the planet's ocean is similar to that of Earth, the net delivery rate of amino acids must be $\sim 10^7-10^8\,\mathrm{kg/yr}$ to attain these global concentrations over Gyr timescales (in the absence of turnover processes). However, it should be noted that the local synthesis of polymerized biomolecules is considerably enhanced due to the presence of thermodynamic cycles that are governed by microenvironmental factors \citep{BGL03,BWD07,DMD15}. Hence, the preceding estimate was based on global considerations, whereas several origin-of-life scenarios posit local regions as the sites of abiogenesis \citep{Deam97,SAB13},  and the latter environments may possess sufficiently high concentrations of prebiotic molecules and other ingredients (e.g. minerals) that enable life to originate \citep{Fer93}.

As mentioned previously, there is also a possibility that life might have originated within the ice layer. On Earth, psychrophiles have evolved a wide range of genotypic and phenotypic characteristics in order to inhabit a diverse array of sea-ice habitats \citep{TD02,DAC06,HA08}. In fact, the manifold biological adaptations displayed by polyextremophilc terrestrial sea-ice microbes have led to suggestions that they could survive in some environments on Europa and Enceladus \citep{GG00,MFE03,LR05,MM17}. Hence, these factors suggest that the possibility of certain niches within the sea-ice being occupied by motile microbial organisms ought not be ruled out \citep{NL16}.

Before proceeding further, we emphasize that the existence of thermodynamically feasible condensation, cyclical processes and polymerization are necessary but not sufficient conditions for life to originate. Hence, although both these processes appear to be likely in the ice envelope, several other requirements - such as the self-assembly of amphiphilic compounds into vesicles \citep{DD02,SZS} - must also be fulfilled. It may therefore be possible that no \emph{single} environment contains all of the ingredients necessary for abiogenesis to occur. In that event, the combined action of multiple environments and mechanisms could, perhaps, collectively facilitate the origin of life \citep{Haz17} on these worlds with subsurface oceans.

\section{Ecosystems in planets with subsurface oceans}\label{SecEco}
Next, we shall explore the feasibility of subsurface oceanic ecosystems, and delineate some of the challenges and limitations that they are likely to face. We will restrict ourselves to only aquatic habitats, as sea-ice environments have been briefly explored earlier. 

\subsection{The biological potential of subsurface ecosystems}\label{SSecBP}
As subsurface oceans do not have access to sunlight, they are not readily capable of supporting photosynthetic organisms. Furthermore, solar radiation is regarded as the most widely available energy source for life on Earth. Hence, the absence of this pathway has often been invoked to argue that the biological potential of icy moons with subsurface oceans is very low relative to Earth \citep{RS83,JS98,GNK99,McCo99,Pas16}. However, several authors have discussed multiple ecosystems that are not dependent on photosynthesis, and we shall examine these possibilities below. 

Before proceeding further, one notable point worth bearing in mind is that the water in these subsurface oceans is \emph{not} guaranteed to be habitable. As noted in Sec. \ref{SecAbio}, there are multiple constraints imposed by temperature, pH, water activity, etc. that can make the conditions impossible for life-as-we-know-it to exist. For instance, it has been pointed out in \citet{PG12} that oxidants delivered to Europa's oceans from the surface could react with sulphides and the oceans would subsequently undergo acidification, and potentially become inimical to life. Among other factors, a highly acidic ocean may disrupt skeletal
biomineralization and induce narcosis \citep{OF05}; a lowered pH due to hypercapnia has been identified as one of the putative factors in driving the devastating Permian-Triassic mass extinction event \citep{KB96,KBP07}. Hence, if life exists in these oceans, it might have evolved either non-standard biomineralization mechanisms and/or resemble acidophiles on Earth \citep{RoMa,BAD07}.

We start with the putative delivery of organics and oxidants from the icy surface to the subsurface oceans. The importance of oxygen has been thoroughly documented for life on Earth \citep{Lane}, and aerobic metabolism provides about an order of magnitude more energy than anaerobic metabolism for the same quantity of raw materials \citep{CGZ05,McC07,KB08}. Hence, the availability of oxygen has been identified as a potentially significant rate-limiting step for the evolution of complex (extra)terrestrial life \citep{MC14,LRP14,CLV15,KN17,CaKa}; the reader should, however, consult \citet{Butt09} and \citet{SZO11} that discuss the subtleties involved. Despite these advantages, it should also be borne in mind that O$_2$ also has severe deleterious effects: it forms superoxides and peroxides that destroy both enzymes and DNA \citep{Im13}. In order to cope with these harmful consequences, bacteria have been compelled to evolve suitable adaptions. 

It was pointed out in Sec. \ref{SSecEnS} that there is a steady flux of energetic particles to the planetary surface from giant planetary magnetospheres, SEPs and GCRs. Let us denote the total flux, which is the sum of the individual three fluxes, by $\Phi_\mathrm{T}$. For Type B moons, we anticipate that $\Phi_\mathrm{T} \approx \Phi_\mathrm{GP}$, while $\Phi_\mathrm{T} \approx \Phi_\mathrm{CR}$ for Type U planets. The basic set-up for the delivery of oxidants to the subsurface ocean is as follows. Ionizing radiation from the aforementioned sources leads to the formation of clathrate hydrates of oxidants \citep{HCCC}, such as H$_2$O$_2$, O$_2$ and CO$_2$, through radiolysis on the surface \citep{JQC03}. Through a combination of gardening and other geological processes, these compounds are buried in the lower layers and eventually delivered to the subsurface ocean, where they can sustain an indirectly radiation-driven ecosystem \citep{Chy00,HCCC}.

As there are multiple steps involved, there is no guarantee that these oxidants would ultimately reach the ocean. Furthermore, there exist several uncertainties regarding the rates of sputtering and gardening. As a result, the estimated rates of O$_2$ delivered to the Europan ocean have ranged from $\sim 10^5$ mol/yr \citep{CP01} to $\sim 10^{11}$ mol/yr \citep{Green10} based on the depth of the oxygenation layer and the concentration of surficial radiolytic products. We can estimate the delivery rate of $ \mathcal{N}_{O_2}$ by following the approach delineated in \citet{HCC07}. Thus, we have
\begin{equation} \label{NOform}
 \mathcal{N}_{O_2} \sim \frac{4\pi R^2 d_g C_0}{\tau_d},
\end{equation}
where $d_g$ is the gardening depth, $C_0$ is the molar concentration (with units mol/m$^3$) of oxidants and $\tau_d$ is the delivery time. It appears reasonable to suggest that $C_0 \propto \Phi$ since a higher particle flux leads to more oxidants deposited on the surface,  and we use $d_g \propto \tau_d^{1/2}$ \citep{CJM01}. Substituting these scaling relations into the above equation and using fiducial values of $\mathcal{N}_{O_2} \sim 10^{9}$ mol/yr for $\tau_d \sim 50$ Myr for Europa \citep{HCC07}, we obtain
\begin{eqnarray}\label{NO2D1}
 &&\mathcal{N}_{O_2} \sim 1.7 \times 10^{10}\,\mathrm{mol/yr}\,\left(\frac{\tau_d}{50\,\mathrm{Myr}}\right)^{-1/2}\left(\frac{R}{R_\oplus}\right)^2 \nonumber \\
 &&\hspace{0.5in} \times \left(\frac{a_m}{4.5 \times 10^{-3}\,\mathrm{AU}}\right)^{-2},
\end{eqnarray}
and we have made use of (\ref{PhiGP}). In contrast, if we assume that there exist resurfacing processes other than gardening, it is more instructive to introduce the variable $\delta = d_g/\tau_d$ and it equals $4$ m/Myr for Europa \citep{Green10}, leading to $\mathcal{N}_{O_2} \sim 10^{11}$ mol/yr. Using this data in (\ref{NOform}), we end up with
\begin{eqnarray}\label{NO2D2}
    &&\mathcal{N}_{O_2} \sim 4.3 \times 10^{11}\,\mathrm{mol/yr}\,\left(\frac{\delta}{1\,\mathrm{m/Myr}}\right)\left(\frac{R}{R_\oplus}\right)^2 \nonumber \\
 &&\hspace{0.5in} \times \left(\frac{a_m}{4.5 \times 10^{-3}\,\mathrm{AU}}\right)^{-2}.
\end{eqnarray}

Next, let us recall that long-lived radionuclides are responsible for generating heat. In addition, they also play an important secondary role: through a combination of alpha, beta and gamma decay processes powered by $^{40}$K, $^{232}$T, $^{235}$U and $^{238}$U, the radiolysis of water leads to the formation of O$_2$ and H$_2$. A combination of these processes may have led to the production of $\sim 2 \times 10^{10}$ mol/yr of H$_2$ and $\sim 10^{10}$ mol/yr of O$_2$ on Earth \citep{DBD91,Drag}. If we make use of the potentially reasonable assumption that the mass of radioactive isotopes is proportional to the mass of the ocean \citep{CH01} and make use of the radionuclide enhancement factor $\Gamma$ introduced earlier, we obtain the following estimates:
\begin{equation} \label{NH2rad}
    \mathcal{N}_{H_2} \sim 5.4 \times 10^{9}\,\mathrm{mol/yr}\,\,\Gamma\left(\frac{R}{R_\oplus}\right)^2 \left(\frac{\mathcal{H}}{1\,\mathrm{km}}\right),
\end{equation}
\begin{equation} \label{NO2rad}
    \mathcal{N}_{O_2} \sim 2.7 \times 10^{9}\,\mathrm{mol/yr}\,\,\Gamma\left(\frac{R}{R_\oplus}\right)^2 \left(\frac{\mathcal{H}}{1\,\mathrm{km}}\right),
\end{equation}
where $\mathcal{H}$ is the ocean depth, and is distinct from the depth of the ice envelope ($H$); for the sake of simplicity, we have assumed $H/R \ll 1$ and $\mathcal{H}/R \ll 1$ but these relations are not wholly accurate for small objects like Enceladus. Using the fact that $\mathcal{H} \approx 26-31$ km for Enceladus \citep{TT16}, we find $\mathcal{N}_{H_2} \approx 2.5 \times 10^8$ mol/yr. This estimate is in excellent agreement with the value $\sim 1-3 \times 10^8$ mol/yr derived by means of a detailed alternative radiolysis model \citep{BGW,WG17}. Although the overall production rates of O$_2$ and H$_2$ are similar, worlds situated in the vicinity of magnetized environments (such as the magnetospheres of giant planets), may eventually develop hemispheric concentration gradients \citep{CH01}.

Hence, we have provided two different channels by which H$_2$ can be synthesized: hydrothermal vents and water radiolysis, whose production rates are given by (\ref{NH2}) and (\ref{NH2rad}) respectively. Similarly, we have identified two mechanisms for the production of O$_2$, namely, the delivery of oxidants from the surface and water radiolysis. The former can be determined from (\ref{NO2D1}) and (\ref{NO2D2}), while the latter is given by (\ref{NO2rad}). As there are several free parameters and processes, it is not easy to identify characteristic values. However, it is clear that the rates of production of oxygen and hydrogen are not very dissimilar, i.e. they differ by $1$-$2$ orders of magnitude in most cases. This feature may imply that an approximate redox balance exists on some of these planets, analogous to Earth and possibly Europa \citep{VHP16}. It will thus be necessary to take into account the long-term redox history in order to properly assess the habitability of Type B and U planets.

By adopting the analysis outlined in \citet{CP01} for the energy-limited case, where O$_2$ is consumed by methanotrophs \citep{RMH17}, we find
\begin{equation}
    \frac{d m_0}{d t} \sim 1.1 \times 10^{9}\,\mathrm{kg/yr}\,\left(\frac{\mathcal{N}_{O_2}}{10^{10}\,\mathrm{mol/yr}}\right),
\end{equation}
with $d m_0/dt$ denoting the rate of production of biomass. Assuming a turnover time of $\sim 10^3$ yr based on studies of Earth's deep biosphere \citep{HJ13},\footnote{However, in extreme (e.g. ice and permafrost) environments at $\sim 230$ K, the turnover time is predicted to be several orders of magnitude higher \citep{PS04}.} we arrive at the following steady-state biomass ($m_c$):
\begin{equation}
    m_c \sim 1.1 \times 10^{12}\,\mathrm{kg}\,\left(\frac{\mathcal{N}_{O_2}}{10^{10}\,\mathrm{mol/yr}}\right),
\end{equation}
and the corresponding number of cells and their rate of production (in cells/yr) can be computed from the fact that each cell is $\sim 2 \times 10^{-17}$ kg. In comparison, the global net primary production (NPP) of the Earth is $\sim 10^{14}$ kg/yr \citep{FBRF} and the total biomass is $\sim 2 \times 10^{15}$ kg \citep{LFC15}. We can also estimate the biomass produced per year due to reductants instead by using the model proposed in \citet{SDM17} for methanogens.\footnote{In this regard, we caution the reader that it is not yet clear as to whether methane may have served as the energy source or a byproduct of early life \citep{RN17}, and the existence of either methanogens or methanotrophs on subsurface worlds is not guaranteed.} Since the biomass produced is proportional to $\mathcal{N}_{H_2}$, we obtain the production rate
\begin{equation}
    \frac{d m_0}{d t} \sim 2 \times 10^7\,\mathrm{kg/yr}\,\left(\frac{\mathcal{N}_{H_2}}{10^{10}\,\mathrm{mol/yr}}\right),
\end{equation}
and the steady-state biomass is
\begin{equation}
     m_c \sim 2 \times 10^{10}\,\mathrm{kg}\,\left(\frac{\mathcal{N}_{H_2}}{10^{10}\,\mathrm{mol/yr}}\right).
\end{equation}

In addition to these putative ecosystems, life may derive energy from electrical currents by means of electron-transfer reactions, where the electrons are supplied from the magnetospheres of giant planets, and this process has been dubbed ``direct electrophy'' \citep{SNVM}. It should be recognized that this near-surface ecosystem would be feasible only for (Type B) moons orbiting giant planets. Using the data tabulated in Fig. 2 and Table 1 of \citet{SNVM}, the maximum steady-state biomass is given by
\begin{equation}
    m_c \sim 10^{11}\,\mathrm{kg}\,\left(\frac{R}{R_\oplus}\right)^{2} \left(\frac{\Phi_e}{10^{10}\,\mathrm{m}^{-2}\,\mathrm{s}^{-1}}\right),
    \end{equation}
where $\Phi_e$ represents the electron number flux (units of $\mathrm{m}^{-2}\,\mathrm{s}^{-1}$) received at the surface of the moon, assuming that the average energy of the particles is approximately $0.5$ MeV. 

The steady-state concentration of cells ($\eta$) in the ocean can be estimated from $m_c$ via
\begin{equation}
    \eta \sim 10^9\,\mathrm{cells/m^3}\,\left(\frac{m_c}{10^{10}\,\mathrm{kg}}\right)\left(\frac{R}{R_\oplus}\right)^{-2}\left(\frac{\mathcal{H}}{1\,\mathrm{km}}\right)^{-1},
\end{equation}
and the concentration in the plumes (if they are present) will be a factor of $\sim 10$ lower. The above value is similar to the concentrations observed in other extreme habitats on Earth, some of which are delineated below.
\begin{itemize}
    \item Subglacial lakes and icy environments, such as Vostok and Gr\'imsv{\"o}tn, are expected to have concentrations of $\sim 10^9-10^{10}$ cells/m$^3$ \citep{PAL99,Pri00,MP06,PM16}, although concentrations as low as $\sim 10^6-10^7$ cells/m$^3$ have been identified for Vostok \citep{DVR08}. 
    \item The concentration of microbes in deep granitic rock groundwater is typically $\sim 10^{10}-10^{12}$ cells/m$^3$ \citep{Ped97,AT05,LWR06}.
    \item In highly oligotrophic habitats, for e.g. the sub seafloor sediments of the North Pacific Gyre, concentrations of $\sim 10^9$ cells/m$^3$ exist \citep{OSKE,RKA12}.  
\end{itemize}

We conclude by analysis by observing that other forms of life that draw upon alternative sources of energy - such as thermal gradients, magnetic fields and gravitation \citep{SMI06,MSM06} - might also exist in Type B and U planets, but we shall not analyze them here since their total biomass is probably lower \citep{SI02}.

\subsection{The availability of nutrients}\label{SSecNut}
The preceding section indicates that the production of biomass can be quite high, albeit some orders of magnitude lower than Earth, when viewed purely from the perspective of energetics. Yet, an important point worth reiterating is that life needs more than just an energy source and a solvent. Other criteria include the availability of elements such as boron and phosphorus. The former has been argued to play a key role in bringing about the RNA world \citep{Org04,RJ12,HL15} via the chemical stabilization of ribose \citep{Sco12}. As a result, the access to borates in the Hadean-Archean environment has been posited as a crucial factor in abiogenesis \citep{GBH}. Phosphorus plays a vital role in nucleic acids, metabolism (through ATP) and membranes \citep{West87,KSPW,NKB13}. In addition, trace metals (e.g. molybdenum) play an essential role in bioinorganic chemistry and span mechanisms such as photosynthesis, respiration and DNA synthesis, and even minimal changes in their concentration can lead to pathological effects \citep{BGLV}; many trace elements are also closely linked with primary marine productivity and macroevolutionary processes \citep{AK02,Anbar}.

For certain classes of planets and moons, serious considerations of habitability should take into account biogeochemical cycles (e.g. carbon, sulphur) as well as their likelihood of functioning over geological timescales \citep{CB16}; this approach has also been deemed necessary for understanding the evolution of life on early Earth \citep{KBS16}. However, the existence of biogeochemical cycles is \emph{not} a universal requirement for habitability - waterworlds represent an interesting counterexample, as shown recently in \citet{KF18}. We will focus on phosphorus herein since it constitutes an essential element in regulating ocean productivity \citep{Foll96,Ty99}.  Phosphorus also plays an important role in cellular and molecular biology as discussed earlier, and has even been referred to as ``the staff of life'' in the context of aquatic ecology \citep{Karl}. On Earth, the intimately intertwined evolution of marine productivity and phosphorus availability during the Archean and Proterozoic aeons has been extensively investigated \citep{BC02,KLA07,Pap10,LS14,KS17}, especially in connection with the apparently coincidental emergence of metazoans and the rise of atmospheric oxygen in the late Neoproterozoic era \citep{PR10,PRW14,KS14,RPO16,RPG17,Kno17} but alternative chronologies have been proposed recently by \citet{MWJ14} and \citet{ZWW16}.

Our choice of phosphorus is therefore particularly relevant since we are essentially studying (subsurface) ocean planets. Hence, maintaining a balance between phosphorus sources and sinks in the ocean is arguably necessary for a sustaining a reasonably copiotrophic (nutrient-rich) biosphere over geologic timescales. On Earth, the two major conventional sources for phosphorus in the ocean are riverine and atmospheric in nature, and the former is greater than the latter by about an order of magnitude, as seen from Table 1 of \citet{BN00}. The notable sinks in the ocean include sedimentation of organic material, precipitation leading to phosphorite formation, and hydrothermal activity. We shall not concern ourselves with the biological recycling of phosphorus in the ocean, despite its undoubted significance across Earth's geologic history, since this process does not constitute (for the most part) a net sink or source in the current epoch \citep{SB13}.

It is evident that the two major sources identified above are not likely to be functional for planets that have subsurface oceans. Amongst the sinks, the first two are closely linked with biological processes, and we set them aside since they cannot be easily estimated. In addition, their relative contribution is dependent on redox conditions; for instance, the organic burial of phosphorus is enhanced for oxic sediments \citep{PM07}. As a result, we are left with only one mechanism - a sink that would quickly deplete phosphorus from the ocean. In order to compute the amount of phosphorus lost per year, we shall invoke that the assumption that the hydrothermal flux of phosphorus removal is constant, which yields
\begin{equation} \label{NPL}
\mathcal{N}_\mathrm{P} \sim -\,3 \times 10^{10}\,\mathrm{mol/yr}\,\left(\frac{R}{R_\oplus}\right)^2
\end{equation}
where $\mathcal{N}_\mathrm{P}$ denotes the rate of phosphorus gain/loss. Here, we have assumed that the area of the seafloor is proportional to $R^2$, and used the fiducial value of $\sim 3 \times 10^{10}$ mol/yr for the overall hydrothermal sink on Earth \citep{WM02}; note that the negative sign in (\ref{NPL}) signifies the depletion of phosphorus. If we assume that the total mass of the phosphorus $\mathcal{M}_P$ present in the ocean is proportional to its volume, we have
\begin{equation}
   \mathcal{M}_\mathrm{P} \sim 8.6 \times 10^{14}\,\mathrm{mol}\,\left(\frac{R}{R_\oplus}\right)^2 \left(\frac{\mathcal{H}}{1\,\mathrm{km}}\right),
\end{equation}
and the normalization is determined from the fact that Earth has $\sim 3.2 \times 10^{15}$ moles of phosphorus \citep{BN00}. Hence, all of the phosphorus in the ocean will be removed over the timescale $\tau_P$ given by
\begin{equation}
  \tau_P \sim \frac{\mathcal{M}_\mathrm{P}}{|\mathcal{N}_\mathrm{P}|} \sim  2.9 \times 10^4\,\mathrm{yr}\,\left(\frac{\mathcal{H}}{1\,\mathrm{km}}\right).
\end{equation}

Hence, the above analysis suggests that the biosphere will be mostly comprised of oligotrophes eventually. However, there exist some possibilities that may mitigate this important issue. We begin by recalling that Earth-like hydrothermal vents are not anticipated to exist on planets with high-pressure ices, where the ocean is sandwiched between the two ice layers. This would exclude the hydrothermal sink mechanism described above. As a result, these worlds may be conducive to hosting biospheres but they would constitute examples of \emph{cycle-independent} habitability - see also \citet{KF18} - distinct from Earth-like planets. From the standpoint of sources, phosphorus may be delivered via cometary impacts and meteorites \citep{PL08,ABB16}, and the amount delivered exogenously can be quite high depending on the cratering rate. Under certain circumstances, low-temperature submarine weathering (seawater-basalt interactions) can possibly function as a source although it has been claimed to be a small sink on Earth \citep{FBL}.

However, there is one process that we have not considered. The complex interplay of resurfacing processes (e.g. ridge formation, dilation) can result in some fraction of the ice envelope being melted into the ocean. This process is partly reminiscent of the deposition of phosphorus into the oceans through ice rafting, and an upper bound of $5\times 10^{10}$ mol/yr for this mechanism was proposed in \citet{Wall10}; see also \citet{RTB06}. If we assume that an ice layer of thickness $d_i$ melts into the ocean in a turnover time of $\tau_i$, we introduce the variable $\delta_i = d_i/\tau_i$ with units of m/Myr. We assume that the concentration of phosphorus in the ice layer is $\mathcal{C}_P$ (in units of mmol/kg), which may be supplied through meteorite and comet impacts. With these assumptions, we obtain
\begin{equation}\label{NPG}
 \mathcal{N}_\mathrm{P} \sim 4.7 \times 10^8\,\mathrm{mol/yr}\,\left(\frac{R}{R_\oplus}\right)^2 \left(\frac{\delta_i}{1\,\mathrm{m/Myr}}\right)\left(\frac{\mathcal{C}_P}{1\,\mathrm{mmol/kg}}\right),
\end{equation}
and this can easily become comparable to (\ref{NPL}) when $\delta_i$ and/or $\mathcal{C}_P$ are sufficiently high. For example, if we choose $\mathcal{C}_P \sim 18$ mmol/kg based on the composition of Earth's continental crust rocks \citep{Fau98} and $\delta_i \sim 4$ m/Myr for Europa \citep{Green10}, we find that (\ref{NPG}) is approximately equal to (\ref{NPL}) for arbitrary values of $R$. However, we caution that (\ref{NPG}) constitutes an upper bound since all of the phosphorus in the melted ice will not be accessible to organisms. More specifically, the phosphorus must be available in the form of chemical compounds that are soluble and active in liquid water, and can therefore be readily used by biota. Furthermore, in equilibrium, the amount of ice melted should be replenished by an equal amount of ice formed via freezing. Hence, it is not clear as to whether (\ref{NPG}) would ultimately serve as a net source or sink. Lastly, we note that the potential source mechanism (\ref{NPG}) is not expected to be valid for planets with surface oceans that also possess negligible subaerial ice coverage, suggesting that these worlds are relatively likely to have oligotrophic biospheres.\footnote{If the rise of oxygen in Earth's atmosphere was due to the growth/oxidation of continents \citep{Kast13,LYM16} or changes in subaerial volcanism \citep{KB07,Holl09,GSA11}, it ought not be easy for oxygen levels to attain sufficiently high concentrations (only insofar these specific mechanisms are concerned) in order for complex life to arise on planets with deep surface oceans. The greater mass of liquid water should also result in the dilution of nutrients and prebiotic molecules, although, at local scales, it may still be feasible to achieve sufficiently high concentrations ostensibly necessary for prebiotic self-assembly and abiogenesis.}

Looking beyond phosphorus, the importance of sulphur in biogeochemistry has been well documented \citep{FFD08,SB13}, and its possible role as an energy source for chemoautotrophs in the iron-sulphur world \citep{Wac90} has also received much attention. Hence, in this context, we note that the role of sulphur cycling has been extensively investigated for Europa \citep{SI08}. The concentration of sulphates in the ocean layer may approach or exceed that of Earth, although lower values cannot be ruled out as seen from Table 1 of \citet{MZ03}; this model was also used to conclude that the likelihood of massive sulphur beds (preventing life near hydrothermal vents) is not negligible. Moreover, since the chemical energy available for sulphate-reducing lifeforms decreases when the pH is lowered \citep{ZS03}, it seems plausible that an acidic ocean with pH $\sim 2.6$ \citep{PG12}, would be unfavourable for these organisms; the pH value of $\sim 2.6$ was obtained by calculating the equilibrium molar concentrations based on the oxidant delivery rates proposed in \citet{Green10}. Acidic environments could also possess challenges for the polymerization of biomolecules and the origin of life \citep{DSR06}. As with the phosphorus cycle, future studies of subsurface oceans should investigate the oceanic source-sink mechanisms for sulphur and the likelihood of sustaining this biogeochemical cycle over geological timescales \citep{ZS04}.

\subsection{The transitions in evolution}\label{SSecEvo}
Although the ``biological complexity'' has increased over time \citep{Carr01}, much remains unknown about the mechanisms responsible for this process, notwithstanding the manifold recent developments in this area \citep{Ada02,MSB10,Koo11,GoWo}. Over the past two decades, motivated by the seminal work of \citet{JMS95}, the approach of modelling evolution as a series of ``evolutionary transitions'' has proven to be valuable in understanding how smaller units agglomerate to form larger structures that are subsequently acted upon by natural selection \citep{Ok06,CS11,OMP16,SM16}. The common features underlying these transitions stem from the organization, storage and transmittance of information \citep{SS97,Woe04,JL14,DaWa}, and the transitions are characterized by increasing complexity, division of labour, and innovations in heredity to name a few \citep{Sza15}. Examples of these transitions, not all of which are universally accepted, include the origin of prokaryotes, multicellularity, eukaryotes, and eusociality \citep{SS95}.\footnote{The number of these transitions that led to noogenesis (the emergence of intelligence) on Earth remains unsettled, but has been suggested to lie between $4$ and $7$ by several authors \citep{Cart83,Wat08,Cart08,ML10}; see also: \url{http://mason.gmu.edu/~rhanson/hardstep.pdf}}
 
We observe that related theoretical frameworks, also entailing the evolution of new species that are increasingly complex, have been explored by several authors, for e.g. the \emph{megatrajectories} of \citet{KB00}, the \emph{singularities} of \citet{deDu}, and the \emph{energy expansions} of \citet{Jud17}. The likes of \citet{Bieri,Mor11,Bog11,Ros13,LSCW,SMB17} have also drawn upon similar approaches to arrive at some general predictions concerning the nature of exo-evolutionary transitions and the likelihood of complex extraterrestrial life.\footnote{The question as to whether these transitions are ``universal'' is indubitably an important one, and depends on the nature and degree of universality of the evolutionary process itself \citep{WFDS}, and the interplay of contingency \citep{Mon71,Gou89,Mayr,Gould02,BBL08} and convergence \citep{Mor03,Verm06,Losos,RPB14}.} We propose that the adoption of the above methodologies could play an important heuristic role in biological analyses of habitability, since they may enable us to understand the likelihood of these transitions on exoplanets. We will offer a few select examples, in the context of Type B and U planets, to qualitatively illustrate how this methodology can be employed.

If we approach evolution from the viewpoint of energy expansions, we find that epoch II, corresponding to anaerobic photosynthesis, is unlikely on planets with subsurface oceans, except for putative ecological niches close to the surface. Naively, one may therefore suggest that epoch III, i.e. the rise of oxygen, is not feasible. However, as we have seen in Sec. \ref{SSecBP}, there exist two potential mechanisms for the supply of O$_2$ on Europa and other planets/moons in similar environments: radiolysis of water, and the delivery of oxidants from the surface (also via radiolysis). The likelihood of epoch IV, which is essentially the manifestation of phagocytosis, cannot either be dismissed or validated \emph{a priori}. However, epoch V, where fire functions as a global energy source, is again unlikely on these planets because not all of the requisite basic ingredients are anticipated to be prevalent.

We now turn our attention to megatrajectories \citep{KB00}, some of which also fall under the category of orthodox evolutionary transitions. When dealing with the earliest stages, for e.g. the step from abiogenesis to the Last Universal Common Ancestor (LUCA), arriving at unambiguous conclusions beyond the identification of the manifold energy sources for life is difficult. Next, we observe that many, but not all, eukaryotes on Earth are reliant on aerobic metabolism \citep{Knoll14,Knoll15}. The issue of oxygen generation has already been discussed in Sec. \ref{SSecBP} but the likelihood of eukaryogenesis cannot be addressed here in further detail. The factors responsible for the origin of eukaryotes have not yet been conclusively identified, although serial/singular endosymbiosis is expected to have played an important role in multiple respects \citep{Sag67,KN05,EM06,deDu07,MGZ15,Lane17}. Hence, the dependency of eukaryogenesis on specific environmental constraints prevalent in subsurface environments cannot be predicted based on our current knowledge.

When we consider the higher evolutionary transitions or megatrajectories, it may be relatively easier to identify the feasibility of these steps. For instance, eusociality (a ``classical'' evolutionary transition) is predominantly terrestrial, and one recent study has tentatively identified the role of nesting as being a precondition for eusociality that is more pronounced on land than on sea \citep{RHM14}. Hence, it seems plausible to some degree, at least for life-as-we-know-it, that not many species would evolve this feature on exoplanets with only oceans. From a related standpoint, land-dwelling organisms have been identified as one of the six megatrajectories since they exhibit traits (and occupy ecospace) inaccessible to aquatic lifeforms. On the whole, there is a distinctive trend favouring the emergence of high-performance innovations on land relative to water - more specifically, 11 out of the last 13 major post-Ordovician breakthroughs appear to have originated on land \citep{Verm17}. Of these 13 innovations, the dispersal of propagules (e.g. spores and seeds) by animals, and the communal construction of dwellings have not yet arisen (or been documented) in oceans.

The question of whether ``forbidden phenotypes'' \citep{Verm15} in water, i.e. external characteristics that are unique to land-based organisms (on Earth), can eventually arise on ocean planets patently lacks a conclusive answer at this stage. Nevertheless, based on the available empirical evidence from Earth elucidated earlier, the majority of the higher evolutionary transitions might not occur on planets with (sub)surface oceans.\footnote{In this context, it is worth pointing out that most planets with $R\gtrsim 1.6 R_\oplus$ are not likely to possess a rocky composition \citep{Rog15,CK17}. Moreover, many exoplanets in the HZ of M-dwarfs could also end up as ocean planets \citep{TI15}.} Although cetaceans have been (controversially) linked with certain ``human'' traits such as culture, consciousness and intelligence \citep{Gri01,RW01,WVS07,MCF07,WR15,DW16,HW17} - see, however, \citet{Ty01,PHP08,RB08,Mang13,Sud13} for alternative perspectives - at this stage, it remains fundamentally unclear as to whether certain attributes including advanced tool construction, mental time travel, recursive thought processes, and \emph{perhaps} syntactical-grammatical language \citep{RD05,SC07,Cor11,BC16,Lal17} may constitute evolutionary innovations that are unique to land. 

\section{Implications for detection and panspermia}\label{SecNumb}
We will briefly explore some of the implications stemming from the likelihood of life on planets with subsurface worlds, and comment on the prospects for detection.

\subsection{Number of planets with potential subsurface oceans}
We begin by introducing some notation. The variable `$N$' denotes the number and `$\mathcal{P}$' signifies the probability. We use subscripts `HZ' and `SO' to distinguish between rocky planets located in the habitable zone and those that could have subsurface oceans, i.e. Type B and U planets introduced in Sec. \ref{SSecHab}. We will introduce the rest of the notation as we proceed further. However, before doing so, let us recall that all planets in the HZ are not guaranteed to have water on the surface; similarly, not all Type B and U planets will actually possess subsurface oceans.

In order to estimate $N_\mathrm{HZ}$, we can use the data from the \emph{Kepler} mission. Statistical studies have yielded fairly disparate results depending on the spectral type of star considered, the limits of the HZ, etc. A summary of these findings for main-sequence stars and white dwarfs can be found in Table 1 of \citet{Kal17} and Section 6 of \citet{VSVE} respectively. We adopt an estimate of $\sim 0.1$ rocky planets in the HZ per host star. This value is slightly on the conservative side, since it is approximately $50\%$ of the corresponding fraction for M-dwarfs  \citep{DC15}, which are the most common type of stars in our Galaxy. Thus, using the fact that there are $\sim 10^{11}$ stars in the Galaxy, we obtain
\begin{equation} \label{NHZ}
   N_\mathrm{HZ} \sim 0.1 \times 10^{11} \sim 10^{10}. 
\end{equation}

Next, let us estimate $N_\mathrm{SO}$, i.e. the number of planets that could host subsurface oceans. This can be done by noting that $N_\mathrm{SO} \approx N_\mathrm{B} + N_\mathrm{U}$, where $N_\mathrm{B}$ and $N_\mathrm{U}$ are the number of bound and free-floating planets respectively. However, it is not easy to estimate how many potential planets with subsurface oceans exist outside the HZ of the host star. Hence, at this stage, we must resort to a variant of the Copernican Principle, also referred to as the Principle of Mediocrity.\footnote{It shares close connections with the Principle of Cosmic Modesty proposed recently \citep{Loeb17}, but their stances are not exactly the same.} 

We will therefore assume that the solar system is not highly atypical, and that the number of potential subsurface worlds with oceans per star is similar to that of the Solar system. Since we are not concerned with the actual existence of subsurface oceans herein (merely the possibility that they could have one), we count the number of objects within the range $200 < R < 6400$ km in the solar system outside the HZ;\footnote{\url{http://www.johnstonsarchive.net/astro/tnoslist.html}} the lower and upper bounds reflect the radius of Enceladus and Earth respectively. Bearing in mind the fact that not all TNOs have been detected, we find that there are $\sim 100$ ``planets'' within the above range. If the lower cutoff is increased to $\sim 500$ km, the number drops to $\sim 25$. Barring Enceladus and possibly Mimas, most of the objects that may possess subsurface oceans have $R \gtrsim 500$ km \citep{Lun17}, and therefore it seems more prudent to use the higher cutoff. With this set of assumptions, we arrive at
\begin{equation} \label{NB}
   N_\mathrm{B} \sim 25 \times 10^{11} \sim 2.5 \times 10^{12}.
\end{equation}

In order to determine $N_\mathrm{U}$, we can adopt two different strategies. The first is to use results from simulations, whereby the ejected number of planets (with different masses) is computed for a wide range of initial planetary and debris disk configurations. Here, we need to impose a lower cutoff once again for free-floating planets that could host subsurface oceans. For sufficiently high volatile inventories, planets similar in size to Europa should be able to retain oceans purely through radiogenic heating \citep{SS03}. From Fig. 1 of \citet{BQ17}, it can be seen that $\sim 100$ planets above this cutoff are ejected from systems with giant planets and $\sim 10$ planets when there are no giant planets. Since giant planets exist only in $\sim 20\%$ of all stellar systems, on average $\sim 30$ planets are ejected per star, thus yielding
\begin{equation} \label{NU1}
   N_\mathrm{U} \sim 30 \times 10^{11} \sim 3 \times 10^{12}. 
\end{equation}
In the above calculation, we have assumed that there exist $\sim 10^{11}$ stars in the Milky Way. In reality, the total number of stars that have existed over the Milky Way's lifetime is higher, especially given that the star formation rate peaked at $z\approx 1.9$ \citep{MD14}, and Type U planets would have been ejected from them.\footnote{Type U planets are in sharp contrast to rocky planets in the conventional HZ, whose habitability will be terminated even prior to the star's death \citep{RCO13}.} However, we shall proceed with the more conservative estimate, namely (\ref{NU1}), in our subsequent analysis.

The second method for inferring $N_\mathrm{U}$ relies upon the very recent discovery of the putative interstellar asteroid `Oumuamua by the Pan-STARRS telescope \citep{MWM17}.\footnote{\url{https://www.nasa.gov/feature/jpl/small-asteroid-or-comet-visits-from-beyond-the-solar-system}} There have been several follow-up studies concerning the structure, origin and travel time of this asteroid \citep{Mam17,GWK17,YZ17,JLR17,BS17,BWF17} and its implications for the formation and architecture of planetary systems \citep{TR17,LB17,RA17,JT17,Cuk18}. We make use of the fact that the density of such objects has been predicted to be $\sim 10^{14}-10^{15}$ pc$^{-3}$ in the solar neighbourhood \citep{PZ17,FJ17,DTT18}. It translates to a value of $\sim 0.01-0.1$ AU$^{-3}$, which is about $1$-$2$ orders of magnitude higher than \citet{MC89}, $2$-$3$ orders of magnitude larger compared to \citet{EJ17} and $6$-$7$ orders of magnitude higher than \citet{MT09}. Collectively, these estimates serve to illustrate the fact that there is significant uncertainty surrounding the number of such objects in the solar neighbourhood \citep{CRG16}.

In order to carry out the order-of-magnitude calculations, we assume that the value specified above serves as the global density in the Galaxy. Upon doing so, we find a total of $\sim 7 \times 10^{25}$ objects in the Milky Way. In order to compute the number of objects with diameter $> 3000$ km, we resort to the Copernican Principle and use Fig. 1 of \citet{BDN05} to formulate an approximate power-law with spectral exponent $\approx -2.5$ but, in reality, there exists significant variability based on the asteroid size, age, etc. that will not be considered here. We note that our choice of the power-law distribution is comparable to the size distribution of elliptic comets, but less steep compared to Kuiper Belt Objects \citep{MT09}. Since the diameter of `Oumuamua is $\sim 0.1$ km, we find that
\begin{equation}
    N_\mathrm{U} \sim 7 \times 10^{25} \left(3 \times 10^4\right)^{-2.5} \sim 4 \times 10^{14},
\end{equation}
which is higher than (\ref{NU1}) by two orders of magnitude; this translates to $\sim 10^3$ Type U (Moon-sized and larger) worlds per star which is approximately equal to the prediction of $\gtrsim 2 \times 10^3$ free-floating objects per star in \citet{DG18}, but is about two orders of magnitude lower than the estimate provided in Fig. 1 of \citet{SBMB}. Nevertheless, we shall adopt the more conservative value, given by (\ref{NU1}), in our subsequent analysis.

Thus, upon adding (\ref{NB}) and (\ref{NU1}), we arrive at $N_\mathrm{SO} \sim 5.5 \times 10^{12}$. From (\ref{NHZ}), one can see that $N_\mathrm{SO}$ is $\sim 10^3$ times higher than $N_\mathrm{HZ}$. Hence, one can pose the question: since planets with subsurface oceans are more common than rocky planets in the HZ, why do we find ourselves on the latter? The reason most likely stems from the fact that ``we'' refers to an intelligent, conscious and technologically sophisticated species. Hence, it is still plausible that the probability of life on these subsurface worlds ($\mathcal{P}_\mathrm{SO}$) is non-negligible, but the likelihood of technological life could instead be selectively lowered on Type U and B planets as discussed in Sec. \ref{SSecEvo}.\footnote{A related question concerning the likelihood of intelligent life on planets around M- and G-type stars was discussed in \citet{Wal11,LBS16,HKW18}, with some potential solutions having been advanced in \citet{DHL17,DLMC,DJL18} and \citet{Man17,MLin}.}

\subsection{On the likelihood of lithopanspermia}\label{SSecPanS}
Lithopanspermia represents the transfer of life by means of rocky material from one object to another \citep{Bur04,Wess10,Wick10,Lin16}. Most studies have tended to focus on either interstellar \citep{N04,BM12} or interplanetary panspermia \citep{Mel88,GB96,GDL05}.\footnote{Panspermia in the Galactic centre \citep{CFL} and in globular clusters \citep{DSR16} can be viewed as a juxtaposition of the interstellar and interplanetary cases because of the close distances between the stars, although the likelihood of planet formation and stable orbits in such environments may be quite low \citep{DKB12,PHC16}.} Here, we will briefly explore the possibility that free-floating planets could seed life on gravitationally bound planets through panspermia \citep{LL18}; a variant of this idea was also discussed in \citet{WW12}. Our subsequent discussion is also applicable, with some slight modifications, to the scenario wherein Type U planets may facilitate the transfer of biomolecules by means of pseudo-panspermia \citep{Org04,Linga}.

We envision a two-step process wherein a free-floating planet is temporarily captured by a star, and then seeds other planets orbiting that star. The total probability $\mathcal{P}_\mathrm{tot}$ for this process is estimated through a Drake-type equation:
\begin{equation}
    \mathcal{P}_\mathrm{tot} = \mathcal{P}_\mathrm{cap} \cdot \mathcal{P}_\mathrm{planet} \cdot \mathcal{P}_{SO} \cdot  \mathcal{P}_{PS},
\end{equation}
where $\mathcal{P}_\mathrm{cap}$ is the capture probability of a free-floating planet by a star in its lifetime, $\mathcal{P}_{SO}$ is the probability that the captured planet already has life, $\mathcal{P}_\mathrm{planet}$ is the number of planets that could host life around that star (but not necessarily inside the HZ), and $\mathcal{P}_\mathrm{PS}$ is the probability of interplanetary panspermia. Note that $\mathcal{P}_\mathrm{planet} \sim 0.1$ if we restrict ourselves only to rocky planets inside the HZ. However, allowing for the possibility of subsurface ocean worlds, we set $\mathcal{P}_\mathrm{planet} \sim 1$. It is not easy to properly assess $\mathcal{P}_\mathrm{cap}$ since it will depend on the velocity dispersion, stellar and planetary masses, inclination angle, etc. Recent simulations undertaken by \citet{GP18} appear to suggest that $\mathcal{P}_\mathrm{cap} \sim 0.01$, but this value is parameter-dependent.

Using these values, we find $\mathcal{P}_\mathrm{tot} \sim 0.01 \cdot \mathcal{P}_{SO} \cdot  \mathcal{P}_{PS}$, and the magnitudes for the remaining two variables are highly uncertain. Since $\mathcal{P}_\mathrm{PS}$ represents the probability of interplanetary panspermia,\footnote{More specifically, it refers to the probability of glaciopanspermia, i.e. life being seeded by ejecta from impacts of icy worlds; a similar process has been conjectured to have occurred in the early Solar system, with Ceres functioning as the source \citep{Hout11}.} its likelihood of occurrence should be relatively higher compared to interstellar panspermia \citep{Mel03}; for instance, \citet{Mil00} found that the transfer of microbes from Mars to Earth was highly probable, and the transfer from Earth/Mars to the Galilean satellites is also possible, especially during the Late Heavy Bombardment \citep{WSH13}. Furthermore, for low-mass stars with closely packed planetary systems, the chances for interplanetary panspermia are boosted by several orders of magnitude compared to our Solar system \citep{Linga}. In contrast, the prospects for interstellar panspermia appear to be much lower \citep{AS05}. Finally, we are left with $\mathcal{P}_{SO}$, and there is no available method to estimate it. If life is discovered on Europa or Enceladus, it will help constrain the probability of life originating on planets/satellites with subsurface oceans. The implications of discovering life elsewhere for the timescale associated with the origin of life have been discussed in \citet{ST12}.

In order to assess the total number of stellar systems that have been seeded with panspermia, we must multiply $\mathcal{P}_\mathrm{tot}$ with $10^{11}$, which is a fairly large number. Hence, even if we choose $\mathcal{P}_{SO} \sim 10^{-3}$ and $\mathcal{P}_\mathrm{PS} \sim 10^{-2}$, we find that $\sim 10^4$ stellar systems could have been seeded with life. These are, of course, fiducial estimates and the actual number of stellar systems seeded can be either much higher ($< 10^9$) or much lower (perhaps equal to zero). Future statistical surveys can constrain the likelihood of panspermia processes by looking for signs of clustering; it was pointed out in \citet{LL15} - see also \citet{Ling16} - that the detection of $\gtrsim 25$ planets with biospheres will enable (under ideal circumstances) a rigorous test of the panspermia hypothesis.

\subsection{The prospects for detection}
We will confine ourselves to discussing the detection of Type U planets herein, since the case for \emph{in situ} exploration of Type B worlds is more straightforward to espouse given the relatively large number of moons and planets with subsurface oceans in our Solar system \citep{NiPa}. A possible difference between Type U worlds and Type B planets/moons within our Solar system is that the initial conditions for their formation might have been different, for e.g. gas-starved vs gas-rich disks. In turn, such distinctions could have important consequences for the subsequent geological, chemical and biological evolution of these worlds, thereby providing a motive for the detection and study of Type U worlds.

From (\ref{NU1}), we see that the number of free-floating planets (with $R \gtrsim 0.3 R_\oplus$) is about $30$ times higher than the total number of stars. Since the nearest star is $\sim 1$ pc away, we suggest that the nearest such object might be located at a distance of $\langle{r}\rangle \sim 0.01-0.1$ pc from the Earth, which translates to $\langle{r}\rangle \sim 2 \times 10^3-2 \times 10^4$ AU. Note that the lower bound is roughly comparable to the inner edge of the Oort cloud and the aphelion of certain TNOs such as Sedna. 

We can estimate the thermal flux from this planet that would be received on Earth using
\begin{equation}
    S_\mathrm{max} \approx 1.5\,\mathrm{mJy}\, \left(\frac{T_s}{40\,\mathrm{K}}\right)^3 \left(\frac{R}{R_\oplus}\right)^2 \left(\frac{\langle{r}\rangle}{2000\,\mathrm{AU}}\right)^{-2},
\end{equation}
where the flux density $S_\mathrm{max}$ has been computed at the black body peak (Wien maximum), with $\lambda_\mathrm{max} \approx 126$ $\mu$m for the characteristic temperature of $40$ K for Type U planets. It can be seen that the value of $\lambda_\mathrm{max}$ is in the far-IR range, and several telescopes are operational at such wavelengths.

The maximum distance at which Earth-sized free-floating planets can be detected is $\sim 830$ AU \citep{AS11} for both LSST \citep{JC09} and PAN-STARRS \citep{Jew03}, suggesting that the characteristic distance of the nearest Type U planet falls below the detection threshold. The maximum sensitivity of the Herschel/PACS instrument appears to be a few mJy \citep{B10,Pog10}, indicating a borderline case. In contrast, the Cornell-Caltech Atacama Telescope (CCAT) has been predicted to reach a sensitivity of $\sim 0.36$ mJy for a wavelength of $200$ $\mu$m,\footnote{\url{http://www.ccatobservatory.org/index.cfm}} which is slightly lower than the value of $S_\mathrm{max}$ obtained above. Hence, it seems plausible that upcoming telescopes could detect such free-floating planets although it is apparent that the result will depend critically on the number density of such objects in the solar neighbourhood. 

The question as to whether any tangible biomarkers, or even the interior composition and structure \citep{VKP18}, pertaining to these planets can be unambiguously identified is not easy to resolve. The major difficulty stems from the fact that there is no atmosphere, since most studies have hitherto focused on atmospheric biosignatures such as oxygen and ozone \citep{Mead17,Gren17}. If these planets emit plumes akin to Enceladus \citep{WLM09,PSH11} and perhaps Europa \citep{SH16,SSM17}, it may be possible to search for biomarkers therein \citep{MP08,JP17} but a significant difficulty arises from the fact that the photon flux received at Earth scales as $\langle{r}\rangle^{-2}$. Hence, \emph{ceteris paribus}, the flux from the nearest Type U planet would be $10^{-5}-10^{-6}$ lower than that of Enceladus unless we serendipitously discover such an object much closer to our planet. Another possibility in the future is to make use of small spacecraft powered by light-sail technology, along the lines of the recently announced \emph{Breakthrough Starshot} project,\footnote{\url{http://breakthroughinitiatives.org/initiative/3}} for carrying out flyby missions \citep{HPL17}. A spacecraft travelling at $20\%$ the speed of light might be able to reach the nearest Type U planet in a span of $\sim 1$ yr.

We conclude by pointing out a promising, and possibly universal, feature of life-as-we-know-it: its propensity to generate thermodynamic disequilibrium \citep{BBB17,BBG17}. This feature has been invoked widely in the context of detecting atmospheric biosignatures \citep{Kal17}, with some caveats \citep{KT16}, ever since the pioneering work by \citet{Led65} and \citet{Love65}. However, the atmosphere is only one component of the Earth that exists in a state of disequilibrium, and it has been recognized that the energy balance of the surface is also profoundly altered by a mature biosphere \citep{LM74,Sch99,LW11,Klei16}.\footnote{In addition to the energy balance, many crucial components of Earth's biosphere, such as productivity, nutrient cycles and even the evolutionary process itself, have been significantly influenced by the emergence of organisms \citep{Lew00,OLF03,PP09,Butt11,LUF15}.} Hence, it remains an open question (and one perhaps worth further consideration) - especially given our rudimentary understanding of planetary bio-regulation mechanisms \citep{Ty13} - as to whether subsurface exolife would be capable of altering the planetary interior to the degree that it can be detected with sufficient precision to distinguish it from false positives \citep{Walk17}.

\section{Conclusions}\label{SecConc}
The goal of this work was to examine the constraints on the habitability of planets and moons with subsurface oceans and an outer ice envelope. Some of our findings are also applicable to planets with deep terrestrial biospheres or ocean planets (with surface water). We began by presenting a simple model of a conducting ice layer, and showed that its thickness was regulated by the mass, surface temperature and the availability of radioactive materials. As a result, we concluded that a wide range of ``planets'' with ice shells of moderate thickness may exist in a diverse array of habitats. 

Although the availability of water is an important constraint, life-as-we-know-it also requires an energy source for both origination and sustenance. Hence, we quantified the energy available from a wide range of sources such as ionizing radiation, exogenous delivery of dust, and radiogenic heating. In each instance, we computed the amount of amino acids that could be produced. However, there are several steps between the formation of prebiotic compounds and the origin of life, and we examined how certain unique properties of ice can play a beneficial role in this regard, especially with respect to the concentration and polymerization of these molecules.

Subsequently, given these energy sources we examined the biological potential of these worlds. We found that a wide variety of mechanisms are capable of supporting biospheres, such as the energy derived from the delivery of oxidants from the surface, hydrothermal vents and the radiolysis of water. Under certain circumstances, a redox balance akin to that of Earth, and perhaps Europa, may exist although its likelihood is low in general. In most cases, we concluded that the rate of biomass production was likely to be several orders of magnitude lower than on Earth. Next, we highlighted the fact that life also requires a steady long-term supply of bioessential nutrients in addition to energy sources, and consequently crucial biogeochemical cycles (e.g. the phosphorus cycle) might face challenges on these planets. We presented a brief sketch of the major evolutionary transitions on Earth, and hypothesized that some of the later (more complex) innovations have a low probability of occurring on worlds with (sub)surface oceans.

We concluded our analysis by presenting heuristic estimates for the total number of planets capable of possessing subsurface oceans that exist in our Galaxy, and found that they are perhaps $\sim 100-1000$ times more common than rocky planets in the HZ of stars. We briefly discussed the possibility that free-floating planets can enable lithopanspermia to occur on an interplanetary level. We explored potential avenues for detecting these planets, and found that the identification of distinctive subsurface biosignatures does not appear to be feasible with current space- and ground-based telescopes.

To summarize, life on (exo)planets with subsurface oceans is likely to face \emph{sui generis} challenges that are not prevalent on Earth. Examples include the lack of an abundant energy source equivalent to sunlight, and the possibility that the biosphere becomes primarily oligotrophic. On the other hand, we have not been able to identify any definitive limiters that can prevent biospheres from emerging and functioning over geologic timescales. As these worlds are likely to be far more abundant than rocky planets in the HZ of stars, we suggest that more effort should focus on modeling and understanding the prospects for life in subsurface oceans. By doing so, we will be able to take one step further towards understanding whether life, especially sentient life, in the Universe is a ``cosmic imperative'' \citep{Duve95,MS07,BaSM16} or a genuinely rare occurrence \citep{Gay64,Mayr85,WaB00,Mor03}.

\section*{Acknowledgments}
We are grateful to our reviewer, Edwin Kite, for his detailed and constructive referee report. ML also wishes to thank Andrew Knoll for the insightful discussions. This work was supported in part by the Breakthrough Prize Foundation for the Starshot Initiative, Harvard University's Faculty of Arts and Sciences, and the Institute for Theory and Computation (ITC) at Harvard University.


\begin{thebibliography}{}

\bibitem[\protect\astroncite{{Abbot} and {Switzer}}{2011}]{AS11}
{Abbot}, D.~S. and {Switzer}, E.~R. (2011).
\newblock {The Steppenwolf: A Proposal for a Habitable Planet in Interstellar
  Space}.
\newblock {\em Astrophys. J. Lett.}, 735(2):L27.

\bibitem[\protect\astroncite{{Abramov} and {Mojzsis}}{2011}]{AM11}
{Abramov}, O. and {Mojzsis}, S.~J. (2011).
\newblock {Abodes for life in carbonaceous asteroids?}
\newblock {\em Icarus}, 213(1):273--279.

\bibitem[\protect\astroncite{{Adam}}{2007}]{Ad07}
{Adam}, Z. (2007).
\newblock {Actinides and Life's Origins}.
\newblock {\em Astrobiology}, 7(6):852--872.

\bibitem[\protect\astroncite{{Adam}}{2016}]{Ada16}
{Adam}, Z.~R. (2016).
\newblock {Temperature oscillations near natural nuclear reactor cores and the
  potential for prebiotic oligomer synthesis}.
\newblock {\em Orig. Life Evol. Biosph.}, 46(2-3):171--187.

\bibitem[\protect\astroncite{{Adam} et~al.}{2018}]{AHC18}
{Adam}, Z.~R., {Hongo}, Y., {Cleaves}, H.~J., {Yi}, R., {Fahrenbach}, A.~C.,
  {Yoda}, I., and {Aono}, M. (2018).
\newblock {Estimating the capacity for production of formamide by radioactive
  minerals on the prebiotic Earth}.
\newblock {\em Sci. Rep.}, 8:265.

\bibitem[\protect\astroncite{{Adami}}{2002}]{Ada02}
{Adami}, C. (2002).
\newblock {What is complexity?}
\newblock {\em BioEssays}, 24(12):1085--1094.

\bibitem[\protect\astroncite{{Adams} and {Spergel}}{2005}]{AS05}
{Adams}, F.~C. and {Spergel}, D.~N. (2005).
\newblock {Lithopanspermia in Star-Forming Clusters}.
\newblock {\em Astrobiology}, 5(4):497--514.

\bibitem[\protect\astroncite{{Agol}}{2011}]{Ag11}
{Agol}, E. (2011).
\newblock {Transit Surveys for Earths in the Habitable Zones of White Dwarfs}.
\newblock {\em Astrophys. J.}, 731(2):L31.

\bibitem[\protect\astroncite{{Akanuma} et~al.}{2013}]{ANY13}
{Akanuma}, S., {Nakajima}, Y., {Yokobori}, S., {Kimura}, M., {Nemoto}, N.,
  {Mase}, T., {Miyazono}, K., {Tanokura}, M., and {Yamagishi}, A. (2013).
\newblock {Experimental evidence for the thermophilicity of ancestral life}.
\newblock {\em Proc. Natl. Acad. Sci. USA}, 110(27):11067--11072.

\bibitem[\protect\astroncite{{Albarr{\'a}n} et~al.}{1987}]{ACC87}
{Albarr{\'a}n}, G., {Collins}, K.~E., and {Collins}, C.~H. (1987).
\newblock {Formation of organic products in self-radiolyzed calcium carbonate}.
\newblock {\em J. Mol. Evol.}, 25(1):12--14.

\bibitem[\protect\astroncite{{Alroy}}{2008}]{Al08}
{Alroy}, J. (2008).
\newblock {Dynamics of origination and extinction in the marine fossil record}.
\newblock {\em Proc. Natl. Acad. Sci. USA}, 105(1):11536--11542.

\bibitem[\protect\astroncite{{Altwegg} et~al.}{2016}]{ABB16}
{Altwegg}, K., {Balsiger}, H., {Bar-Nun}, A., {Berthelier}, J.-J., {Bieler},
  A., {Bochsler}, P., {Briois}, C., {Calmonte}, U., {Combi}, M.~R., {Cottin},
  H., {De Keyser}, J., {Dhooghe}, F., {Fiethe}, B., {Fuselier}, S.~A., {Gasc},
  S., {Gombosi}, T.~I., {Hansen}, K.~C., {Haessig}, M., {Ja ckel}, A., {Kopp},
  E., {Korth}, A., {Le Roy}, L., {Mall}, U., {Marty}, B., {Mousis}, O., {Owen},
  T., {Reme}, H., {Rubin}, M., {Semon}, T., {Tzou}, C.-Y., {Waite}, J.~H., and
  {Wurz}, P. (2016).
\newblock {Prebiotic chemicals--amino acid and phosphorus--in the coma of comet
  67P/Churyumov-Gerasimenko}.
\newblock {\em Sci. Adv.}, 2(5):e1600285.

\bibitem[\protect\astroncite{{Amend} et~al.}{2013}]{ALM13}
{Amend}, J.-P., {LaRowe}, D.-E., {McCollom}, T.-M., and {Shock}, E.-L. (2013).
\newblock {The energetics of organic synthesis inside and outside the cell}.
\newblock {\em Phil. Trans. R. Soc. B}, 368(1622):20120255.

\bibitem[\protect\astroncite{{Amend} and {Teske}}{2005}]{AT05}
{Amend}, J.~P. and {Teske}, A. (2005).
\newblock {Expanding frontiers in deep subsurface microbiology}.
\newblock {\em Palaeogeogr. Palaeoclimatol. Palaeoecol.}, 219(1):131--155.

\bibitem[\protect\astroncite{{Anbar}}{2008}]{Anbar}
{Anbar}, A.~D. (2008).
\newblock {Elements and Evolution}.
\newblock {\em Science}, 322(5907):1481--1483.

\bibitem[\protect\astroncite{{Anbar} and {Knoll}}{2002}]{AK02}
{Anbar}, A.~D. and {Knoll}, A.~H. (2002).
\newblock {Proterozoic Ocean Chemistry and Evolution: A Bioinorganic Bridge?}
\newblock {\em Science}, 297(5584):1137--1143.

\bibitem[\protect\astroncite{{Attwater} et~al.}{2013}]{AWH13}
{Attwater}, J., {Wochner}, A., and {Holliger}, P. (2013).
\newblock {In-ice evolution of RNA polymerase ribozyme activity}.
\newblock {\em Nat. Chem.}, 5(12):1011--1018.

\bibitem[\protect\astroncite{{Attwater} et~al.}{2010}]{AWPC}
{Attwater}, J., {Wochner}, A., {Pinheiro}, V.~B., {Coulson}, A., and
  {Holliger}, P. (2010).
\newblock {Ice as a protocellular medium for RNA replication}.
\newblock {\em Nat. Commun.}, 1:76.

\bibitem[\protect\astroncite{{Aubrey} et~al.}{2009}]{ACB09}
{Aubrey}, A.~D., {Cleaves}, H.~J., and {Bada}, J.~L. (2009).
\newblock {The Role of Submarine Hydrothermal Systems in the Synthesis of Amino
  Acids}.
\newblock {\em Orig. Life Evol. Biosph.}, 39(2):91--108.

\bibitem[\protect\astroncite{{Baaske} et~al.}{2007}]{BWD07}
{Baaske}, P., {Weinert}, F.~M., {Duhr}, S., {Lemke}, K.~H., {Russell}, M.~J.,
  and {Braun}, D. (2007).
\newblock {Extreme accumulation of nucleotides in simulated hydrothermal pore
  systems}.
\newblock {\em Proc. Natl. Acad. Sci. USA}, 104(22):9346--9351.

\bibitem[\protect\astroncite{{Bada}}{2004}]{Bad04}
{Bada}, J.~L. (2004).
\newblock {How life began on Earth: a status report}.
\newblock {\em Earth Planet. Sci. Lett.}, 226(1-2):1--15.

\bibitem[\protect\astroncite{{Bada}}{2013}]{Ba13}
{Bada}, J.~L. (2013).
\newblock {New insights into prebiotic chemistry from Stanley Miller's spark
  discharge experiments}.
\newblock {\em Chem. Soc. Rev.}, 42(5):2186--2196.

\bibitem[\protect\astroncite{{Bada} and {Lazcano}}{2002}]{BL02}
{Bada}, J.~L. and {Lazcano}, A. (2002).
\newblock {Some Like It Hot, But Not the First Biomolecules}.
\newblock {\em Science}, 296(5575):1982--1983.

\bibitem[\protect\astroncite{{Badescu}}{2011}]{Bad11}
{Badescu}, V. (2011).
\newblock {Free-floating planets as potential seats for aqueous and non-aqueous
  life}.
\newblock {\em Icarus}, 216(2):485--491.

\bibitem[\protect\astroncite{{Bains}}{2004}]{Bains}
{Bains}, W. (2004).
\newblock {Many Chemistries Could Be Used to Build Living Systems}.
\newblock {\em Astrobiology}, 4(2):137--167.

\bibitem[\protect\astroncite{{Bains} and {Schulze-Makuch}}{2016}]{BaSM16}
{Bains}, W. and {Schulze-Makuch}, D. (2016).
\newblock {The Cosmic Zoo: The (Near) Inevitability of the Evolution of
  Complex, Macroscopic Life}.
\newblock {\em Life}, 6(3):25.

\bibitem[\protect\astroncite{{Baker-Austin} and {Dopson}}{2007}]{BAD07}
{Baker-Austin}, C. and {Dopson}, M. (2007).
\newblock {Life in acid: pH homeostasis in acidophiles}.
\newblock {\em Trends Microbiol.}, 15(4):165--171.

\bibitem[\protect\astroncite{{Balbi} and {Tombesi}}{2017}]{BT17}
{Balbi}, A. and {Tombesi}, F. (2017).
\newblock {The habitability of the Milky Way during the active phase of its
  central supermassive black hole}.
\newblock {\em Sci. Rep.}, 7:16626.

\bibitem[\protect\astroncite{{Ball}}{2008}]{Ball08}
{Ball}, P. (2008).
\newblock {Water as an Active Constituent in Cell Biology}.
\newblock {\em Chem. Rev.}, 108(1):74--108.

\bibitem[\protect\astroncite{{Ball} and {Hallsworth}}{2015}]{BH15}
{Ball}, P. and {Hallsworth}, J.~E. (2015).
\newblock {Water structure and chaotropicity: their uses, abuses and biological
  implications}.
\newblock {\em Phys. Chem. Chem. Phys.}, 17(13):8297--8305.

\bibitem[\protect\astroncite{{Bannister} et~al.}{2017}]{BS17}
{Bannister}, M.~T., {Schwamb}, M.~E., {Fraser}, W.~C., {Marsset}, M.,
  {Fitzsimmons}, A., {Benecchi}, S.~D., {Lacerda}, P., {Pike}, R.~E.,
  {Kavelaars}, J.~J., {Smith}, A.~B., {Stewart}, S.~O., {Wang}, S.-Y., and
  {Lehner}, M.~J. (2017).
\newblock {Col-OSSOS: Colors of the Interstellar Planetesimal 1I/`Oumuamua}.
\newblock {\em Astrophys. J. Lett.}, 851(2):L38.

\bibitem[\protect\astroncite{{Baraffe} et~al.}{2014}]{BCFS}
{Baraffe}, I., {Chabrier}, G., {Fortney}, J., and {Sotin}, C. (2014).
\newblock {Planetary Internal Structures}.
\newblock {\em Protostars and Planets VI}, pages 763--786.

\bibitem[\protect\astroncite{{Baratta} et~al.}{2002}]{BLP02}
{Baratta}, G.~A., {Leto}, G., and {Palumbo}, M.~E. (2002).
\newblock {A comparison of ion irradiation and UV photolysis of CH$_{4}$ and
  CH$_{3}$OH}.
\newblock {\em Astron. Astrophys.}, 384:343--349.

\bibitem[\protect\astroncite{{Barclay} et~al.}{2017}]{BQ17}
{Barclay}, T., {Quintana}, E.~V., {Raymond}, S.~N., and {Penny}, M.~T. (2017).
\newblock {The Demographics of Rocky Free-floating Planets and their
  Detectability by WFIRST}.
\newblock {\em Astrophys. J.}, 841(2):86.

\bibitem[\protect\astroncite{{Barge} et~al.}{2017}]{BBB17}
{Barge}, L.~M., {Branscomb}, E., {Brucato}, J.~R., {Cardoso}, S.~S.~S.,
  {Cartwright}, J.~H.~E., {Danielache}, S.~O., {Galante}, D., {Kee}, T.~P.,
  {Miguel}, Y., {Mojzsis}, S., {Robinson}, K.~J., {Russell}, M.~J.,
  {Simoncini}, E., and {Sobron}, P. (2017).
\newblock {Thermodynamics, Disequilibrium, Evolution: Far-From-Equilibrium
  Geological and Chemical Considerations for Origin-Of-Life Research}.
\newblock {\em Orig. Life Evol. Biosph.}, 47(1):39--56.

\bibitem[\protect\astroncite{{Barnes} and {Heller}}{2013}]{BH13}
{Barnes}, R. and {Heller}, R. (2013).
\newblock {Habitable Planets Around White and Brown Dwarfs: The Perils of a
  Cooling Primary}.
\newblock {\em Astrobiology}, 13(3):279--291.

\bibitem[\protect\astroncite{{Baross} and {Hoffman}}{1985}]{BH85}
{Baross}, J.~A. and {Hoffman}, S.~E. (1985).
\newblock {Submarine hydrothermal vents and associated gradient environments as
  sites for the origin and evolution of life}.
\newblock {\em Orig. Life Evol. Biosph.}, 15(4):327--345.

\bibitem[\protect\astroncite{{Barr} et~al.}{2017}]{BDK17}
{Barr}, A.~C., {Dobos}, V., and {Kiss}, L.~L. (2017).
\newblock {Interior Structures and Tidal Heating in the TRAPPIST-1 Planets}.
\newblock {\em Astron. Astrophys. (arXiv:1712.05641)}.

\bibitem[\protect\astroncite{{Barr} and {McKinnon}}{2007}]{BMK07}
{Barr}, A.~C. and {McKinnon}, W.~B. (2007).
\newblock {Convection in Enceladus' ice shell: Conditions for initiation}.
\newblock {\em Geophys. Res. Lett.}, 34(9):L09202.

\bibitem[\protect\astroncite{{Barr} and {Showman}}{2009}]{BS09}
{Barr}, A.~C. and {Showman}, A.~P. (2009).
\newblock {Heat Transfer in Europa's Icy Shell}.
\newblock In {Pappalardo}, R.~T., {McKinnon}, W.~B., and {Khurana}, K.~K.,
  editors, {\em Europa}, pages 405--430. Univ. of Arizona Press.

\bibitem[\protect\astroncite{{Bartels-Rausch} et~al.}{2012}]{BBC}
{Bartels-Rausch}, T., {Bergeron}, V., {Cartwright}, J.~H.~E., {Escribano}, R.,
  {Finney}, J.~L., {Grothe}, H., {Guti\'errez}, P.~J., {Haapala}, J., {Kuhs},
  W.~F., {Pettersson}, J.~B.~C., {Price}, S.~D., {Sainz-D\'iaz}, C.~I.,
  {Stokes}, D.~J., {Strazzulla}, G., {Thomson}, E.~S., {Trinks}, H., and
  {Uras-Aytemiz}, N. (2012).
\newblock {Ice structures, patterns, and processes: A view across the
  icefields}.
\newblock {\em Rev. Mod. Phys.}, 84:885--944.

\bibitem[\protect\astroncite{{Behroozi} and {Peeples}}{2015}]{BP15}
{Behroozi}, P. and {Peeples}, M.~S. (2015).
\newblock {On the history and future of cosmic planet formation}.
\newblock {\em Mon. Not. R. Astron. Soc.}, 454(2):1811--1817.

\bibitem[\protect\astroncite{{Belbruno} et~al.}{2012}]{BM12}
{Belbruno}, E., {Moro-Mart{\'{\i}}n}, A., {Malhotra}, R., and {Savransky}, D.
  (2012).
\newblock {Chaotic Exchange of Solid Material Between Planetary Systems:
  Implications for Lithopanspermia}.
\newblock {\em Astrobiology}, 12(8):754--774.

\bibitem[\protect\astroncite{{Bellissent-Funel} et~al.}{2016}]{BHH16}
{Bellissent-Funel}, M.-C., {Hassanali}, A., {Havenith}, M., {Henchman}, R.,
  {Pohl}, P., {Sterpone}, F., {van der Spoel}, D., {Xu}, Y., and {Garcia},
  A.~E. (2016).
\newblock {Water Determines the Structure and Dynamics of Proteins}.
\newblock {\em Chem. Rev.}, 116(13):7673--7697.

\bibitem[\protect\astroncite{{Belloche} et~al.}{2008}]{BM08}
{Belloche}, A., {Menten}, K.~M., {Comito}, C., {M{\"u}ller}, H.~S.~P.,
  {Schilke}, P., {Ott}, J., {Thorwirth}, S., and {Hieret}, C. (2008).
\newblock {Detection of amino acetonitrile in Sgr B2(N)}.
\newblock {\em Astron. Astrophys.}, 482(1):179--196.

\bibitem[\protect\astroncite{{Belloche} et~al.}{2013}]{BM13}
{Belloche}, A., {M{\"u}ller}, H.~S.~P., {Menten}, K.~M., {Schilke}, P., and
  {Comito}, C. (2013).
\newblock {Complex organic molecules in the interstellar medium: IRAM 30 m line
  survey of Sagittarius B2(N) and (M)}.
\newblock {\em Astron. Astrophys.}, 559:A47.

\bibitem[\protect\astroncite{{Benitez-Nelson}}{2000}]{BN00}
{Benitez-Nelson}, C.~R. (2000).
\newblock {The biogeochemical cycling of phosphorus in marine systems}.
\newblock {\em Earth Sci. Rev.}, 51(1):109--135.

\bibitem[\protect\astroncite{{Benner} et~al.}{2012}]{BKC12}
{Benner}, S.~A., {Kim}, H.-J., and {Carrigan}, M.~A. (2012).
\newblock {Asphalt, Water, and the Prebiotic Synthesis of Ribose,
  Ribonucleosides, and RNA}.
\newblock {\em Acc. Chem. Res.}, 45(12):2025--2034.

\bibitem[\protect\astroncite{{Benner} et~al.}{2010}]{BKKR}
{Benner}, S.~A., {Kim}, H.-J., {Kim}, M.-J., and {Ricardo}, A. (2010).
\newblock {Planetary Organic Chemistry and the Origins of Biomolecules}.
\newblock {\em Cold Spring Harb. Perspect. Biol.}, 2(7):a003467.

\bibitem[\protect\astroncite{{Benner} et~al.}{2004}]{BRC04}
{Benner}, S.~A., {Ricardo}, A., and {Carrigan}, M.~A. (2004).
\newblock {Is there a common chemical model for life in the universe?}
\newblock {\em Curr. Opin. Chem. Biol.}, 8(6):672--689.

\bibitem[\protect\astroncite{{Bennett} and {Kaiser}}{2007}]{BK07}
{Bennett}, C.~J. and {Kaiser}, R.~I. (2007).
\newblock {On the Formation of Glycolaldehyde (HCOCH$_{2}$OH) and Methyl
  Formate (HCOOCH$_{3}$) in Interstellar Ice Analogs}.
\newblock {\em Astrophys. J.}, 661(2):899--909.

\bibitem[\protect\astroncite{{Bergin} et~al.}{2015}]{BBC15}
{Bergin}, E.~A., {Blake}, G.~A., {Ciesla}, F., {Hirschmann}, M.~M., and {Li},
  J. (2015).
\newblock {Tracing the ingredients for a habitable earth from interstellar
  space through planet formation}.
\newblock {\em Proc. Natl. Acad. Sci. USA}, 112(29):8965--8970.

\bibitem[\protect\astroncite{{Berta} et~al.}{2010}]{B10}
{Berta}, S., {Magnelli}, B., {Lutz}, D., {Altieri}, B., {Aussel}, H.,
  {Andreani}, P., {Bauer}, O., {Bongiovanni}, A., {Cava}, A., {Cepa}, J.,
  {Cimatti}, A., {Daddi}, E., {Dominguez}, H., {Elbaz}, D., {Feuchtgruber}, H.,
  {F{\"o}rster Schreiber}, N.~M., {Genzel}, R., {Gruppioni}, C., {Katterloher},
  R., {Magdis}, G., {Maiolino}, R., {Nordon}, R., {P{\'e}rez Garc{\'{\i}}a},
  A.~M., {Poglitsch}, A., {Popesso}, P., {Pozzi}, F., {Riguccini}, L.,
  {Rodighiero}, G., {Saintonge}, A., {Santini}, P., {Sanchez-Portal}, M.,
  {Shao}, L., {Sturm}, E., {Tacconi}, L.~J., {Valtchanov}, I., {Wetzstein}, M.,
  and {Wieprecht}, E. (2010).
\newblock {Dissecting the cosmic infra-red background with Herschel/PEP}.
\newblock {\em Astron. Astrophys.}, 518:L30.

\bibitem[\protect\astroncite{{Bertini} et~al.}{1994}]{BGLV}
{Bertini}, I., {Gray}, H.~B., {Lippard}, S.~J., and {Valentine}, J.~S. (1994).
\newblock {\em {Bioinorganic Chemistry}}.
\newblock University Science Books.

\bibitem[\protect\astroncite{{Berwick} and {Chomsky}}{2016}]{BC16}
{Berwick}, R.~C. and {Chomsky}, N. (2016).
\newblock {\em {Why Only Us: Language and Evolution}}.
\newblock The MIT Press.

\bibitem[\protect\astroncite{{Bialy} et~al.}{2015}]{BSL15}
{Bialy}, S., {Sternberg}, A., and {Loeb}, A. (2015).
\newblock {Water Formation During the Epoch of First Metal Enrichment}.
\newblock {\em Astrophys. J. Lett.}, 804(2):L29.

\bibitem[\protect\astroncite{{Bieri}}{1964}]{Bieri}
{Bieri}, R. (1964).
\newblock {Huminoids on other planets?}
\newblock {\em Am. Sci.}, 52(4):452--458.

\bibitem[\protect\astroncite{{Bjerrum} and {Canfield}}{2002}]{BC02}
{Bjerrum}, C.~J. and {Canfield}, D.~E. (2002).
\newblock {Ocean productivity before about 1.9Gyr ago limited by phosphorus
  adsorption onto iron oxides}.
\newblock {\em Nature}, 417(6885):159--162.

\bibitem[\protect\astroncite{{Blount} et~al.}{2008}]{BBL08}
{Blount}, Z.~D., {Borland}, C.~Z., and {Lenski}, R.~E. (2008).
\newblock {Historical contingency and the evolution of a key innovation in an
  experimental population of Escherichia coli}.
\newblock {\em Proc. Natl. Acad. Sci. USA}, 105(23):7899--7906.

\bibitem[\protect\astroncite{{Bogonovich}}{2011}]{Bog11}
{Bogonovich}, M. (2011).
\newblock {Intelligence's likelihood and evolutionary time frame}.
\newblock {\em Int. J. Astrobiol.}, 10(2):113--122.

\bibitem[\protect\astroncite{{Bolin} et~al.}{2018}]{BWF17}
{Bolin}, B.~T., {Weaver}, H.~A., {Fernandez}, Y.~R., {Lisse}, C.~M.,
  {Huppenkothen}, D., {Jones}, R.~L., {Juri{\'c}}, M., {Moeyens}, J.,
  {Schambeau}, C.~A., {Slater}, C.~T., {Ivezi{\'c}}, {\v Z}., and {Connolly},
  A.~J. (2018).
\newblock {APO Time-resolved Color Photometry of Highly Elongated Interstellar
  Object 1I/`Oumuamua}.
\newblock {\em Astrophys. J. Lett.}, 852(1):L2.

\bibitem[\protect\astroncite{{Bolton} et~al.}{2015}]{BB15}
{Bolton}, S.~J., {Bagenal}, F., {Blanc}, M., {Cassidy}, T., {Chan{\'e}}, E.,
  {Jackman}, C., {Jia}, X., {Kotova}, A., {Krupp}, N., {Milillo}, A.,
  {Plainaki}, C., {Smith}, H.~T., and {Waite}, H. (2015).
\newblock {Jupiter's Magnetosphere: Plasma Sources and Transport}.
\newblock {\em Space Sci. Rev.}, 192(1-4):209--236.

\bibitem[\protect\astroncite{{Bondi}}{1952}]{Bond52}
{Bondi}, H. (1952).
\newblock {On spherically symmetrical accretion}.
\newblock {\em Mon. Not. R. Astron. Soc.}, 112(2):195--204.

\bibitem[\protect\astroncite{{Bottke} et~al.}{2005}]{BDN05}
{Bottke}, W.~F., {Durda}, D.~D., {Nesvorn{\'y}}, D., {Jedicke}, R.,
  {Morbidelli}, A., {Vokrouhlick{\'y}}, D., and {Levison}, H. (2005).
\newblock {The fossilized size distribution of the main asteroid belt}.
\newblock {\em Icarus}, 175(1):111--140.

\bibitem[\protect\astroncite{{Bouquet} et~al.}{2017}]{BGW}
{Bouquet}, A., {Glein}, C.~R., {Wyrick}, D., and {Waite}, J.~H. (2017).
\newblock {Alternative Energy: Production of H$_{2}$ by Radiolysis of Water in
  the Rocky Cores of Icy Bodies}.
\newblock {\em Astrophys. J. Lett.}, 840(1):L8.

\bibitem[\protect\astroncite{{Branscomb} et~al.}{2017}]{BBG17}
{Branscomb}, E., {Biancalani}, T., {Goldenfeld}, N., and {Russell}, M. (2017).
\newblock {Escapement mechanisms and the conversion of disequilibria; the
  engines of creation}.
\newblock {\em Phys. Rep.}, 677:1--60.

\bibitem[\protect\astroncite{{Braun} et~al.}{2003}]{BGL03}
{Braun}, D., {Goddard}, N.~L., and {Libchaber}, A. (2003).
\newblock {Exponential DNA Replication by Laminar Convection}.
\newblock {\em Phys. Rev. Lett.}, 91(15):158103.

\bibitem[\protect\astroncite{{Bromm}}{2013}]{Bro13}
{Bromm}, V. (2013).
\newblock {Formation of the first stars}.
\newblock {\em Rep. Prog. Phys.}, 76(11):112901.

\bibitem[\protect\astroncite{{Buchhave} et~al.}{2014}]{BB14}
{Buchhave}, L.~A., {Bizzarro}, M., {Latham}, D.~W., {Sasselov}, D., {Cochran},
  W.~D., {Endl}, M., {Isaacson}, H., {Juncher}, D., and {Marcy}, G.~W. (2014).
\newblock {Three regimes of extrasolar planet radius inferred from host star
  metallicities}.
\newblock {\em Nature}, 509(7502):593--595.

\bibitem[\protect\astroncite{{Buchhave} et~al.}{2012}]{BL12}
{Buchhave}, L.~A., {Latham}, D.~W., {Johansen}, A., {Bizzarro}, M., {Torres},
  G., {Rowe}, J.~F., {Batalha}, N.~M., {Borucki}, W.~J., {Brugamyer}, E.,
  {Caldwell}, C., {Bryson}, S.~T., {Ciardi}, D.~R., {Cochran}, W.~D., {Endl},
  M., {Esquerdo}, G.~A., {Ford}, E.~B., {Geary}, J.~C., {Gilliland}, R.~L.,
  {Hansen}, T., {Isaacson}, H., {Laird}, J.~B., {Lucas}, P.~W., {Marcy}, G.~W.,
  {Morse}, J.~A., {Robertson}, P., {Shporer}, A., {Stefanik}, R.~P., {Still},
  M., and {Quinn}, S.~N. (2012).
\newblock {An abundance of small exoplanets around stars with a wide range of
  metallicities}.
\newblock {\em Nature}, 486(7403):375--377.

\bibitem[\protect\astroncite{{Budin} et~al.}{2009}]{BBS09}
{Budin}, I., {Bruckner}, R.~J., and {Szostak}, J.~W. (2009).
\newblock {Formation of Protocell-like Vesicles in a Thermal Diffusion Column}.
\newblock {\em J. Am. Chem. Soc.}, 131(28):9628--9629.

\bibitem[\protect\astroncite{{Budin} and {Szostak}}{2010}]{BS10}
{Budin}, I. and {Szostak}, J.~W. (2010).
\newblock {Expanding Roles for Diverse Physical Phenomena During the Origin of
  Life}.
\newblock {\em Annu. Rev. Biophys.}, 39:245--263.

\bibitem[\protect\astroncite{{Burcar} et~al.}{2015}]{BBT15}
{Burcar}, B.~T., {Barge}, L.~M., {Trail}, D., {Watson}, E.~B., {Russell},
  M.~J., and {McGown}, L.~B. (2015).
\newblock {RNA Oligomerization in Laboratory Analogues of Alkaline Hydrothermal
  Vent Systems}.
\newblock {\em Astrobiology}, 15(7):509--522.

\bibitem[\protect\astroncite{{Burchell}}{2004}]{Bur04}
{Burchell}, M.~J. (2004).
\newblock {Panspermia today}.
\newblock {\em Int. J. Astrobiol.}, 3(2):73--80.

\bibitem[\protect\astroncite{{Butterfield}}{2009}]{Butt09}
{Butterfield}, N.~J. (2009).
\newblock {Oxygen, animals and oceanic ventilation: an alternative view}.
\newblock {\em Geobiology}, 7(1):1--7.

\bibitem[\protect\astroncite{{Butterfield}}{2011}]{Butt11}
{Butterfield}, N.~J. (2011).
\newblock {Animals and the invention of the Phanerozoic Earth system}.
\newblock {\em Trends Ecol. Evol.}, 26(2):81--87.

\bibitem[\protect\astroncite{{Calcott} and {Sterelny}}{2011}]{CS11}
{Calcott}, B. and {Sterelny}, K. (2011).
\newblock {\em {The Major Transitions in Evolution Revisited}}.
\newblock The MIT Press.

\bibitem[\protect\astroncite{{Callahan} et~al.}{2011}]{CSC11}
{Callahan}, M.~P., {Smith}, K.~E., {Cleaves}, H.~J., {Ruzicka}, J., {Stern},
  J.~C., {Glavin}, D.~P., {House}, C.~H., and {Dworkin}, J.~P. (2011).
\newblock {Carbonaceous meteorites contain a wide range of extraterrestrial
  nucleobases}.
\newblock {\em Proc. Natl. Acad. Sci. USA}, 108(34):13995--13998.

\bibitem[\protect\astroncite{{Cantine} and {Fournier}}{2018}]{CF17}
{Cantine}, M.~D. and {Fournier}, G.~P. (2018).
\newblock {Environmental Adaptation from the Origin of Life to the Last
  Universal Common Ancestor}.
\newblock {\em Orig. Life Evol. Biosph.}, 48(1):35--54.

\bibitem[\protect\astroncite{{Carroll}}{2001}]{Carr01}
{Carroll}, S.~B. (2001).
\newblock {Chance and necessity: the evolution of morphological complexity and
  diversity}.
\newblock {\em Nature}, 409(6823):1102--1109.

\bibitem[\protect\astroncite{{Carter}}{1983}]{Cart83}
{Carter}, B. (1983).
\newblock {The Anthropic Principle and its Implications for Biological
  Evolution}.
\newblock {\em Phil. Trans. R. Soc. A}, 310(1512):347--363.

\bibitem[\protect\astroncite{{Carter}}{2008}]{Cart08}
{Carter}, B. (2008).
\newblock {Five- or six-step scenario for evolution?}
\newblock {\em Int. J. Astrobiol.}, 7(2):177--182.

\bibitem[\protect\astroncite{{Cassidy} et~al.}{2010}]{CC10}
{Cassidy}, T., {Coll}, P., {Raulin}, F., {Carlson}, R.~W., {Johnson}, R.~E.,
  {Loeffler}, M.~J., {Hand}, K.~P., and {Baragiola}, R.~A. (2010).
\newblock {Radiolysis and Photolysis of Icy Satellite Surfaces: Experiments and
  Theory}.
\newblock {\em Space Sci. Rev.}, 153(1-4):299--315.

\bibitem[\protect\astroncite{{Castillo-Rogez} and {McCord}}{2010}]{CRM10}
{Castillo-Rogez}, J.~C. and {McCord}, T.~B. (2010).
\newblock {Ceres' evolution and present state constrained by shape data}.
\newblock {\em Icarus}, 205(2):443--459.

\bibitem[\protect\astroncite{{Catling} et~al.}{2005}]{CGZ05}
{Catling}, D.~C., {Glein}, C.~R., {Zahnle}, K.~J., and {McKay}, C.~P. (2005).
\newblock {Why O$_{2}$ Is Required by Complex Life on Habitable Planets and the
  Concept of Planetary ``Oxygenation Time''}.
\newblock {\em Astrobiology}, 5(3):415--438.

\bibitem[\protect\astroncite{{Catling} and {Kasting}}{2017}]{CaKa}
{Catling}, D.~C. and {Kasting}, J.~F. (2017).
\newblock {\em {Atmospheric Evolution on Inhabited and Lifeless Worlds}}.
\newblock Cambridge Univ. Press.

\bibitem[\protect\astroncite{{Charette} and {Smith}}{2010}]{CS10}
{Charette}, M.~A. and {Smith}, W.~H.~F. (2010).
\newblock {The Volume of Earth's Ocean}.
\newblock {\em Oceanography}, 23(2):112--114.

\bibitem[\protect\astroncite{{Chen} et~al.}{2018}]{CFL}
{Chen}, H., {Forbes}, J.~C., and {Loeb}, A. (2018).
\newblock {Habitable Evaporated Cores and the Occurrence of Panspermia Near the
  Galactic Center}.
\newblock {\em Astrophys. J. Lett.}, 855(1):L1.

\bibitem[\protect\astroncite{{Chen} and {Kipping}}{2017}]{CK17}
{Chen}, J. and {Kipping}, D. (2017).
\newblock {Probabilistic Forecasting of the Masses and Radii of Other Worlds}.
\newblock {\em Astrophys. J.}, 834(1):17.

\bibitem[\protect\astroncite{{Chen} et~al.}{2015}]{CLV15}
{Chen}, X., {Ling}, H.-F., {Vance}, D., {Shields-Zhou}, G.~A., {Zhu}, M.,
  {Poulton}, S.~W., {Och}, L.~M., {Jiang}, S.-Y., {Li}, D., {Cremonese}, L.,
  and {Archer}, C. (2015).
\newblock {Rise to modern levels of ocean oxygenation coincided with the
  Cambrian radiation of animals}.
\newblock {\em Nat. Commun.}, 6:7142.

\bibitem[\protect\astroncite{{Chivian} et~al.}{2008}]{CB08}
{Chivian}, D., {Brodie}, E.~L., {Alm}, E.~J., {Culley}, D.~E., {Dehal}, P.~S.,
  {DeSantis}, T.~Z., {Gihring}, T.~M., {Lapidus}, A., {Lin}, L.-H., {Lowry},
  S.~R., {Moser}, D.~P., {Richardson}, P.~M., {Southam}, G., {Wanger}, G.,
  {Pratt}, L.~M., {Andersen}, G.~L., {Hazen}, T.~C., {Brockman}, F.~J.,
  {Arkin}, A.~P., and {Onstott}, T.~C. (2008).
\newblock {Environmental Genomics Reveals a Single-Species Ecosystem Deep
  Within Earth}.
\newblock {\em Science}, 322(5899):275--278.

\bibitem[\protect\astroncite{{Choblet} et~al.}{2017a}]{CTCB17}
{Choblet}, G., {Tobie}, G., {Sotin}, C., {B{\v e}hounkov{\'a}}, M., {{\v
  C}adek}, O., {Postberg}, F., and {Sou{\v c}ek}, O. (2017a).
\newblock {Powering prolonged hydrothermal activity inside Enceladus}.
\newblock {\em Nat. Astron.}, 1:841--847.

\bibitem[\protect\astroncite{{Choblet} et~al.}{2017b}]{CTS17}
{Choblet}, G., {Tobie}, G., {Sotin}, C., {Kalousov{\'a}}, K., and {Grasset}, O.
  (2017b).
\newblock {Heat transport in the high-pressure ice mantle of large icy moons}.
\newblock {\em Icarus}, 285:252--262.

\bibitem[\protect\astroncite{{Choukroun} and {Grasset}}{2007}]{CG07}
{Choukroun}, M. and {Grasset}, O. (2007).
\newblock {Thermodynamic model for water and high-pressure ices up to 2.2 GPa
  and down to the metastable domain}.
\newblock {\em J. Chem. Phys.}, 127(12):124506--124506.

\bibitem[\protect\astroncite{{Chyba} and {Sagan}}{1992}]{CS92}
{Chyba}, C. and {Sagan}, C. (1992).
\newblock {Endogenous production, exogenous delivery and impact-shock synthesis
  of organic molecules: an inventory for the origins of life}.
\newblock {\em Nature}, 355(6356):125--132.

\bibitem[\protect\astroncite{{Chyba}}{2000}]{Chy00}
{Chyba}, C.~F. (2000).
\newblock {Energy for microbial life on Europa}.
\newblock {\em Nature}, 403(6768):381--382.

\bibitem[\protect\astroncite{{Chyba} and {Hand}}{2001}]{CH01}
{Chyba}, C.~F. and {Hand}, K.~P. (2001).
\newblock {Life Without Photosynthesis}.
\newblock {\em Science}, 292(5524):2026--2027.

\bibitem[\protect\astroncite{{Chyba} and {Phillips}}{2001}]{CP01}
{Chyba}, C.~F. and {Phillips}, C.~B. (2001).
\newblock {Possible ecosystems and the search for life on Europa}.
\newblock {\em Proc. Natl. Acad. Sci. USA}, 98(3):801--804.

\bibitem[\protect\astroncite{{Chyba} and {Phillips}}{2002}]{CP02}
{Chyba}, C.~F. and {Phillips}, C.~B. (2002).
\newblock {Europa as an Abode of Life}.
\newblock {\em Orig. Life Evol. Biosph.}, 32(1):47--67.

\bibitem[\protect\astroncite{{Ciesla} et~al.}{2015}]{CMPA15}
{Ciesla}, F.~J., {Mulders}, G.~D., {Pascucci}, I., and {Apai}, D. (2015).
\newblock {Volatile Delivery to Planets from Water-rich Planetesimals around
  Low Mass Stars}.
\newblock {\em Astrophys. J.}, 804(1):9.

\bibitem[\protect\astroncite{{Ciesla} and {Sandford}}{2012}]{CS12}
{Ciesla}, F.~J. and {Sandford}, S.~A. (2012).
\newblock {Organic Synthesis via Irradiation and Warming of Ice Grains in the
  Solar Nebula}.
\newblock {\em Science}, 336(6080):452--454.

\bibitem[\protect\astroncite{{Cleeves} et~al.}{2014}]{CBA14}
{Cleeves}, L.~I., {Bergin}, E.~A., {Alexander}, C.~M.~O.~., {Du}, F.,
  {Graninger}, D., {{\"O}berg}, K.~I., and {Harries}, T.~J. (2014).
\newblock {The ancient heritage of water ice in the solar system}.
\newblock {\em Science}, 345(6204):1590--1593.

\bibitem[\protect\astroncite{{Cockell}}{2014}]{Cock14}
{Cockell}, C.~S. (2014).
\newblock {Habitable worlds with no signs of life}.
\newblock {\em Phil. Trans. R. Soc. A}, 372(2014):20130082--20130082.

\bibitem[\protect\astroncite{{Cockell} et~al.}{2016}]{CB16}
{Cockell}, C.~S., {Bush}, T., {Bryce}, C., {Direito}, S., {Fox-Powell}, M.,
  {Harrison}, J.~P., {Lammer}, H., {Landenmark}, H., {Martin-Torres}, J.,
  {Nicholson}, N., {Noack}, L., {O'Malley-James}, J., {Payler}, S.~J.,
  {Rushby}, A., {Samuels}, T., {Schwendner}, P., {Wadsworth}, J., and
  {Zorzano}, M.~P. (2016).
\newblock {Habitability: A Review}.
\newblock {\em Astrobiology}, 16(1):89--117.

\bibitem[\protect\astroncite{{Combe} et~al.}{2016}]{CMT16}
{Combe}, J.-P., {McCord}, T.~B., {Tosi}, F., {Ammannito}, E., {Carrozzo},
  F.~G., {De Sanctis}, M.~C., {Raponi}, A., {Byrne}, S., {Landis}, M.~E.,
  {Hughson}, K.~H.~G., {Raymond}, C.~A., and {Russell}, C.~T. (2016).
\newblock {Detection of local H$_{2}$O exposed at the surface of Ceres}.
\newblock {\em Science}, 353(6303):aaf3010.

\bibitem[\protect\astroncite{{Cook} et~al.}{2016}]{CRG16}
{Cook}, N.~V., {Ragozzine}, D., {Granvik}, M., and {Stephens}, D.~C. (2016).
\newblock {Realistic Detectability of Close Interstellar Comets}.
\newblock {\em Astrophys. J.}, 825(1):51.

\bibitem[\protect\astroncite{{Cooper} et~al.}{2001}]{CJM01}
{Cooper}, J.~F., {Johnson}, R.~E., {Mauk}, B.~H., {Garrett}, H.~B., and
  {Gehrels}, N. (2001).
\newblock {Energetic Ion and Electron Irradiation of the Icy Galilean
  Satellites}.
\newblock {\em Icarus}, 149(1):133--159.

\bibitem[\protect\astroncite{{Corballis}}{2011}]{Cor11}
{Corballis}, M.~C. (2011).
\newblock {\em {The Recursive Mind: The Origins of Human Language, Thought, and
  Civilization}}.
\newblock Princeton Univ. Press.

\bibitem[\protect\astroncite{{Cravens}}{2004}]{Crav04}
{Cravens}, T.~E. (2004).
\newblock {\em {Physics of Solar System Plasmas}}.
\newblock Cambridge Univ. Press.

\bibitem[\protect\astroncite{{{\'C}uk}}{2018}]{Cuk18}
{{\'C}uk}, M. (2018).
\newblock {1I/`Oumuamua as a Tidal Disruption Fragment from a Binary Star
  System}.
\newblock {\em Astrophys. J. Lett.}, 852:L15.

\bibitem[\protect\astroncite{{Cunningham} et~al.}{2007}]{CJ07}
{Cunningham}, M.~R., {Jones}, P.~A., {Godfrey}, P.~D., {Cragg}, D.~M., {Bains},
  I., {Burton}, M.~G., {Calisse}, P., {Crighton}, N.~H.~M., {Curran}, S.~J.,
  {Davis}, T.~M., {Dempsey}, J.~T., {Fulton}, B., {Hidas}, M.~G., {Hill}, T.,
  {Kedziora-Chudczer}, L., {Minier}, V., {Pracy}, M.~B., {Purcell}, C.,
  {Shobbrook}, J., and {Travouillon}, T. (2007).
\newblock {A search for propylene oxide and glycine in Sagittarius B2 (LMH) and
  Orion}.
\newblock {\em Mon. Not. R. Astron. Soc.}, 376(3):1201--1210.

\bibitem[\protect\astroncite{{Da Silva} et~al.}{2015}]{DMD15}
{Da Silva}, L., {Maurel}, M.-C., and {Deamer}, D. (2015).
\newblock {Salt-Promoted Synthesis of RNA-like Molecules in Simulated
  Hydrothermal Conditions}.
\newblock {\em J. Mol. Evol.}, 80(2):86--97.

\bibitem[\protect\astroncite{{Dai} and {Guerras}}{2018}]{DG18}
{Dai}, X. and {Guerras}, E. (2018).
\newblock {Probing Extragalactic Planets Using Quasar Microlensing}.
\newblock {\em Astrophys. J. Lett.}, 853(2):L27.

\bibitem[\protect\astroncite{{D'Amico} et~al.}{2006}]{DAC06}
{D'Amico}, S., {Collins}, T., {Marx}, J.-C., {Feller}, G., and {Gerday}, C.
  (2006).
\newblock {Psychrophilic microorganisms: challenges for life}.
\newblock {\em EMBO reports}, 7(4):385--389.

\bibitem[\protect\astroncite{{Dartnell}}{2011}]{Dart11}
{Dartnell}, L.~R. (2011).
\newblock {Ionizing Radiation and Life}.
\newblock {\em Astrobiology}, 11(6):551--582.

\bibitem[\protect\astroncite{{Davies} and {Walker}}{2016}]{DaWa}
{Davies}, P.~C.~W. and {Walker}, S.~I. (2016).
\newblock {The hidden simplicity of biology}.
\newblock {\em Rep. Prog. Phys.}, 79(10):102601.

\bibitem[\protect\astroncite{{de Duve}}{1995}]{Duve95}
{de Duve}, C. (1995).
\newblock {\em {Vital Dust: Life as a Cosmic Imperative}}.
\newblock Basic Books.

\bibitem[\protect\astroncite{{De Duve}}{2005}]{deDu}
{De Duve}, C. (2005).
\newblock {\em {Singularities: Landmarks on the Pathways of Life}}.
\newblock Cambridge Univ. Press.

\bibitem[\protect\astroncite{{de Duve}}{2007}]{deDu07}
{de Duve}, C. (2007).
\newblock {The origin of eukaryotes: a reappraisal}.
\newblock {\em Nat. Rev. Genet.}, 8(5):395--403.

\bibitem[\protect\astroncite{{de Juan Ovelar} et~al.}{2012}]{DKB12}
{de Juan Ovelar}, M., {Kruijssen}, J.~M.~D., {Bressert}, E., {Testi}, L.,
  {Bastian}, N., and {C{\'a}novas}, H. (2012).
\newblock {Can habitable planets form in clustered environments?}
\newblock {\em Astron. Astrophys.}, 546:L1.

\bibitem[\protect\astroncite{{de Sanctis} et~al.}{2016}]{DS16}
{de Sanctis}, M.~C., {Raponi}, A., {Ammannito}, E., {Ciarniello}, M., {Toplis},
  M.~J., {McSween}, H.~Y., {Castillo-Rogez}, J.~C., {Ehlmann}, B.~L.,
  {Carrozzo}, F.~G., {Marchi}, S., {Tosi}, F., {Zambon}, F., {Capaccioni}, F.,
  {Capria}, M.~T., {Fonte}, S., {Formisano}, M., {Frigeri}, A., {Giardino}, M.,
  {Longobardo}, A., {Magni}, G., {Palomba}, E., {McFadden}, L.~A., {Pieters},
  C.~M., {Jaumann}, R., {Schenk}, P., {Mugnuolo}, R., {Raymond}, C.~A., and
  {Russell}, C.~T. (2016).
\newblock {Bright carbonate deposits as evidence of aqueous alteration on (1)
  Ceres}.
\newblock {\em Nature}, 536(7614):54--57.

\bibitem[\protect\astroncite{{De Waal}}{2016}]{DW16}
{De Waal}, F. (2016).
\newblock {\em {Are We Smart Enough to Know How Smart Animals Are?}}
\newblock W. W. Norton \& Company.

\bibitem[\protect\astroncite{{Deamer} et~al.}{2002}]{DD02}
{Deamer}, D., {Dworkin}, J.~P., {Sandford}, S.~A., {Bernstein}, M.~P., and
  {Allamandola}, L.~J. (2002).
\newblock {The First Cell Membranes}.
\newblock {\em Astrobiology}, 2(4):371--381.

\bibitem[\protect\astroncite{{Deamer} et~al.}{2006}]{DSR06}
{Deamer}, D., {Singaram}, S., {Rajamani}, S., {Kompanichenko}, V., and
  {Guggenheim}, S. (2006).
\newblock {Self-assembly processes in the prebiotic environment}.
\newblock {\em Phil. Trans. R. Soc. B}, 361(1474):1809--1818.

\bibitem[\protect\astroncite{{Deamer} and {Weber}}{2010}]{DW10}
{Deamer}, D. and {Weber}, A.~L. (2010).
\newblock {Bioenergetics and Life's Origins}.
\newblock {\em Cold Spring Harb. Perspect. Biol.}, 2(2):a004929.

\bibitem[\protect\astroncite{{Deamer}}{1997}]{Deam97}
{Deamer}, D.~W. (1997).
\newblock {The first living systems: a bioenergetic perspective}.
\newblock {\em Microbiol. Mol. Biol. Rev.}, 61(2):239--261.

\bibitem[\protect\astroncite{{Debes} and {Sigurdsson}}{2007}]{DS07}
{Debes}, J.~H. and {Sigurdsson}, S. (2007).
\newblock {The Survival Rate of Ejected Terrestrial Planets with Moons}.
\newblock {\em Astrophys. J. Lett.}, 668(2):L167--L170.

\bibitem[\protect\astroncite{{Delaye} and {Lazcano}}{2005}]{DL05}
{Delaye}, L. and {Lazcano}, A. (2005).
\newblock {Prebiological evolution and the physics of the origin of life}.
\newblock {\em Phys. Life Rev.}, 2(1):47--64.

\bibitem[\protect\astroncite{{D'Elia} et~al.}{2008}]{DVR08}
{D'Elia}, T., {Veerapaneni}, R., and {Rogers}, S.~O. (2008).
\newblock {Isolation of Microbes from Lake Vostok Accretion Ice}.
\newblock {\em Appl. Environ. Microbiol.}, 74(15):4962--4965.

\bibitem[\protect\astroncite{{Di Stefano} and {Ray}}{2016}]{DSR16}
{Di Stefano}, R. and {Ray}, A. (2016).
\newblock {Globular Clusters as Cradles of Life and Advanced Civilizations}.
\newblock {\em Astrophys. J.}, 827(1):54.

\bibitem[\protect\astroncite{{Do} et~al.}{2018}]{DTT18}
{Do}, A., {Tucker}, M.~A., and {Tonry}, J. (2018).
\newblock {Interstellar Interlopers: Number Density and Origins of
  'Oumuamua-like Objects}.
\newblock {\em Astrophys. J. Lett.}, 855(1):L10.

\bibitem[\protect\astroncite{{Dong} et~al.}{2017a}]{DHL17}
{Dong}, C., {Huang}, Z., {Lingam}, M., {T{\'o}th}, G., {Gombosi}, T., and
  {Bhattacharjee}, A. (2017a).
\newblock {The Dehydration of Water Worlds via Atmospheric Losses}.
\newblock {\em Astrophys. J. Lett.}, 847(1):L4.

\bibitem[\protect\astroncite{{Dong} et~al.}{2018}]{DJL18}
{Dong}, C., {Jin}, M., {Lingam}, M., {Airapetian}, V.~S., {Ma}, Y., and {van
  der Holst}, B. (2018).
\newblock {Atmospheric escape from the TRAPPIST-1 planets and implications for
  habitability}.
\newblock {\em Proc. Natl. Acad. Sci. USA}, 115(2):260--265.

\bibitem[\protect\astroncite{{Dong} et~al.}{2017b}]{DLMC}
{Dong}, C., {Lingam}, M., {Ma}, Y., and {Cohen}, O. (2017b).
\newblock {Is Proxima Centauri b Habitable? A Study of Atmospheric Loss}.
\newblock {\em Astrophys. J. Lett.}, 837(2):L26.

\bibitem[\protect\astroncite{{Dragani{\'c}}}{2005}]{Drag}
{Dragani{\'c}}, I.~G. (2005).
\newblock {Radiolysis of water: a look at its origin and occurrence in the
  nature}.
\newblock {\em Rad. Phys. Chem.}, 72(2-3):181--186.

\bibitem[\protect\astroncite{{Dragani{\'c}} et~al.}{1991}]{DBD91}
{Dragani{\'c}}, I.~G., {Bjergbakke}, E., {Dragani{\'c}}, Z.~D., and {Sehested},
  K. (1991).
\newblock {Decomposition of ocean waters by potassium-40 radiation 3800 Ma ago
  as a source of oxygen and oxidizing species}.
\newblock {\em Precambrian Res.}, 52(3-4):337--345.

\bibitem[\protect\astroncite{{Dragani{\'c}} et~al.}{1983}]{DDA}
{Dragani{\'c}}, I.~G., {Dragani{\'c}}, Z.~D., and {Altiparmakov}, D. (1983).
\newblock {Natural nuclear reactors and ionizing radiation in the Precambrian}.
\newblock {\em Precambrian Res.}, 20(2-4):283--298.

\bibitem[\protect\astroncite{{Dressing} and {Charbonneau}}{2015}]{DC15}
{Dressing}, C.~D. and {Charbonneau}, D. (2015).
\newblock {The Occurrence of Potentially Habitable Planets Orbiting M Dwarfs
  Estimated from the Full Kepler Dataset and an Empirical Measurement of the
  Detection Sensitivity}.
\newblock {\em Astrophys. J.}, 807(1):45.

\bibitem[\protect\astroncite{{Dyson}}{1999}]{Dys99}
{Dyson}, F. (1999).
\newblock {\em {Origins of Life}}.
\newblock Cambridge Univ. Press.

\bibitem[\protect\astroncite{{Dyson}}{2003}]{Dys03}
{Dyson}, F.~J. (2003).
\newblock {Looking for life in unlikely places: reasons why planets may not be
  the best places to look for life}.
\newblock {\em Int. J. Astrobiol.}, 2(2):103--110.

\bibitem[\protect\astroncite{{Edgar}}{2004}]{Ed04}
{Edgar}, R. (2004).
\newblock {A review of Bondi-Hoyle-Lyttleton accretion}.
\newblock {\em New Astron. Rev.}, 48(10):843--859.

\bibitem[\protect\astroncite{{Ehrenreich} and {Cassan}}{2007}]{EC07}
{Ehrenreich}, D. and {Cassan}, A. (2007).
\newblock {Are extrasolar oceans common throughout the Galaxy?}
\newblock {\em Astron. Nachr.}, 328(8):789--792.

\bibitem[\protect\astroncite{{Ehrenreich} et~al.}{2006}]{EL06}
{Ehrenreich}, D., {Lecavelier des Etangs}, A., {Beaulieu}, J.-P., and
  {Grasset}, O. (2006).
\newblock {On the Possible Properties of Small and Cold Extrasolar Planets: Is
  OGLE 2005-BLG-390Lb Entirely Frozen?}
\newblock {\em Astrophys. J.}, 651(1):535--543.

\bibitem[\protect\astroncite{{Elsila} et~al.}{2007}]{ED07}
{Elsila}, J.~E., {Dworkin}, J.~P., {Bernstein}, M.~P., {Martin}, M.~P., and
  {Sandford}, S.~A. (2007).
\newblock {Mechanisms of Amino Acid Formation in Interstellar Ice Analogs}.
\newblock {\em Astrophys. J.}, 660(1):911--918.

\bibitem[\protect\astroncite{{Embley} and {Martin}}{2006}]{EM06}
{Embley}, T.~M. and {Martin}, W. (2006).
\newblock {Eukaryotic evolution, changes and challenges}.
\newblock {\em Nature}, 440(7084):623--630.

\bibitem[\protect\astroncite{{Engelhardt} et~al.}{2017}]{EJ17}
{Engelhardt}, T., {Jedicke}, R., {Vere{\v s}}, P., {Fitzsimmons}, A.,
  {Denneau}, L., {Beshore}, E., and {Meinke}, B. (2017).
\newblock {An Observational Upper Limit on the Interstellar Number Density of
  Asteroids and Comets}.
\newblock {\em Astron. J.}, 153(3):133.

\bibitem[\protect\astroncite{{Fagents}}{2003}]{Fa03}
{Fagents}, S.~A. (2003).
\newblock {Considerations for effusive cryovolcanism on Europa: The
  post-Galileo perspective}.
\newblock {\em J. Geophys. Res. E}, 108(E12):13--1.

\bibitem[\protect\astroncite{{Falkowski} et~al.}{2008}]{FFD08}
{Falkowski}, P.~G., {Fenchel}, T., and {Delong}, E.~F. (2008).
\newblock {The Microbial Engines That Drive Earth's Biogeochemical Cycles}.
\newblock {\em Science}, 320(5879):1034--1039.

\bibitem[\protect\astroncite{{Faure}}{1998}]{Fau98}
{Faure}, G. (1998).
\newblock {\em {Principles and Applications of Geochemistry}}.
\newblock Prentice-Hall.

\bibitem[\protect\astroncite{{Feng} and {Jones}}{2018}]{FJ17}
{Feng}, F. and {Jones}, H.~R.~A. (2018).
\newblock {`Oumuamua as a Messenger from the Local Association}.
\newblock {\em Astrophys. J. Lett.}, 852(2):L27.

\bibitem[\protect\astroncite{{Ferri{\`e}re}}{2001}]{Ferr01}
{Ferri{\`e}re}, K.~M. (2001).
\newblock {The interstellar environment of our galaxy}.
\newblock {\em Rev. Mod. Phys.}, 73(4):1031--1066.

\bibitem[\protect\astroncite{{Ferris}}{1993}]{Fer93}
{Ferris}, J.~P. (1993).
\newblock {Catalysis and prebiotic RNA synthesis}.
\newblock {\em Orig. Life Evol. Biosph.}, 23(5-6):307--315.

\bibitem[\protect\astroncite{{Feynman} et~al.}{1993}]{FSWG}
{Feynman}, J., {Spitale}, G., {Wang}, J., and {Gabriel}, S. (1993).
\newblock {Interplanetary proton fluence model - JPL 1991}.
\newblock {\em J. Geophys. Res.}, 98(A8):13281--13294.

\bibitem[\protect\astroncite{{Field} et~al.}{1998}]{FBRF}
{Field}, C.~B., {Behrenfeld}, M.~J., {Randerson}, J.~T., and {Falkowski}, P.
  (1998).
\newblock {Primary Production of the Biosphere: Integrating Terrestrial and
  Oceanic Components}.
\newblock {\em Science}, 281(5374):237--240.

\bibitem[\protect\astroncite{{F{\"o}llmi}}{1996}]{Foll96}
{F{\"o}llmi}, K.~B. (1996).
\newblock {The phosphorus cycle, phosphogenesis and marine phosphate-rich
  deposits}.
\newblock {\em Earth Sci. Rev.}, 40(1-2):55--124.

\bibitem[\protect\astroncite{{Forbes} and {Loeb}}{2017}]{FL17}
{Forbes}, J.~C. and {Loeb}, A. (2017).
\newblock {Evaporation of planetary atmospheres due to XUV illumination by
  quasars}.
\newblock {\em submitted to Mon. Not. R. Astron. Soc. (arXiv:1705.06741)}.

\bibitem[\protect\astroncite{{Forgan} et~al.}{2017}]{FDCL}
{Forgan}, D., {Dayal}, P., {Cockell}, C., and {Libeskind}, N. (2017).
\newblock {Evaluating galactic habitability using high-resolution cosmological
  simulations of galaxy formation}.
\newblock {\em Int. J. Astrobiol.}, 16(1):60--73.

\bibitem[\protect\astroncite{{Fox-Powell} et~al.}{2016}]{FHCC}
{Fox-Powell}, M.~G., {Hallsworth}, J.~E., {Cousins}, C.~R., and {Cockell},
  C.~S. (2016).
\newblock {Ionic Strength Is a Barrier to the Habitability of Mars}.
\newblock {\em Astrobiology}, 16(6):427--442.

\bibitem[\protect\astroncite{{France} et~al.}{2013}]{FF13}
{France}, K., {Froning}, C.~S., {Linsky}, J.~L., {Roberge}, A., {Stocke},
  J.~T., {Tian}, F., {Bushinsky}, R., {D{\'e}sert}, J.-M., {Mauas}, P.,
  {Vieytes}, M., and {Walkowicz}, L.~M. (2013).
\newblock {The Ultraviolet Radiation Environment around M dwarf Exoplanet Host
  Stars}.
\newblock {\em Astrophys. J.}, 763(2):149.

\bibitem[\protect\astroncite{{Froelich} et~al.}{1982}]{FBL}
{Froelich}, P.~N., {Bender}, M.~L., and {Luedtke}, N.~A. (1982).
\newblock {The marine phosphorus cycle}.
\newblock {\em Am. J. Sci.}, 282(4):474--511.

\bibitem[\protect\astroncite{{Fry}}{2000}]{Fry}
{Fry}, I. (2000).
\newblock {\em {The Emergence of Life on Earth}}.
\newblock Rutgers Univ. Press.

\bibitem[\protect\astroncite{{Fu} et~al.}{2010}]{FOCS}
{Fu}, R., {O'Connell}, R.~J., and {Sasselov}, D.~D. (2010).
\newblock {The Interior Dynamics of Water Planets}.
\newblock {\em Astrophys. J.}, 708(2):1326--1334.

\bibitem[\protect\astroncite{{Furukawa} et~al.}{2015}]{FNS15}
{Furukawa}, Y., {Nakazawa}, H., {Sekine}, T., {Kobayashi}, T., and {Kakegawa},
  T. (2015).
\newblock {Nucleobase and amino acid formation through impacts of meteorites on
  the early ocean}.
\newblock {\em Earth Planet. Sci. Lett.}, 429:216--222.

\bibitem[\protect\astroncite{{Gaidos} et~al.}{2005}]{GDD05}
{Gaidos}, E., {Deschenes}, B., {Dundon}, L., {Fagan}, K., {Menviel-Hessler},
  L., {Moskovitz}, N., and {Workman}, M. (2005).
\newblock {Beyond the Principle of Plentitude: A Review of Terrestrial Planet
  Habitability}.
\newblock {\em Astrobiology}, 5(2):100--126.

\bibitem[\protect\astroncite{{Gaidos} et~al.}{2017}]{GWK17}
{Gaidos}, E., {Williams}, J., and {Kraus}, A. (2017).
\newblock {Origin of Interstellar Object A/2017 U1 in a Nearby Young Stellar
  Association?}
\newblock {\em Research Notes of the AAS}, 1(1):13.

\bibitem[\protect\astroncite{{Gaidos} et~al.}{1999}]{GNK99}
{Gaidos}, E.~J., {Nealson}, K.~H., and {Kirschvink}, J.~L. (1999).
\newblock {Life in Ice-Covered Oceans}.
\newblock {\em Science}, 284(5420):1631--1633.

\bibitem[\protect\astroncite{{Gaillard} et~al.}{2011}]{GSA11}
{Gaillard}, F., {Scaillet}, B., and {Arndt}, N.~T. (2011).
\newblock {Atmospheric oxygenation caused by a change in volcanic degassing
  pressure}.
\newblock {\em Nature}, 478(7368):229--232.

\bibitem[\protect\astroncite{{Garz{\'o}n} and {Garz{\'o}n}}{2001}]{GG01}
{Garz{\'o}n}, L. and {Garz{\'o}n}, M.~L. (2001).
\newblock {Radioactivity as a Significant Energy Source in Prebiotic
  Synthesis}.
\newblock {\em Orig. Life Evol. Biosph.}, 31(1-2):3--13.

\bibitem[\protect\astroncite{{Genzel} et~al.}{2010}]{GEG10}
{Genzel}, R., {Eisenhauer}, F., and {Gillessen}, S. (2010).
\newblock {The Galactic Center massive black hole and nuclear star cluster}.
\newblock {\em Rev. Mod. Phys.}, 82(4):3121--3195.

\bibitem[\protect\astroncite{{Gerakines} et~al.}{2004}]{GMH04}
{Gerakines}, P.~A., {Moore}, M.~H., and {Hudson}, R.~L. (2004).
\newblock {Ultraviolet photolysis and proton irradiation of astrophysical ice
  analogs containing hydrogen cyanide}.
\newblock {\em Icarus}, 170(1):202--213.

\bibitem[\protect\astroncite{{Gillon} et~al.}{2017}]{GTD17}
{Gillon}, M., {Triaud}, A.~H.~M.~J., {Demory}, B.-O., {Jehin}, E., {Agol}, E.,
  {Deck}, K.~M., {Lederer}, S.~M., {de Wit}, J., {Burdanov}, A., {Ingalls},
  J.~G., {Bolmont}, E., {Leconte}, J., {Raymond}, S.~N., {Selsis}, F.,
  {Turbet}, M., {Barkaoui}, K., {Burgasser}, A., {Burleigh}, M.~R., {Carey},
  S.~J., {Chaushev}, A., {Copperwheat}, C.~M., {Delrez}, L., {Fernandes},
  C.~S., {Holdsworth}, D.~L., {Kotze}, E.~J., {Van Grootel}, V., {Almleaky},
  Y., {Benkhaldoun}, Z., {Magain}, P., and {Queloz}, D. (2017).
\newblock {Seven temperate terrestrial planets around the nearby ultracool
  dwarf star TRAPPIST-1}.
\newblock {\em Nature}, 542(7642):456--460.

\bibitem[\protect\astroncite{{Gladman} et~al.}{2005}]{GDL05}
{Gladman}, B., {Dones}, L., {Levison}, H.~F., and {Burns}, J.~A. (2005).
\newblock {Impact Seeding and Reseeding in the Inner Solar System}.
\newblock {\em Astrobiology}, 5(4):483--496.

\bibitem[\protect\astroncite{{Gladman} et~al.}{1996}]{GB96}
{Gladman}, B.~J., {Burns}, J.~A., {Duncan}, M., {Lee}, P., and {Levison}, H.~F.
  (1996).
\newblock {The Exchange of Impact Ejecta Between Terrestrial Planets}.
\newblock {\em Science}, 271(5254):1387--1392.

\bibitem[\protect\astroncite{{Glein} et~al.}{2015}]{GBW15}
{Glein}, C.~R., {Baross}, J.~A., and {Waite}, J.~H. (2015).
\newblock {The pH of Enceladus' ocean}.
\newblock {\em Geochim. Cosmochim. Acta}, 162:202--219.

\bibitem[\protect\astroncite{{Gold}}{1992}]{Go92}
{Gold}, T. (1992).
\newblock {The Deep, Hot Biosphere}.
\newblock {\em Proc. Natl. Acad. Sci. USA}, 89(13):6045--6049.

\bibitem[\protect\astroncite{{Goldenfeld} and {Woese}}{2011}]{GoWo}
{Goldenfeld}, N. and {Woese}, C. (2011).
\newblock {Life is Physics: Evolution as a Collective Phenomenon Far From
  Equilibrium}.
\newblock {\em Annu. Rev. Condens. Matter Phys.}, 2:375--399.

\bibitem[\protect\astroncite{{Gonzalez}}{2005}]{Gon05}
{Gonzalez}, G. (2005).
\newblock {Habitable Zones in the Universe}.
\newblock {\em Orig. Life Evol. Biosph.}, 35(6):555--606.

\bibitem[\protect\astroncite{{Gould}}{1989}]{Gou89}
{Gould}, S.~J. (1989).
\newblock {\em {Wonderful Life: The Burgess Shale and the Nature of History}}.
\newblock W.~W.~Norton \& Co.

\bibitem[\protect\astroncite{{Gould}}{2002}]{Gould02}
{Gould}, S.~J. (2002).
\newblock {\em {The Structure of Evolutionary Theory}}.
\newblock Harvard Univ. Press.

\bibitem[\protect\astroncite{{Goulinski} and {Ribak}}{2018}]{GP18}
{Goulinski}, N. and {Ribak}, E.~N. (2018).
\newblock {Capture of free-floating planets by planetary systems}.
\newblock {\em Mon. Not. R. Astron. Soc.}, 473(2):1589--1595.

\bibitem[\protect\astroncite{{Grant} et~al.}{2017}]{GGH17}
{Grant}, P.~R., {Grant}, B.~R., {Huey}, R.~B., {Johnson}, M.~T.~J., {Knoll},
  A.~H., and {Schmitt}, J. (2017).
\newblock {Evolution caused by extreme events}.
\newblock {\em Phil. Trans. R. Soc. B}, 372(1723):20160146.

\bibitem[\protect\astroncite{{Greenberg}}{2010}]{Green10}
{Greenberg}, R. (2010).
\newblock {Transport Rates of Radiolytic Substances into Europa's Ocean:
  Implications for the Potential Origin and Maintenance of Life}.
\newblock {\em Astrobiology}, 10(3):275--283.

\bibitem[\protect\astroncite{{Greenberg} et~al.}{2000}]{GG00}
{Greenberg}, R., {Geissler}, P., {Tufts}, B.~R., and {Hoppa}, G.~V. (2000).
\newblock {Habitability of Europa's crust: The role of tidal-tectonic
  processes}.
\newblock {\em J. Geophys. Res.}, 105(E7):17551--17562.

\bibitem[\protect\astroncite{{Grenfell}}{2017}]{Gren17}
{Grenfell}, J.~L. (2017).
\newblock {A review of exoplanetary biosignatures}.
\newblock {\em Phys. Rep.}, 713:1--17.

\bibitem[\protect\astroncite{{Grew} et~al.}{2011}]{GBH}
{Grew}, E.~S., {Bada}, J.~L., and {Hazen}, R.~M. (2011).
\newblock {Borate Minerals and Origin of the RNA World}.
\newblock {\em Orig. Life Evol. Biosph.}, 41(4):307--316.

\bibitem[\protect\astroncite{{Griffin}}{2001}]{Gri01}
{Griffin}, D.~R. (2001).
\newblock {\em {Animal Minds: Beyond Cognition to Consciousness}}.
\newblock The Univ. of Chicago Press.

\bibitem[\protect\astroncite{{Gruen} et~al.}{1994}]{GG94}
{Gruen}, E., {Gustafson}, B., {Mann}, I., {Baguhl}, M., {Morfill}, G.~E.,
  {Staubach}, P., {Taylor}, A., and {Zook}, H.~A. (1994).
\newblock {Interstellar dust in the heliosphere}.
\newblock {\em Astron. Astrophys.}, 286:915--924.

\bibitem[\protect\astroncite{{Hand} et~al.}{2007}]{HCC07}
{Hand}, K.~P., {Carlson}, R.~W., and {Chyba}, C.~F. (2007).
\newblock {Energy, Chemical Disequilibrium, and Geological Constraints on
  Europa}.
\newblock {\em Astrobiology}, 7(6):1006--1022.

\bibitem[\protect\astroncite{{Hand} et~al.}{2006}]{HCCC}
{Hand}, K.~P., {Chyba}, C.~F., {Carlson}, R.~W., and {Cooper}, J.~F. (2006).
\newblock {Clathrate Hydrates of Oxidants in the Ice Shell of Europa}.
\newblock {\em Astrobiology}, 6(3):463--482.

\bibitem[\protect\astroncite{{Haqq-Misra} et~al.}{2018}]{HKW18}
{Haqq-Misra}, J., {Kopparapu}, R.~K., and {Wolf}, E.~T. (2018).
\newblock {Why do we find ourselves around a yellow star instead of a red
  star?}
\newblock {\em Int. J. Astrobiol.}, 17(1):77--86.

\bibitem[\protect\astroncite{{Hazen}}{2017}]{Haz17}
{Hazen}, R.~M. (2017).
\newblock {Chance, necessity and the origins of life: a physical sciences
  perspective}.
\newblock {\em Phil. Trans. R. Soc. A}, 375(2109):20160353.

\bibitem[\protect\astroncite{{Hazen} and {Sverjensky}}{2010}]{HS10}
{Hazen}, R.~M. and {Sverjensky}, D.~A. (2010).
\newblock {Mineral Surfaces, Geochemical Complexities, and the Origins of
  Life}.
\newblock {\em Cold Spring Harb. Perspect. Biol.}, 2(5):a002162.

\bibitem[\protect\astroncite{{Hein} et~al.}{2017}]{HPL17}
{Hein}, A.~M., {Perakis}, N., {Long}, K.~F., {Crowl}, A., {Eubanks}, M.,
  {Kennedy}, III, R.~G., and {Osborne}, R. (2017).
\newblock {Project Lyra: Sending a Spacecraft to 1I/'Oumuamua (former A/2017
  U1), the Interstellar Asteroid}.
\newblock {\em ArXiv e-prints (arXiv:1711.03155)}.

\bibitem[\protect\astroncite{{Herbst} and {van Dishoeck}}{2009}]{HVD13}
{Herbst}, E. and {van Dishoeck}, E.~F. (2009).
\newblock {Complex Organic Interstellar Molecules}.
\newblock {\em Annu. Rev. Astron. Astrophys.}, 47:427--480.

\bibitem[\protect\astroncite{{Higgs} and {Lehman}}{2015}]{HL15}
{Higgs}, P.~G. and {Lehman}, N. (2015).
\newblock {The RNA World: molecular cooperation at the origins of life}.
\newblock {\em Nat. Rev. Genet.}, 16:7--17.

\bibitem[\protect\astroncite{{Hodson} et~al.}{2008}]{HA08}
{Hodson}, A., {Anesio}, A.~M., {Tranter}, M., {Fountain}, A., {Osborn}, M.,
  {Priscu}, J., {Laybourn-Parry}, J., and {Sattler}, B. (2008).
\newblock {Glacial Ecosystems}.
\newblock {\em Ecol. Monographs}, 78(1):41--67.

\bibitem[\protect\astroncite{{Hoehler}}{2004}]{Ho04}
{Hoehler}, T.~M. (2004).
\newblock {Biological energy requirements as quantitative boundary conditions
  for life in the subsurface}.
\newblock {\em Geobiology}, 2(4):205--215.

\bibitem[\protect\astroncite{{Hoehler}}{2007}]{Ho07}
{Hoehler}, T.~M. (2007).
\newblock {An Energy Balance Concept for Habitability}.
\newblock {\em Astrobiology}, 7(6):824--838.

\bibitem[\protect\astroncite{{Hoehler} et~al.}{2007}]{HAS07}
{Hoehler}, T.~M., {Amend}, J.~P., and {Shock}, E.~L. (2007).
\newblock {A ``Follow the Energy'' Approach for Astrobiology}.
\newblock {\em Astrobiology}, 7(6):819--823.

\bibitem[\protect\astroncite{{Hoehler} and {J{\o}rgensen}}{2013}]{HJ13}
{Hoehler}, T.~M. and {J{\o}rgensen}, B.~B. (2013).
\newblock {Microbial life under extreme energy limitation}.
\newblock {\em Nat. Rev. Microbiol.}, 11(2):83--94.

\bibitem[\protect\astroncite{{Holland}}{2009}]{Holl09}
{Holland}, H.~D. (2009).
\newblock {Why the atmosphere became oxygenated: A proposal}.
\newblock {\em Geochim. Cosmochim. Acta}, 73(18):5241--5255.

\bibitem[\protect\astroncite{{Hollis} et~al.}{2000}]{HLJ00}
{Hollis}, J.~M., {Lovas}, F.~J., and {Jewell}, P.~R. (2000).
\newblock {Interstellar Glycolaldehyde: The First Sugar}.
\newblock {\em Astrophys. J. Lett.}, 540(2):L107--L110.

\bibitem[\protect\astroncite{{Holm} et~al.}{2015}]{HOM15}
{Holm}, N.~G., {Oze}, C., {Mousis}, O., {Waite}, J.~H., and
  {Guilbert-Lepoutre}, A. (2015).
\newblock {Serpentinization and the Formation of H$_2$ and CH$_4$ on Celestial
  Bodies (Planets, Moons, Comets)}.
\newblock {\em Astrobiology}, 15(7):587--600.

\bibitem[\protect\astroncite{{Houtkooper}}{2011}]{Hout11}
{Houtkooper}, J.~M. (2011).
\newblock {Glaciopanspermia: Seeding the terrestrial planets with life?}
\newblock {\em Planet. Space Sci.}, 59(10):1107--1111.

\bibitem[\protect\astroncite{{Hoyle} and {Lyttleton}}{1939}]{HL39}
{Hoyle}, F. and {Lyttleton}, R.~A. (1939).
\newblock {The effect of interstellar matter on climatic variation}.
\newblock {\em Proc. Cam. Phil. Soc.}, 35(3):405--415.

\bibitem[\protect\astroncite{{Hsu} et~al.}{2015}]{HP15}
{Hsu}, H.-W., {Postberg}, F., {Sekine}, Y., {Shibuya}, T., {Kempf}, S.,
  {Hor{\'a}nyi}, M., {Juh{\'a}sz}, A., {Altobelli}, N., {Suzuki}, K., {Masaki},
  Y., {Kuwatani}, T., {Tachibana}, S., {Sirono}, S.-I., {Moragas-Klostermeyer},
  G., and {Srama}, R. (2015).
\newblock {Ongoing hydrothermal activities within Enceladus}.
\newblock {\em Nature}, 519(7542):207--210.

\bibitem[\protect\astroncite{{Hudson} et~al.}{2008}]{HM08}
{Hudson}, R.~L., {Moore}, M.~H., {Dworkin}, J.~P., {Martin}, M.~P., and
  {Pozun}, Z.~D. (2008).
\newblock {Amino Acids from Ion-Irradiated Nitrile-Containing Ices}.
\newblock {\em Astrobiology}, 8(4):771--779.

\bibitem[\protect\astroncite{{Hussmann} et~al.}{2010}]{HCL10}
{Hussmann}, H., {Choblet}, G., {Lainey}, V., {Matson}, D.~L., {Sotin}, C.,
  {Tobie}, G., and {van Hoolst}, T. (2010).
\newblock {Implications of Rotation, Orbital States, Energy Sources, and Heat
  Transport for Internal Processes in Icy Satellites}.
\newblock {\em Space Sci. Rev.}, 153(1-4):317--348.

\bibitem[\protect\astroncite{{Hussmann} et~al.}{2006}]{HSS06}
{Hussmann}, H., {Sohl}, F., and {Spohn}, T. (2006).
\newblock {Subsurface oceans and deep interiors of medium-sized outer planet
  satellites and large trans-neptunian objects}.
\newblock {\em Icarus}, 185(1):258--273.

\bibitem[\protect\astroncite{{Iess} et~al.}{2012}]{IJ12}
{Iess}, L., {Jacobson}, R.~A., {Ducci}, M., {Stevenson}, D.~J., {Lunine},
  J.~I., {Armstrong}, J.~W., {Asmar}, S.~W., {Racioppa}, P., {Rappaport},
  N.~J., and {Tortora}, P. (2012).
\newblock {The Tides of Titan}.
\newblock {\em Science}, 337(6093):457--459.

\bibitem[\protect\astroncite{{Imlay}}{2013}]{Im13}
{Imlay}, J.~A. (2013).
\newblock {The molecular mechanisms and physiological consequences of oxidative
  stress: lessons from a model bacterium}.
\newblock {\em Nat. Rev. Microbiol.}, 11(7):443--454.

\bibitem[\protect\astroncite{{Jablonka} and {Lamb}}{2014}]{JL14}
{Jablonka}, E. and {Lamb}, M.~J. (2014).
\newblock {\em {Evolution in Four Dimensions}}.
\newblock The MIT Press.

\bibitem[\protect\astroncite{{Jackson} et~al.}{2018}]{JT17}
{Jackson}, A.~P., {Tamayo}, D., {Hammond}, N., {Ali-Dib}, M., and {Rein}, H.
  (2018).
\newblock {Ejection of rocky and icy material from binary star systems:
  Implications for the origin and composition of 1I/`Oumuamua}.
\newblock {\em Mon. Not. R. Astron. Soc. Lett. (arXiv:1712.04435)}.

\bibitem[\protect\astroncite{{Jakosky} and {Shock}}{1998}]{JS98}
{Jakosky}, B.~M. and {Shock}, E.~L. (1998).
\newblock {The biological potential of Mars, the early Earth, and Europa}.
\newblock {\em J. Geophys. Res.}, 103(E8):19359--19364.

\bibitem[\protect\astroncite{{Jewitt}}{2003}]{Jew03}
{Jewitt}, D. (2003).
\newblock {Project Pan-STARRS and the Outer Solar System}.
\newblock {\em Earth Moon and Planets}, 92(1):465--476.

\bibitem[\protect\astroncite{{Jewitt} et~al.}{2017}]{JLR17}
{Jewitt}, D., {Luu}, J., {Rajagopal}, J., {Kotulla}, R., {Ridgway}, S., {Liu},
  W., and {Augusteijn}, T. (2017).
\newblock {Interstellar Interloper 1I/2017 U1: Observations from the NOT and
  WIYN Telescopes}.
\newblock {\em Astrophys. J. Lett.}, 850(2):L36.

\bibitem[\protect\astroncite{{Johansen} et~al.}{2009}]{JYM}
{Johansen}, A., {Youdin}, A., and {Mac Low}, M.-M. (2009).
\newblock {Particle Clumping and Planetesimal Formation Depend Strongly on
  Metallicity}.
\newblock {\em Astrophys. J. Lett.}, 704(2):L75--L79.

\bibitem[\protect\astroncite{{Johnson} and {Apps}}{2009}]{JA09}
{Johnson}, J.~A. and {Apps}, K. (2009).
\newblock {On the Metal Richness of M Dwarfs with Planets}.
\newblock {\em Astrophys. J.}, 699(2):933--937.

\bibitem[\protect\astroncite{{Johnson} and {Li}}{2012}]{JL12}
{Johnson}, J.~L. and {Li}, H. (2012).
\newblock {The First Planets: The Critical Metallicity for Planet Formation}.
\newblock {\em Astrophys. J.}, 751(2):81.

\bibitem[\protect\astroncite{{Johnson} et~al.}{2003}]{JQC03}
{Johnson}, R.~E., {Quickenden}, T.~I., {Cooper}, P.~D., {McKinley}, A.~J., and
  {Freeman}, C.~G. (2003).
\newblock {The Production of Oxidants in Europa's Surface}.
\newblock {\em Astrobiology}, 3(4):823--850.

\bibitem[\protect\astroncite{{Jones} and {Lineweaver}}{2010}]{JL10}
{Jones}, E.~G. and {Lineweaver}, C.~H. (2010).
\newblock {To What Extent Does Terrestrial Life - ``Follow The Water''?}
\newblock {\em Astrobiology}, 10(3):349--361.

\bibitem[\protect\astroncite{{Jones} et~al.}{2012}]{JBC}
{Jones}, P.~A., {Burton}, M.~G., {Cunningham}, M.~R., {Requena-Torres}, M.~A.,
  {Menten}, K.~M., {Schilke}, P., {Belloche}, A., {Leurini}, S.,
  {Mart{\'{\i}}n-Pintado}, J., {Ott}, J., and {Walsh}, A.~J. (2012).
\newblock {Spectral imaging of the Central Molecular Zone in multiple 3-mm
  molecular lines}.
\newblock {\em Mon. Not. R. Astron. Soc.}, 419(4):2961--2986.

\bibitem[\protect\astroncite{{Jones} et~al.}{2009}]{JC09}
{Jones}, R.~L., {Chesley}, S.~R., {Connolly}, A.~J., {Harris}, A.~W., {Ivezic},
  Z., {Knezevic}, Z., {Kubica}, J., {Milani}, A., and {Trilling}, D.~E. (2009).
\newblock {Solar System Science with LSST}.
\newblock {\em Earth Moon and Planets}, 105(2-4):101--105.

\bibitem[\protect\astroncite{{Judge}}{2017}]{JP17}
{Judge}, P. (2017).
\newblock {A Novel Strategy to Seek Biosignatures at Enceladus and Europa}.
\newblock {\em Astrobiology}, 17(9):852--861.

\bibitem[\protect\astroncite{{Judson}}{2017}]{Jud17}
{Judson}, O.~P. (2017).
\newblock {The energy expansions of evolution}.
\newblock {\em Nat. Ecol. Evol.}, 1:0138.

\bibitem[\protect\astroncite{{Kalousov{\'a}} et~al.}{2018}]{KSC18}
{Kalousov{\'a}}, K., {Sotin}, C., {Choblet}, G., {Tobie}, G., and {Grasset}, O.
  (2018).
\newblock {Two-phase convection in Ganymede's high-pressure ice layer -
  Implications for its geological evolution}.
\newblock {\em Icarus}, 299:133--147.

\bibitem[\protect\astroncite{{Kaltenegger}}{2017}]{Kal17}
{Kaltenegger}, L. (2017).
\newblock {How to Characterize Habitable Worlds and Signs of Life}.
\newblock {\em Annu. Rev. Astron. Astrophys.}, 55:433--485.

\bibitem[\protect\astroncite{{Kamerlin} et~al.}{2013}]{KSPW}
{Kamerlin}, S.~C.~L., {Sharma}, P.~K., {Prasad}, R.~B., and {Warshel}, A.
  (2013).
\newblock {Why nature really chose phosphate}.
\newblock {\em Q. Rev. Biophys.}, 46(1):1--132.

\bibitem[\protect\astroncite{{Kargel} et~al.}{2000}]{KK00}
{Kargel}, J.~S., {Kaye}, J.~Z., {Head}, J.~W., {Marion}, G.~M., {Sassen}, R.,
  {Crowley}, J.~K., {Ballesteros}, O.~P., {Grant}, S.~A., and {Hogenboom},
  D.~L. (2000).
\newblock {Europa's Crust and Ocean: Origin, Composition, and the Prospects for
  Life}.
\newblock {\em Icarus}, 148(1):226--265.

\bibitem[\protect\astroncite{{Karl}}{2000}]{Karl}
{Karl}, D.~M. (2000).
\newblock {Aquatic ecology: Phosphorus, the staff of life}.
\newblock {\em Nature}, 406(6791):31--33.

\bibitem[\protect\astroncite{{Kasting}}{2013}]{Kast13}
{Kasting}, J.~F. (2013).
\newblock {What caused the rise of atmospheric O$_2$?}
\newblock {\em Chem. Geol.}, 362:13--25.

\bibitem[\protect\astroncite{{Kasting} et~al.}{1993}]{KWR93}
{Kasting}, J.~F., {Whitmire}, D.~P., and {Reynolds}, R.~T. (1993).
\newblock {Habitable Zones around Main Sequence Stars}.
\newblock {\em Icarus}, 101(1):108--128.

\bibitem[\protect\astroncite{{Kattenhorn} and {Prockter}}{2014}]{KP14}
{Kattenhorn}, S.~A. and {Prockter}, L.~M. (2014).
\newblock {Evidence for subduction in the ice shell of Europa}.
\newblock {\em Nat. Geosci.}, 7(10):762--767.

\bibitem[\protect\astroncite{{Kebukawa} et~al.}{2017}]{KCT17}
{Kebukawa}, Y., {Chan}, Q.~H.~S., {Tachibana}, S., {Kobayashi}, K., and
  {Zolensky}, M.~E. (2017).
\newblock {One-pot synthesis of amino acid precursors with insoluble organic
  matter in planetesimals with aqueous activity}.
\newblock {\em Sci. Adv.}, 3(3):e1602093.

\bibitem[\protect\astroncite{{Kim} and {Kaiser}}{2011}]{KK11}
{Kim}, Y.~S. and {Kaiser}, R.~I. (2011).
\newblock {On the Formation of Amines (RNH$_{2}$) and the Cyanide Anion
  (CN$^{-}$) in Electron-irradiated Ammonia-hydrocarbon Interstellar Model
  Ices}.
\newblock {\em Astrophys. J.}, 729(1):68.

\bibitem[\protect\astroncite{{Kimura} and {Kitadai}}{2015}]{KK15}
{Kimura}, J. and {Kitadai}, N. (2015).
\newblock {Polymerization of Building Blocks of Life on Europa and Other Icy
  Moons}.
\newblock {\em Astrobiology}, 15(6):430--441.

\bibitem[\protect\astroncite{{Kipp} and {St{\"u}eken}}{2017}]{KS17}
{Kipp}, M.~A. and {St{\"u}eken}, E.~E. (2017).
\newblock {Biomass recycling and Earth's early phosphorus cycle}.
\newblock {\em Sci. Adv.}, 3(11):eaao4795.

\bibitem[\protect\astroncite{{Kite} and {Ford}}{2018}]{KF18}
{Kite}, E.~S. and {Ford}, E.~B. (2018).
\newblock {Habitability of exoplanet waterworlds}.
\newblock {\em submitted to Astrophys. J. (arXiv:1801.00748)}.

\bibitem[\protect\astroncite{{Kleidon}}{2016}]{Klei16}
{Kleidon}, A. (2016).
\newblock {\em {Thermodynamic Foundations of the Earth System}}.
\newblock Cambridge Univ. Press.

\bibitem[\protect\astroncite{{Knoll}}{2014}]{Knoll14}
{Knoll}, A.~H. (2014).
\newblock {Paleobiological Perspectives on Early Eukaryotic Evolution}.
\newblock {\em Cold Spring Harb. Perspect. Biol.}, 6(1):a016121.

\bibitem[\protect\astroncite{{Knoll}}{2015}]{Knoll15}
{Knoll}, A.~H. (2015).
\newblock {\em {Life on a Young Planet: The First Three Billion Years of
  Evolution on Earth}}.
\newblock Princeton Science Library. Princeton Univ. Press.

\bibitem[\protect\astroncite{{Knoll}}{2017}]{Kno17}
{Knoll}, A.~H. (2017).
\newblock {Food for early animal evolution}.
\newblock {\em Nature}, 548(7669):528--530.

\bibitem[\protect\astroncite{{Knoll} and {Bambach}}{2000}]{KB00}
{Knoll}, A.~H. and {Bambach}, R.~K. (2000).
\newblock {Directionality in the history of life: diffusion from the left wall
  or repeated scaling of the right?}
\newblock {\em Paleobiology}, 26(sp4):1--14.

\bibitem[\protect\astroncite{{Knoll} et~al.}{1996}]{KB96}
{Knoll}, A.~H., {Bambach}, R.~K., {Canfield}, D.~E., and {Grotzinger}, J.~P.
  (1996).
\newblock {Comparative Earth History and Late Permian Mass Extinction}.
\newblock {\em Science}, 273(5274):452--457.

\bibitem[\protect\astroncite{{Knoll} et~al.}{2007}]{KBP07}
{Knoll}, A.~H., {Bambach}, R.~K., {Payne}, J.~L., {Pruss}, S., and {Fischer},
  W.~W. (2007).
\newblock {Paleophysiology and end-Permian mass extinction}.
\newblock {\em Earth Planet. Sci. Lett.}, 256(3-4):295--313.

\bibitem[\protect\astroncite{{Knoll} et~al.}{2016}]{KBS16}
{Knoll}, A.~H., {Bergmann}, K.~D., and {Strauss}, J.~V. (2016).
\newblock {Life: the first two billion years}.
\newblock {\em Phil. Trans. R. Soc. B}, 371(1707):20150493.

\bibitem[\protect\astroncite{{Knoll} and {Nowak}}{2017}]{KN17}
{Knoll}, A.~H. and {Nowak}, M.~A. (2017).
\newblock {The timetable of evolution}.
\newblock {\em Sci. Adv.}, 3(5):e1603076.

\bibitem[\protect\astroncite{{Knoll} and {Sperling}}{2014}]{KS14}
{Knoll}, A.~H. and {Sperling}, E.~A. (2014).
\newblock {Oxygen and animals in Earth history}.
\newblock {\em Proc. Natl. Acad. Sci. USA}, 111(11):3907--3908.

\bibitem[\protect\astroncite{{Kobayashi} et~al.}{1998}]{KK98}
{Kobayashi}, K., {Kaneko}, T., {Saito}, T., and {Oshima}, T. (1998).
\newblock {Amino Acid Formation in Gas Mixtures by High Energy Particle
  Irradiation}.
\newblock {\em Orig. Life Evol. Biosph.}, 28(2):155--165.

\bibitem[\protect\astroncite{{Kobayashi} et~al.}{1995}]{KK95}
{Kobayashi}, K., {Kasamatsu}, T., {Kaneko}, T., {Koike}, J., {Oshima}, T.,
  {Saito}, T., {Yamamoto}, T., and {Yanagawa}, H. (1995).
\newblock {Formation of amino acid precursors in cometary ice environments by
  cosmic radiation}.
\newblock {\em Adv. Space Res.}, 16(2):21--26.

\bibitem[\protect\astroncite{{Koch} and {Britton}}{2008}]{KB08}
{Koch}, L.~G. and {Britton}, S.~L. (2008).
\newblock {Aerobic metabolism underlies complexity and capacity}.
\newblock {\em J. Physiol.}, 586(1):83--95.

\bibitem[\protect\astroncite{{Kocsis} and {Loeb}}{2014}]{KL14}
{Kocsis}, B. and {Loeb}, A. (2014).
\newblock {Menus for Feeding Black Holes}.
\newblock {\em Space Sci. Rev.}, 183(1-4):163--187.

\bibitem[\protect\astroncite{{Konhauser} et~al.}{2007}]{KLA07}
{Konhauser}, K.~O., {Lalonde}, S.~V., {Amskold}, L., and {Holland}, H.~D.
  (2007).
\newblock {Was There Really an Archean Phosphate Crisis?}
\newblock {\em Science}, 315(5816):1234.

\bibitem[\protect\astroncite{{Koonin}}{2011}]{Koo11}
{Koonin}, E.~V. (2011).
\newblock {\em {The Logic of Chance: The Nature and Origin of Biological
  Evolution}}.
\newblock FT Press.

\bibitem[\protect\astroncite{{Korenaga}}{2008}]{Kor08}
{Korenaga}, J. (2008).
\newblock {Plate tectonics, flood basalts and the evolution of Earth’s
  oceans}.
\newblock {\em Terra Nova}, 20(6):419--439.

\bibitem[\protect\astroncite{{Korenaga}}{2017}]{Kor17}
{Korenaga}, J. (2017).
\newblock {Pitfalls in modeling mantle convection with internal heat
  production}.
\newblock {\em J. Geophys. Res. B}, 122(5):4064--4085.

\bibitem[\protect\astroncite{{Kreysing} et~al.}{2015}]{KKLB}
{Kreysing}, M., {Keil}, L., {Lanzmich}, S., and {Braun}, D. (2015).
\newblock {Heat flux across an open pore enables the continuous replication and
  selection of oligonucleotides towards increasing length}.
\newblock {\em Nat. Chem.}, 7(3):203--208.

\bibitem[\protect\astroncite{{Krissansen-Totton} et~al.}{2016}]{KT16}
{Krissansen-Totton}, J., {Bergsman}, D.~S., and {Catling}, D.~C. (2016).
\newblock {On Detecting Biospheres from Chemical Thermodynamic Disequilibrium
  in Planetary Atmospheres}.
\newblock {\em Astrobiology}, 16(1):39--67.

\bibitem[\protect\astroncite{{Kr{\"u}ger} et~al.}{2015}]{KS15}
{Kr{\"u}ger}, H., {Strub}, P., {Gr{\"u}n}, E., and {Sterken}, V.~J. (2015).
\newblock {Sixteen Years of Ulysses Interstellar Dust Measurements in the Solar
  System. I. Mass Distribution and Gas-to-dust Mass Ratio}.
\newblock {\em Astrophys. J.}, 812(2):139.

\bibitem[\protect\astroncite{{Kua} and {Bada}}{2011}]{KL11}
{Kua}, J. and {Bada}, J.~L. (2011).
\newblock {Primordial Ocean Chemistry and its Compatibility with the RNA
  World}.
\newblock {\em Orig. Life Evol. Biosph.}, 41(6):553--558.

\bibitem[\protect\astroncite{{Kuan} et~al.}{2003}]{KC03}
{Kuan}, Y.-J., {Charnley}, S.~B., {Huang}, H.-C., {Tseng}, W.-L., and {Kisiel},
  Z. (2003).
\newblock {Interstellar Glycine}.
\newblock {\em Astrophys. J.}, 593(2):848--867.

\bibitem[\protect\astroncite{{Kump} and {Barley}}{2007}]{KB07}
{Kump}, L.~R. and {Barley}, M.~E. (2007).
\newblock {Increased subaerial volcanism and the rise of atmospheric oxygen 2.5
  billion years ago}.
\newblock {\em Nature}, 448(7157):1033--1036.

\bibitem[\protect\astroncite{{K{\"u}ppers} et~al.}{2014}]{KOR14}
{K{\"u}ppers}, M., {O'Rourke}, L., {Bockel{\'e}e-Morvan}, D., {Zakharov}, V.,
  {Lee}, S., {von Allmen}, P., {Carry}, B., {Teyssier}, D., {Marston}, A.,
  {M{\"u}ller}, T., {Crovisier}, J., {Barucci}, M.~A., and {Moreno}, R. (2014).
\newblock {Localized sources of water vapour on the dwarf planet (1)Ceres}.
\newblock {\em Nature}, 505(7484):525--527.

\bibitem[\protect\astroncite{{Kutschera} and {Niklas}}{2005}]{KN05}
{Kutschera}, U. and {Niklas}, K.~J. (2005).
\newblock {Endosymbiosis, cell evolution, and speciation}.
\newblock {\em Theor. Biosci.}, 124(1):1--24.

\bibitem[\protect\astroncite{{Laakso} and {Schrag}}{2014}]{LS14}
{Laakso}, T.~A. and {Schrag}, D.~P. (2014).
\newblock {Regulation of atmospheric oxygen during the Proterozoic}.
\newblock {\em Earth Planet. Sci. Lett.}, 388:81--91.

\bibitem[\protect\astroncite{{Laland}}{2017}]{Lal17}
{Laland}, K.~N. (2017).
\newblock {\em {Darwin's Unfinished Symphony: How Culture Made the Human
  Mind}}.
\newblock Princeton Univ. Press.

\bibitem[\protect\astroncite{{Laland} et~al.}{2015}]{LUF15}
{Laland}, K.~N., {Uller}, T., {Feldman}, M.~W., {Sterelny}, K., {M{\"u}ller},
  G.~B., {Moczek}, A., {Jablonka}, E., and {Odling-Smee}, J. (2015).
\newblock {The extended evolutionary synthesis: its structure, assumptions and
  predictions}.
\newblock {\em Proc. Royal Soc. B}, 282(1813):20151019.

\bibitem[\protect\astroncite{{Lamadrid} et~al.}{2017}]{LR17}
{Lamadrid}, H.~M., {Rimstidt}, J.~D., {Schwarzenbach}, E.~M., {Klein}, F.,
  {Ulrich}, S., {Dolocan}, A., and {Bodnar}, R.~J. (2017).
\newblock {Effect of water activity on rates of serpentinization of olivine}.
\newblock {\em Nat. Commun.}, 8:16107.

\bibitem[\protect\astroncite{{Lambert}}{2008}]{Lam08}
{Lambert}, J.-F. (2008).
\newblock {Adsorption and Polymerization of Amino Acids on Mineral Surfaces: A
  Review}.
\newblock {\em Orig. Life Evol. Biosph.}, 38(3):211--242.

\bibitem[\protect\astroncite{{Lammer} et~al.}{2009}]{Lam09}
{Lammer}, H., {Bredeh{\"o}ft}, J.~H., {Coustenis}, A., {Khodachenko}, M.~L.,
  {Kaltenegger}, L., {Grasset}, O., {Prieur}, D., {Raulin}, F., {Ehrenfreund},
  P., {Yamauchi}, M., {Wahlund}, J.-E., {Grie{\ss}meier}, J.-M., {Stangl}, G.,
  {Cockell}, C.~S., {Kulikov}, Y.~N., {Grenfell}, J.~L., and {Rauer}, H.
  (2009).
\newblock {What makes a planet habitable?}
\newblock {\em Astron. Astrophys. Rev.}, 17(2):181--249.

\bibitem[\protect\astroncite{{Landenmark} et~al.}{2015}]{LFC15}
{Landenmark}, H.~K.~E., {Forgan}, D.~H., and {Cockell}, C.~S. (2015).
\newblock {An Estimate of the Total DNA in the Biosphere}.
\newblock {\em PLOS Biol.}, 13(6):e1002168.

\bibitem[\protect\astroncite{{Lane}}{2002}]{Lane}
{Lane}, N. (2002).
\newblock {\em {Oxygen: The molecule that made the World}}.
\newblock Oxford Univ. Press.

\bibitem[\protect\astroncite{{Lane}}{2017}]{Lane17}
{Lane}, N. (2017).
\newblock {Serial endosymbiosis or singular event at the origin of eukaryotes?}
\newblock {\em J. Theor. Biol.}, 434:58--67.

\bibitem[\protect\astroncite{{LaRowe} and {Helgeson}}{2006}]{LH06}
{LaRowe}, D.~E. and {Helgeson}, H.~C. (2006).
\newblock {Biomolecules in hydrothermal systems: Calculation of the standard
  molal thermodynamic properties of nucleic-acid bases, nucleosides, and
  nucleotides at elevated temperatures and pressures}.
\newblock {\em Geochim. Cosmochim. Acta}, 70(18):4680--4724.

\bibitem[\protect\astroncite{{Laughlin} and {Adams}}{2000}]{LA00}
{Laughlin}, G. and {Adams}, F.~C. (2000).
\newblock {The Frozen Earth: Binary Scattering Events and the Fate of the Solar
  System}.
\newblock {\em Icarus}, 145(2):614--627.

\bibitem[\protect\astroncite{{Laughlin} and {Batygin}}{2017}]{LB17}
{Laughlin}, G. and {Batygin}, K. (2017).
\newblock {On the Consequences of the Detection of an Interstellar Asteroid}.
\newblock {\em Research Notes of the AAS}, 1(1):43.

\bibitem[\protect\astroncite{{Lazcano} and {Miller}}{1994}]{LM94}
{Lazcano}, A. and {Miller}, S.~L. (1994).
\newblock {How long did it take for life to begin and evolve to cyanobacteria?}
\newblock {\em J. Mol. Evol.}, 39(6):546--554.

\bibitem[\protect\astroncite{{Lederberg}}{1965}]{Led65}
{Lederberg}, J. (1965).
\newblock {Signs of Life: Criterion-System of Exobiology}.
\newblock {\em Nature}, 207(4992):9--13.

\bibitem[\protect\astroncite{{Lee} et~al.}{2016}]{LYM16}
{Lee}, C.-T.~A., {Yeung}, L.~Y., {McKenzie}, N.~R., {Yokoyama}, Y., {Ozaki},
  K., and {Lenardic}, A. (2016).
\newblock {Two-step rise of atmospheric oxygen linked to the growth of
  continents}.
\newblock {\em Nat. Geosci.}, 9(6):417--424.

\bibitem[\protect\astroncite{{L{\'e}ger} et~al.}{2004}]{LS04}
{L{\'e}ger}, A., {Selsis}, F., {Sotin}, C., {Guillot}, T., {Despois}, D.,
  {Mawet}, D., {Ollivier}, M., {Lab{\`e}que}, A., {Valette}, C., {Brachet}, F.,
  {Chazelas}, B., and {Lammer}, H. (2004).
\newblock {A new family of planets? ``Ocean-Planets''}.
\newblock {\em Icarus}, 169(2):499--504.

\bibitem[\protect\astroncite{{Leliwa-Kopysty{\'n}ski} et~al.}{2002}]{LMN}
{Leliwa-Kopysty{\'n}ski}, J., {Maruyama}, M., and {Nakajima}, T. (2002).
\newblock {The Water-Ammonia Phase Diagram up to 300 MPa: Application to Icy
  Satellites}.
\newblock {\em Icarus}, 159(2):518--528.

\bibitem[\protect\astroncite{{Lenardic} and {Crowley}}{2012}]{LC12}
{Lenardic}, A. and {Crowley}, J.~W. (2012).
\newblock {On the Notion of Well-defined Tectonic Regimes for Terrestrial
  Planets in this Solar System and Others}.
\newblock {\em Astrophys. J.}, 755(2):132.

\bibitem[\protect\astroncite{{Lenton} and {Watson}}{2011}]{LW11}
{Lenton}, T. and {Watson}, A.~J. (2011).
\newblock {\em {Revolutions that Made the Earth}}.
\newblock Oxford Univ. Press.

\bibitem[\protect\astroncite{{Lepper} et~al.}{2018}]{LWP18}
{Lepper}, C.~P., {Williams}, M.~A.~K., {Penny}, D., {Edwards}, P.~J.~B., and
  {Jameson}, G.~B. (2018).
\newblock {Effects of Pressure and pH on the Hydrolysis of Cytosine:
  Implications for Nucleotide Stability around Deep-Sea Black Smokers}.
\newblock {\em ChemBioChem}.

\bibitem[\protect\astroncite{{Levin} et~al.}{2017}]{LSCW}
{Levin}, S.~R., {Scott}, T.~W., {Cooper}, H.~S., and {West}, S.~A. (2017).
\newblock {Darwin's aliens}.
\newblock {\em Int. J. Astrobiol.}, pages 1--9.

\bibitem[\protect\astroncite{{Levy} and {Miller}}{1998}]{LM98}
{Levy}, M. and {Miller}, S.~L. (1998).
\newblock {The Stability of the RNA Bases: Implications for the Origin of
  Life}.
\newblock {\em Proc. Natl. Acad. Sci. USA}, 95(14):7933--7938.

\bibitem[\protect\astroncite{{Levy} et~al.}{2000}]{LM00}
{Levy}, M., {Miller}, S.~L., {Brinton}, K., and {Bada}, J.~L. (2000).
\newblock {Prebiotic Synthesis of Adenine and Amino Acids Under Europa-like
  Conditions}.
\newblock {\em Icarus}, 145(2):609--613.

\bibitem[\protect\astroncite{{Lewontin}}{2000}]{Lew00}
{Lewontin}, R.~C. (2000).
\newblock {\em {The Triple Helix: Gene, Organism, and Environment}}.
\newblock Harvard Univ. Press.

\bibitem[\protect\astroncite{{Lin} and {Loeb}}{2015}]{LL15}
{Lin}, H.~W. and {Loeb}, A. (2015).
\newblock {Statistical Signatures of Panspermia in Exoplanet Surveys}.
\newblock {\em Astrophys. J. Lett.}, 810(5):L3.

\bibitem[\protect\astroncite{{Lin} et~al.}{2006}]{LWR06}
{Lin}, L.-H., {Wang}, P.-L., {Rumble}, D., {Lippmann-Pipke}, J., {Boice}, E.,
  {Pratt}, L.~M., {Lollar}, B.~S., {Brodie}, E.~L., {Hazen}, T.~C., {Andersen},
  G.~L., {DeSantis}, T.~Z., {Moser}, D.~P., {Kershaw}, D., and {Onstott}, T.~C.
  (2006).
\newblock {Long-Term Sustainability of a High-Energy, Low-Diversity Crustal
  Biome}.
\newblock {\em Science}, 314(5798):479--482.

\bibitem[\protect\astroncite{{Lineweaver}}{2001}]{Lin01}
{Lineweaver}, C.~H. (2001).
\newblock {An Estimate of the Age Distribution of Terrestrial Planets in the
  Universe: Quantifying Metallicity as a Selection Effect}.
\newblock {\em Icarus}, 151(2):307--313.

\bibitem[\protect\astroncite{{Lineweaver} and {Davis}}{2002}]{LD02}
{Lineweaver}, C.~H. and {Davis}, T.~M. (2002).
\newblock {Does the Rapid Appearance of Life on Earth Suggest that Life Is
  Common in the Universe?}
\newblock {\em Astrobiology}, 2(3):293--304.

\bibitem[\protect\astroncite{{Lineweaver} et~al.}{2004}]{LFG04}
{Lineweaver}, C.~H., {Fenner}, Y., and {Gibson}, B.~K. (2004).
\newblock {The Galactic Habitable Zone and the Age Distribution of Complex Life
  in the Milky Way}.
\newblock {\em Science}, 303(5654):59--62.

\bibitem[\protect\astroncite{{Lingam}}{2016a}]{Lin16}
{Lingam}, M. (2016a).
\newblock {Analytical approaches to modelling panspermia - beyond the
  mean-field paradigm}.
\newblock {\em Mon. Not. R. Astron. Soc.}, 455(3):2792--2803.

\bibitem[\protect\astroncite{{Lingam}}{2016b}]{Ling16}
{Lingam}, M. (2016b).
\newblock {Interstellar Travel and Galactic Colonization: Insights from
  Percolation Theory and the Yule Process}.
\newblock {\em Astrobiology}, 16(6):418--426.

\bibitem[\protect\astroncite{{Lingam} et~al.}{2018}]{LDF18}
{Lingam}, M., {Dong}, C., {Fang}, X., {Jakosky}, B.~M., and {Loeb}, A. (2018).
\newblock {The Propitious Role of Solar Energetic Particles in the Origin of
  Life}.
\newblock {\em Astrophys. J.}, 853(1):10.

\bibitem[\protect\astroncite{{Lingam} and {Loeb}}{2017a}]{Linga}
{Lingam}, M. and {Loeb}, A. (2017a).
\newblock {Enhanced interplanetary panspermia in the TRAPPIST-1 system}.
\newblock {\em Proc. Natl. Acad. Sci. USA}, 114(26):6689--6693.

\bibitem[\protect\astroncite{{Lingam} and {Loeb}}{2017b}]{Ling17}
{Lingam}, M. and {Loeb}, A. (2017b).
\newblock {Implications of tides for life on exoplanets}.
\newblock {\em Astrobiology (arXiv:1707.04594)}.

\bibitem[\protect\astroncite{{Lingam} and {Loeb}}{2017c}]{LL17}
{Lingam}, M. and {Loeb}, A. (2017c).
\newblock {Is Life Most Likely Around Sun-like Stars?}
\newblock {\em submitted to J. Cosmol. Astropart. Phys. (arXiv:1710.11134)}.

\bibitem[\protect\astroncite{{Lingam} and {Loeb}}{2017d}]{Man17}
{Lingam}, M. and {Loeb}, A. (2017d).
\newblock {Reduced Diversity of Life around Proxima Centauri and TRAPPIST-1}.
\newblock {\em Astrophys. J. Lett.}, 846(2):L21.

\bibitem[\protect\astroncite{{Lingam} and {Loeb}}{2017e}]{MLin}
{Lingam}, M. and {Loeb}, A. (2017e).
\newblock {Risks for Life on Habitable Planets from Superflares of Their Host
  Stars}.
\newblock {\em Astrophys. J.}, 848(1):41.

\bibitem[\protect\astroncite{{Lingam} and {Loeb}}{2018a}]{LL18}
{Lingam}, M. and {Loeb}, A. (2018a).
\newblock {Implications of Captured Interstellar Objects for Panspermia and
  Extraterrestrial Life}.
\newblock {\em submitted to Astron. J. (arXiv:1801.10254)}.

\bibitem[\protect\astroncite{{Lingam} and {Loeb}}{2018b}]{LiLo17}
{Lingam}, M. and {Loeb}, A. (2018b).
\newblock {Physical constraints on the likelihood of life on exoplanets}.
\newblock {\em Int. J. Astrobiol.}, 17(2):116--126.

\bibitem[\protect\astroncite{{Linsky} et~al.}{2013}]{LFA13}
{Linsky}, J.~L., {France}, K., and {Ayres}, T. (2013).
\newblock {Computing Intrinsic LY{$\alpha$} Fluxes of F5 V to M5 V Stars}.
\newblock {\em Astrophys. J.}, 766(2):69.

\bibitem[\protect\astroncite{{Lipps} and {Rieboldt}}{2005}]{LR05}
{Lipps}, J.~H. and {Rieboldt}, S. (2005).
\newblock {Habitats and taphonomy of Europa}.
\newblock {\em Icarus}, 177(2):515--527.

\bibitem[\protect\astroncite{{Loeb}}{2014}]{Loeb14}
{Loeb}, A. (2014).
\newblock {The habitable epoch of the early Universe}.
\newblock {\em Int. J. Astrobiol.}, 13(4):337--339.

\bibitem[\protect\astroncite{{Loeb}}{2017}]{Loeb17}
{Loeb}, A. (2017).
\newblock {Cosmic Modesty}.
\newblock {\em Sci. Am. (arXiv:1706.05959)}.

\bibitem[\protect\astroncite{{Loeb} et~al.}{2016}]{LBS16}
{Loeb}, A., {Batista}, R.~A., and {Sloan}, D. (2016).
\newblock {Relative likelihood for life as a function of cosmic time}.
\newblock {\em J. Cosmol. Astropart. Phys.}, 8:040.

\bibitem[\protect\astroncite{{Loeb} and {Furlanetto}}{2013}]{LF13}
{Loeb}, A. and {Furlanetto}, S.~R. (2013).
\newblock {\em {The First Galaxies in the Universe}}.
\newblock Princeton Univ. Press.

\bibitem[\protect\astroncite{{Loeb} and {Maoz}}{2013}]{LM13}
{Loeb}, A. and {Maoz}, D. (2013).
\newblock {Detecting biomarkers in habitable-zone earths transiting white
  dwarfs}.
\newblock {\em Mon. Not. R. Astron. Soc. Lett.}, 432(1):L11--L15.

\bibitem[\protect\astroncite{{Lollar} et~al.}{2014}]{LO14}
{Lollar}, B.~S., {Onstott}, T.~C., {Lacrampe-Couloume}, G., and {Ballentine},
  C.~J. (2014).
\newblock {The contribution of the Precambrian continental lithosphere to
  global H$_{2}$ production}.
\newblock {\em Nature}, 516(7531):379--382.

\bibitem[\protect\astroncite{{Lopez} et~al.}{2005}]{LSD05}
{Lopez}, B., {Schneider}, J., and {Danchi}, W.~C. (2005).
\newblock {Can Life Develop in the Expanded Habitable Zones around Red Giant
  Stars?}
\newblock {\em Astrophys. J.}, 627(2):974--985.

\bibitem[\protect\astroncite{{Lorenz} et~al.}{1997}]{LL97}
{Lorenz}, R.~D., {Lunine}, J.~I., and {McKay}, C.~P. (1997).
\newblock {Titan under a red giant sun: A new kind of 'habitable' moon}.
\newblock {\em Geophys. Res. Lett.}, 24(22):2905--2908.

\bibitem[\protect\astroncite{{Losos}}{2011}]{Losos}
{Losos}, J.~B. (2011).
\newblock {Convergence, Adaptation, and Constraint}.
\newblock {\em Evolution}, 65(7):1827--1840.

\bibitem[\protect\astroncite{{Lovelock}}{1965}]{Love65}
{Lovelock}, J.~E. (1965).
\newblock {A Physical Basis for Life Detection Experiments}.
\newblock {\em Nature}, 207(4997):568--570.

\bibitem[\protect\astroncite{{Lovelock} and {Margulis}}{1974}]{LM74}
{Lovelock}, J.~E. and {Margulis}, L. (1974).
\newblock {Atmospheric homeostasis by and for the biosphere: The gaia
  hypothesis}.
\newblock {\em Tellus}, 26(1-2):2--10.

\bibitem[\protect\astroncite{{Luger} and {Barnes}}{2015}]{LB15}
{Luger}, R. and {Barnes}, R. (2015).
\newblock {Extreme Water Loss and Abiotic O2Buildup on Planets Throughout the
  Habitable Zones of M Dwarfs}.
\newblock {\em Astrobiology}, 15(2):119--143.

\bibitem[\protect\astroncite{{Luisi}}{2016}]{Lu16}
{Luisi}, P.~L. (2016).
\newblock {\em The Emergence of Life: From Chemical Origins to Synthetic
  Biology}.
\newblock Cambridge Univ. Press.

\bibitem[\protect\astroncite{{Lunine}}{2010}]{Lun10}
{Lunine}, J.~I. (2010).
\newblock {Titan and habitable planets around M-dwarfs}.
\newblock {\em Faraday Discussions}, 147:405--418.

\bibitem[\protect\astroncite{{Lunine}}{2017}]{Lun17}
{Lunine}, J.~I. (2017).
\newblock {Ocean worlds exploration}.
\newblock {\em Acta Astronaut.}, 131:123--130.

\bibitem[\protect\astroncite{{Lyons} et~al.}{2014}]{LRP14}
{Lyons}, T.~W., {Reinhard}, C.~T., and {Planavsky}, N.~J. (2014).
\newblock {The rise of oxygen in Earth's early ocean and atmosphere}.
\newblock {\em Nature}, 506(7488):307--315.

\bibitem[\protect\astroncite{{Madau} and {Dickinson}}{2014}]{MD14}
{Madau}, P. and {Dickinson}, M. (2014).
\newblock {Cosmic Star-Formation History}.
\newblock {\em Annu. Rev. Astron. Astrophys.}, 52:415--486.

\bibitem[\protect\astroncite{{Mader} et~al.}{2006}]{MP06}
{Mader}, H.~M., {Pettitt}, M.~E., {Wadham}, J.~L., {Wolff}, E.~W., and
  {Parkes}, R.~J. (2006).
\newblock {Subsurface ice as a microbial habitat}.
\newblock {\em Geology}, 34(3):169--172.

\bibitem[\protect\astroncite{{Malamud} and {Prialnik}}{2013}]{MP13}
{Malamud}, U. and {Prialnik}, D. (2013).
\newblock {Modeling serpentinization: Applied to the early evolution of
  Enceladus and Mimas}.
\newblock {\em Icarus}, 225(1):763--774.

\bibitem[\protect\astroncite{{Mamajek}}{2017}]{Mam17}
{Mamajek}, E. (2017).
\newblock {Kinematics of the Interstellar Vagabond 1I/`Oumuamua (A/2017 U1)}.
\newblock {\em Research Notes of the AAS}, 1(1):21.

\bibitem[\protect\astroncite{{Manger}}{2013}]{Mang13}
{Manger}, P.~R. (2013).
\newblock {Questioning the interpretations of behavioral observations of
  cetaceans: Is there really support for a special intellectual status for this
  mammalian order?}
\newblock {\em Neuroscience}, 250:664--696.

\bibitem[\protect\astroncite{{Mann} and {Kimura}}{2000}]{MK00}
{Mann}, I. and {Kimura}, H. (2000).
\newblock {Interstellar dust properties derived from mass density, mass
  distribution, and flux rates in the heliosphere}.
\newblock {\em J. Geophys. Res.}, 105(A5):10317--10328.

\bibitem[\protect\astroncite{{Marino} et~al.}{2007}]{MCF07}
{Marino}, L., {Connor}, R.~C., {Fordyce}, R.~E., {Herman}, L.~M., {Hof}, P.~R.,
  {Lefebvre}, L., {Lusseau}, D., {McCowan}, B., {Nimchinsky}, E.~A., {Pack},
  A.~A., {Rendell}, L., {Reidenberg}, J.~S., {Reiss}, D., {Uhen}, M.~D., {Van
  der Gucht}, E., and {Whitehead}, H. (2007).
\newblock {Cetaceans Have Complex Brains for Complex Cognition}.
\newblock {\em PLOS Biol.}, 5(5):e139.

\bibitem[\protect\astroncite{{Marion} et~al.}{2003}]{MFE03}
{Marion}, G.~M., {Fritsen}, C.~H., {Eicken}, H., and {Payne}, M.~C. (2003).
\newblock {The Search for Life on Europa: Limiting Environmental Factors,
  Potential Habitats, and Earth Analogues}.
\newblock {\em Astrobiology}, 3(4):785--811.

\bibitem[\protect\astroncite{{Martin} and {McMinn}}{2018}]{MM17}
{Martin}, A. and {McMinn}, A. (2018).
\newblock {Sea ice, extremophiles and life on extra-terrestrial ocean worlds}.
\newblock {\em Int. J. Astrobiol.}, 17(1):1--16.

\bibitem[\protect\astroncite{{Martin} et~al.}{2008}]{MB08}
{Martin}, W., {Baross}, J., {Kelley}, D., and {Russell}, M.~J. (2008).
\newblock {Hydrothermal vents and the origin of life}.
\newblock {\em Nat. Rev. Microbiol.}, 6(11):805--814.

\bibitem[\protect\astroncite{{Martin} and {Russell}}{2007}]{MR07}
{Martin}, W. and {Russell}, M.~J. (2007).
\newblock {On the origin of biochemistry at an alkaline hydrothermal vent}.
\newblock {\em Phil. Trans. R. Soc. B}, 362(1486):1887--1926.

\bibitem[\protect\astroncite{{Martin} et~al.}{2015}]{MGZ15}
{Martin}, W.~F., {Garg}, S., and {Zimorski}, V. (2015).
\newblock {Endosymbiotic theories for eukaryote origin}.
\newblock {\em Phil. Trans. R. Soc. B}, 370(1678):20140330.

\bibitem[\protect\astroncite{{Martini}}{2004}]{Mar04}
{Martini}, P. (2004).
\newblock {QSO Lifetimes}.
\newblock In {Ho}, L.~C., editor, {\em Coevolution of Black Holes and
  Galaxies}, volume~1 of {\em Carnegie Observatories Astrophysics Series},
  pages 169--185.

\bibitem[\protect\astroncite{{Martins} et~al.}{2013}]{MPG13}
{Martins}, Z., {Price}, M.~C., {Goldman}, N., {Sephton}, M.~A., and {Burchell},
  M.~J. (2013).
\newblock {Shock synthesis of amino acids from impacting cometary and icy
  planet surface analogues}.
\newblock {\em Nat. Geosci.}, 6(12):1045--1049.

\bibitem[\protect\astroncite{{Mashian} and {Loeb}}{2016}]{MaLo}
{Mashian}, N. and {Loeb}, A. (2016).
\newblock {CEMP stars: possible hosts to carbon planets in the early Universe}.
\newblock {\em Mon. Not. R. Astron. Soc.}, 460(3):2482--2491.

\bibitem[\protect\astroncite{{Materese} et~al.}{2017}]{MNS17}
{Materese}, C.~K., {Nuevo}, M., and {Sandford}, S.~A. (2017).
\newblock {The Formation of Nucleobases from the Ultraviolet Photoirradiation
  of Purine in Simple Astrophysical Ice Analogues}.
\newblock {\em Astrobiology}, 17(8):761--770.

\bibitem[\protect\astroncite{{Mayr}}{1985}]{Mayr85}
{Mayr}, E. (1985).
\newblock {The probability of extraterrestrial intelligent life}.
\newblock In {Regis}, E., editor, {\em Extraterrestrials. Science and Alien
  Intelligence}, pages 23--30. Cambridge Univ. Press.

\bibitem[\protect\astroncite{{Mayr}}{2001}]{Mayr}
{Mayr}, E. (2001).
\newblock {\em {What Evolution Is}}.
\newblock Basic Books.

\bibitem[\protect\astroncite{{McCabe} and {Lucas}}{2010}]{ML10}
{McCabe}, M. and {Lucas}, H. (2010).
\newblock {On the origin and evolution of life in the Galaxy}.
\newblock {\em Int. J. Astrobiol.}, 9(4):217--226.

\bibitem[\protect\astroncite{{McCollom}}{1999}]{McCo99}
{McCollom}, T.~M. (1999).
\newblock {Methanogenesis as a potential source of chemical energy for primary
  biomass production by autotrophic organisms in hydrothermal systems on
  Europa}.
\newblock {\em J. Geophys. Res.}, 104(E12):30729--30742.

\bibitem[\protect\astroncite{{McCollom}}{2007}]{McC07}
{McCollom}, T.~M. (2007).
\newblock {Geochemical Constraints on Sources of Metabolic Energy for
  Chemolithoautotrophy in Ultramafic-Hosted Deep-Sea Hydrothermal Systems}.
\newblock {\em Astrobiology}, 7(6):933--950.

\bibitem[\protect\astroncite{{McCollom}}{2013}]{Mill13}
{McCollom}, T.~M. (2013).
\newblock {Miller-Urey and Beyond: What Have We Learned About Prebiotic Organic
  Synthesis Reactions in the Past 60 Years?}
\newblock {\em Annu. Rev. Earth Planet. Sci.}, 41:207--229.

\bibitem[\protect\astroncite{{McCollom} and {Seewald}}{2007}]{MCS07}
{McCollom}, T.~M. and {Seewald}, J.~S. (2007).
\newblock {Abiotic Synthesis of Organic Compounds in Deep-Sea Hydrothermal
  Environments}.
\newblock {\em Chem. Rev.}, 107(2):382--401.

\bibitem[\protect\astroncite{{McGlynn} and {Chapman}}{1989}]{MC89}
{McGlynn}, T.~A. and {Chapman}, R.~D. (1989).
\newblock {On the nondetection of extrasolar comets}.
\newblock {\em Astrophys. J. Lett.}, 346:L105--L108.

\bibitem[\protect\astroncite{{McKay} et~al.}{2008}]{MP08}
{McKay}, C.~P., {Porco}, C.~C., {Altheide}, T., {Davis}, W.~L., and {Kral},
  T.~A. (2008).
\newblock {The Possible Origin and Persistence of Life on Enceladus and
  Detection of Biomarkers in the Plume}.
\newblock {\em Astrobiology}, 8(5):909--919.

\bibitem[\protect\astroncite{{McKinnon}}{2006}]{McK06}
{McKinnon}, W.~B. (2006).
\newblock {On convection in ice I shells of outer Solar System bodies, with
  detailed application to Callisto}.
\newblock {\em Icarus}, 183(2):435--450.

\bibitem[\protect\astroncite{{McKinnon} and {Zolensky}}{2003}]{MZ03}
{McKinnon}, W.~B. and {Zolensky}, M.~E. (2003).
\newblock {Sulfate Content of Europa's Ocean and Shell: Evolutionary
  Considerations and Some Geological and Astrobiological Implications}.
\newblock {\em Astrobiology}, 3(4):879--897.

\bibitem[\protect\astroncite{{McMahon} et~al.}{2013}]{MOM13}
{McMahon}, S., {O'Malley-James}, J., and {Parnell}, J. (2013).
\newblock {Circumstellar habitable zones for deep terrestrial biospheres}.
\newblock {\em Planet. Space Sci.}, 85:312--318.

\bibitem[\protect\astroncite{{McShea} and {Brandon}}{2010}]{MSB10}
{McShea}, D.~W. and {Brandon}, R.~N. (2010).
\newblock {\em {Biology's First Law}}.
\newblock The Univ. of Chicago Press.

\bibitem[\protect\astroncite{{Meadows}}{2017}]{Mead17}
{Meadows}, V.~S. (2017).
\newblock {Reflections on O$_{2}$ as a Biosignature in Exoplanetary
  Atmospheres}.
\newblock {\em Astrobiology}, 17(10):1022--1052.

\bibitem[\protect\astroncite{{Meech} et~al.}{2017}]{MWM17}
{Meech}, K.~J., {Weryk}, R., {Micheli}, M., {Kleyna}, J.~T., {Hainaut}, O.~R.,
  {Jedicke}, R., {Wainscoat}, R.~J., {Chambers}, K.~C., {Keane}, J.~V.,
  {Petric}, A., {Denneau}, L., {Magnier}, E., {Berger}, T., {Huber}, M.~E.,
  {Flewelling}, H., {Waters}, C., {Schunova-Lilly}, E., and {Chastel}, S.
  (2017).
\newblock {A brief visit from a red and extremely elongated interstellar
  asteroid}.
\newblock {\em Nature}, 552(7685):378--381.

\bibitem[\protect\astroncite{{Melosh}}{1988}]{Mel88}
{Melosh}, H.~J. (1988).
\newblock {The rocky road to panspermia}.
\newblock {\em Nature}, 332(6166):687--688.

\bibitem[\protect\astroncite{{Melosh}}{2003}]{Mel03}
{Melosh}, H.~J. (2003).
\newblock {Exchange of Meteorites (and Life?) Between Stellar Systems}.
\newblock {\em Astrobiology}, 3(1):207--215.

\bibitem[\protect\astroncite{{Melott} and {Thomas}}{2011}]{MT11}
{Melott}, A.~L. and {Thomas}, B.~C. (2011).
\newblock {Astrophysical Ionizing Radiation and Earth: A Brief Review and
  Census of Intermittent Intense Sources}.
\newblock {\em Astrobiology}, 11(4):343--361.

\bibitem[\protect\astroncite{{Michalski} et~al.}{2018}]{MOM18}
{Michalski}, J.~R., {Onstott}, T.~C., {Mojzsis}, S.~J., {Mustard}, J., {Chan},
  Q.~H.~S., {Niles}, P.~B., and {Johnson}, S.~S. (2018).
\newblock {The Martian subsurface as a potential window into the origin of
  life}.
\newblock {\em Nat. Geosci.}, 11(1):21--26.

\bibitem[\protect\astroncite{{Mileikowsky} et~al.}{2000}]{Mil00}
{Mileikowsky}, C., {Cucinotta}, F.~A., {Wilson}, J.~W., {Gladman}, B.,
  {Horneck}, G., {Lindegren}, L., {Melosh}, J., {Rickman}, H., {Valtonen}, M.,
  and {Zheng}, J.~Q. (2000).
\newblock {Natural Transfer of Viable Microbes in Space. 1. From Mars to Earth
  and Earth to Mars}.
\newblock {\em Icarus}, 145(2):391--427.

\bibitem[\protect\astroncite{{Miller} and {Lazcano}}{1995}]{ML95}
{Miller}, S.~L. and {Lazcano}, A. (1995).
\newblock {The origin of life--did it occur at high temperatures?}
\newblock {\em J. Mol. Evol.}, 41(6):689--692.

\bibitem[\protect\astroncite{{Mills} and {Canfield}}{2014}]{MC14}
{Mills}, D.~B. and {Canfield}, D.~E. (2014).
\newblock {Oxygen and animal evolution: Did a rise of atmospheric oxygen
  ``trigger'' the origin of animals?}
\newblock {\em BioEssays}, 36(12):1145--1155.

\bibitem[\protect\astroncite{{Mills} et~al.}{2014}]{MWJ14}
{Mills}, D.~B., {Ward}, L.~M., {Jones}, C., {Sweeten}, B., {Forth}, M.,
  {Treusch}, A.~H., and {Canfield}, D.~E. (2014).
\newblock {Oxygen requirements of the earliest animals}.
\newblock {\em Proc. Natl. Acad. Sci. USA}, 111(11):4168--4172.

\bibitem[\protect\astroncite{{Mitri} et~al.}{2014}]{MM14}
{Mitri}, G., {Meriggiola}, R., {Hayes}, A., {Lefevre}, A., {Tobie}, G.,
  {Genova}, A., {Lunine}, J.~I., and {Zebker}, H. (2014).
\newblock {Shape, topography, gravity anomalies and tidal deformation of
  Titan}.
\newblock {\em Icarus}, 236:169--177.

\bibitem[\protect\astroncite{{Miyakawa} et~al.}{2002a}]{MJCM}
{Miyakawa}, S., {Cleaves}, H.~J., and {Miller}, S.~L. (2002a).
\newblock {The Cold Origin of Life: A. Implications Based On The Hydrolytic
  Stabilities Of Hydrogen Cyanide And Formamide}.
\newblock {\em Orig. Life Evol. Biosph.}, 32(3):195--208.

\bibitem[\protect\astroncite{{Miyakawa} et~al.}{2002b}]{MiCMi}
{Miyakawa}, S., {Cleaves}, H.~J., and {Miller}, S.~L. (2002b).
\newblock {The Cold Origin of Life: B. Implications Based on Pyrimidines and
  Purines Produced From Frozen Ammonium Cyanide Solutions}.
\newblock {\em Orig. Life Evol. Biosph.}, 32(3):209--218.

\bibitem[\protect\astroncite{{Monnard} and {Szostak}}{2008}]{MS08}
{Monnard}, P.-A. and {Szostak}, J.~W. (2008).
\newblock {Metal-ion catalyzed polymerization in the eutectic phase in
  water--ice: A possible approach to template-directed RNA polymerization}.
\newblock {\em J. Inorg. Biochem.}, 102(5):1104--1111.

\bibitem[\protect\astroncite{{Monod}}{1971}]{Mon71}
{Monod}, J. (1971).
\newblock {\em {Chance and Necessity: An Essay on the Natural Philosophy of
  Modern Biology}}.
\newblock Alfred A. Knopf, New York.

\bibitem[\protect\astroncite{{Moore} et~al.}{2017}]{ML17}
{Moore}, W.~B., {Lenardic}, A., {Jellinek}, A.~M., {Johnson}, C.~L.,
  {Goldblatt}, C., and {Lorenz}, R.~D. (2017).
\newblock {How habitable zones and super-Earths lead us astray}.
\newblock {\em Nature Astronomy}, 1:0043.

\bibitem[\protect\astroncite{{Moro-Mart{\'{\i}}n} et~al.}{2009}]{MT09}
{Moro-Mart{\'{\i}}n}, A., {Turner}, E.~L., and {Loeb}, A. (2009).
\newblock {Will the Large Synoptic Survey Telescope Detect Extra-Solar
  Planetesimals Entering the Solar System?}
\newblock {\em Astrophys. J.}, 704(1):733--742.

\bibitem[\protect\astroncite{{Morowitz} and {Smith}}{2007}]{MS07}
{Morowitz}, H. and {Smith}, E. (2007).
\newblock {Energy flow and the organization of life}.
\newblock {\em Complexity}, 13(1):51--59.

\bibitem[\protect\astroncite{{Morris}}{2003}]{Mor03}
{Morris}, S.~C. (2003).
\newblock {\em {Life's Solution: Inevitable Humans in a Lonely Universe}}.
\newblock Cambridge Univ. Press.

\bibitem[\protect\astroncite{{Morris}}{2011}]{Mor11}
{Morris}, S.~C. (2011).
\newblock {Predicting what extra-terrestrials will be like: and preparing for
  the worst}.
\newblock {\em Phil. Trans. R. Soc. A}, 369(1936):555--571.

\bibitem[\protect\astroncite{{Morrison} and {Gowanlock}}{2015}]{MG15}
{Morrison}, I.~S. and {Gowanlock}, M.~G. (2015).
\newblock {Extending Galactic Habitable Zone Modeling to Include the Emergence
  of Intelligent Life}.
\newblock {\em Astrobiology}, 15(8):683--696.

\bibitem[\protect\astroncite{{Mu{\~n}oz Caro} et~al.}{2002}]{MC02}
{Mu{\~n}oz Caro}, G.~M., {Meierhenrich}, U.~J., {Schutte}, W.~A., {Barbier},
  B., {Arcones Segovia}, A., {Rosenbauer}, H., {Thiemann}, W.~H.-P., {Brack},
  A., and {Greenberg}, J.~M. (2002).
\newblock {Amino acids from ultraviolet irradiation of interstellar ice
  analogues}.
\newblock {\em Nature}, 416(6879):403--406.

\bibitem[\protect\astroncite{{Mulders} et~al.}{2015}]{MCMP}
{Mulders}, G.~D., {Ciesla}, F.~J., {Min}, M., and {Pascucci}, I. (2015).
\newblock {The Snow Line in Viscous Disks around Low-mass Stars: Implications
  for Water Delivery to Terrestrial Planets in the Habitable Zone}.
\newblock {\em Astrophys. J.}, 807(1):9.

\bibitem[\protect\astroncite{{Muller} and {Schulze-Makuch}}{2006}]{MSM06}
{Muller}, A.~W.~J. and {Schulze-Makuch}, D. (2006).
\newblock {Thermal Energy and the Origin of Life}.
\newblock {\em Orig. Life Evol. Biosph.}, 36(2):177--189.

\bibitem[\protect\astroncite{{Mumma} and {Charnley}}{2011}]{Mum11}
{Mumma}, M.~J. and {Charnley}, S.~B. (2011).
\newblock {The Chemical Composition of Comets--Emerging Taxonomies and Natal
  Heritage}.
\newblock {\em Annu. Rev. Astron. Astrophys.}, 49:471--524.

\bibitem[\protect\astroncite{{Mutschler} et~al.}{2015}]{MWH15}
{Mutschler}, H., {Wochner}, A., and {Holliger}, P. (2015).
\newblock {Freeze--thaw cycles as drivers of complex ribozyme assembly}.
\newblock {\em Nat. Chem.}, 7(6):502--508.

\bibitem[\protect\astroncite{{Nadeau} et~al.}{2016}]{NL16}
{Nadeau}, J., {Lindensmith}, C., {Deming}, J.~W., {Fernandez}, V.~I., and
  {Stocker}, R. (2016).
\newblock {Microbial Morphology and Motility as Biosignatures for Outer Planet
  Missions}.
\newblock {\em Astrobiology}, 16(10):755--774.

\bibitem[\protect\astroncite{{Napier}}{2004}]{N04}
{Napier}, W.~M. (2004).
\newblock {A mechanism for interstellar panspermia}.
\newblock {\em Mon. Not. R. Astron. Soc.}, 348(1):46--51.

\bibitem[\protect\astroncite{{Neveu} et~al.}{2017}]{NDCR}
{Neveu}, M., {Desch}, S.~J., and {Castillo-Rogez}, J.~C. (2017).
\newblock {Aqueous geochemistry in icy world interiors: Equilibrium fluid,
  rock, and gas compositions, and fate of antifreezes and radionuclides}.
\newblock {\em Geochim. Cosmochim. Acta}, 212:324--371.

\bibitem[\protect\astroncite{{Neveu} et~al.}{2013}]{NKB13}
{Neveu}, M., {Kim}, H.-J., and {Benner}, S.~A. (2013).
\newblock {The ``Strong'' RNA World Hypothesis: Fifty Years Old}.
\newblock {\em Astrobiology}, 13(4):391--403.

\bibitem[\protect\astroncite{{Nimmo} and {Pappalardo}}{2016}]{NiPa}
{Nimmo}, F. and {Pappalardo}, R.~T. (2016).
\newblock {Ocean worlds in the outer solar system}.
\newblock {\em J. Geophys. Res. E}, 121(8):1378--1399.

\bibitem[\protect\astroncite{{Noack} et~al.}{2016}]{NH16}
{Noack}, L., {H{\"o}ning}, D., {Rivoldini}, A., {Heistracher}, C., {Zimov}, N.,
  {Journaux}, B., {Lammer}, H., {Van Hoolst}, T., and {Bredeh{\"o}ft}, J.~H.
  (2016).
\newblock {Water-rich planets: How habitable is a water layer deeper than on
  Earth?}
\newblock {\em Icarus}, 277:215--236.

\bibitem[\protect\astroncite{{Nuevo} et~al.}{2008}]{NAB08}
{Nuevo}, M., {Auger}, G., {Blanot}, D., and {D'Hendecourt}, L. (2008).
\newblock {A Detailed Study of the Amino Acids Produced from the Vacuum UV
  Irradiation of Interstellar Ice Analogs}.
\newblock {\em Orig. Life Evol. Biosph.}, 38(1):37--56.

\bibitem[\protect\astroncite{{Nuevo} et~al.}{2012}]{NM12}
{Nuevo}, M., {Milam}, S.~N., and {Sandford}, S.~A. (2012).
\newblock {Nucleobases and Prebiotic Molecules in Organic Residues Produced
  from the Ultraviolet Photo-Irradiation of Pyrimidine in NH$_3$ and
  H$_2$O+NH$_3$ Ices}.
\newblock {\em Astrobiology}, 12(4):295--314.

\bibitem[\protect\astroncite{{Oberbeck} and {Fogleman}}{1989}]{OF89}
{Oberbeck}, V.~R. and {Fogleman}, G. (1989).
\newblock {Estimates of the maximum time required to originate life}.
\newblock {\em Orig. Life Evol. Biosph.}, 19(6):549--560.

\bibitem[\protect\astroncite{{{\"O}berg}}{2016}]{Ob16}
{{\"O}berg}, K.~I. (2016).
\newblock {Photochemistry and Astrochemistry: Photochemical Pathways to
  Interstellar Complex Organic Molecules}.
\newblock {\em Chem. Rev}, 116(17):9631--9663.

\bibitem[\protect\astroncite{{{\"O}berg} et~al.}{2009}]{OG09}
{{\"O}berg}, K.~I., {Garrod}, R.~T., {van Dishoeck}, E.~F., and {Linnartz}, H.
  (2009).
\newblock {Formation rates of complex organics in UV irradiated CH\_3OH-rich
  ices. I. Experiments}.
\newblock {\em Astron. Astrophys.}, 504(3):891--913.

\bibitem[\protect\astroncite{{Odling-Smee} et~al.}{2003}]{OLF03}
{Odling-Smee}, F.~J., {Laland}, K.~N., and {Feldman}, M.~W. (2003).
\newblock {\em {Niche Construction: The Neglected Process in Evolution}}.
\newblock Number~37 in Monographs in Population Biology. Princeton Univ. Press.

\bibitem[\protect\astroncite{{Okasha}}{2006}]{Ok06}
{Okasha}, S. (2006).
\newblock {\em {Evolution and the Levels of Selection}}.
\newblock Oxford Univ. Press.

\bibitem[\protect\astroncite{{O'Malley} and {Powell}}{2016}]{OMP16}
{O'Malley}, M.~A. and {Powell}, R. (2016).
\newblock {Major problems in evolutionary transitions: how a metabolic
  perspective can enrich our understanding of macroevolution}.
\newblock {\em Biol. Philos.}, 31(2):159--189.

\bibitem[\protect\astroncite{{Orcutt} et~al.}{2011}]{OSKE}
{Orcutt}, B.~N., {Sylvan}, J.~B., {Knab}, N.~J., and {Edwards}, K.~J. (2011).
\newblock {Microbial Ecology of the Dark Ocean above, at, and below the
  Seafloor}.
\newblock {\em Microbiol. Mol. Biol. Rev.}, 75(2):361--422.

\bibitem[\protect\astroncite{{Orgel}}{1998}]{Org98}
{Orgel}, L.~E. (1998).
\newblock {The Origin of Life - How Long did it Take?}
\newblock {\em Orig. Life Evol. Biosph.}, 28(1):91--96.

\bibitem[\protect\astroncite{{Orgel}}{2004}]{Org04}
{Orgel}, L.~E. (2004).
\newblock {Prebiotic Chemistry and the Origin of the RNA World}.
\newblock {\em Crit. Rev. Biochem. Mol. Biol.}, 39(2):99--123.

\bibitem[\protect\astroncite{{Orr} et~al.}{2005}]{OF05}
{Orr}, J.~C., {Fabry}, V.~J., {Aumont}, O., {Bopp}, L., {Doney}, S.~C.,
  {Feely}, R.~A., {Gnanadesikan}, A., {Gruber}, N., {Ishida}, A., {Joos}, F.,
  {Key}, R.~M., {Lindsay}, K., {Maier-Reimer}, E., {Matear}, R., {Monfray}, P.,
  {Mouchet}, A., {Najjar}, R.~G., {Plattner}, G.-K., {Rodgers}, K.~B.,
  {Sabine}, C.~L., {Sarmiento}, J.~L., {Schlitzer}, R., {Slater}, R.~D.,
  {Totterdell}, I.~J., {Weirig}, M.-F., {Yamanaka}, Y., and {Yool}, A. (2005).
\newblock {Anthropogenic ocean acidification over the twenty-first century and
  its impact on calcifying organisms}.
\newblock {\em Nature}, 437(7059):681--686.

\bibitem[\protect\astroncite{{Orzechowska} et~al.}{2007}]{OGJ07}
{Orzechowska}, G.~E., {Goguen}, J.~D., {Johnson}, P.~V., {Tsapin}, A., and
  {Kanik}, I. (2007).
\newblock {Ultraviolet photolysis of amino acids in a 100 K water ice matrix:
  Application to the outer Solar System bodies}.
\newblock {\em Icarus}, 187(2):584--591.

\bibitem[\protect\astroncite{{Papaloizou} and {Terquem}}{2006}]{PT06}
{Papaloizou}, J.~C.~B. and {Terquem}, C. (2006).
\newblock {Planet formation and migration}.
\newblock {\em Rep. Prog. Phys.}, 69(1):119--180.

\bibitem[\protect\astroncite{{Papineau}}{2010}]{Pap10}
{Papineau}, D. (2010).
\newblock {Global Biogeochemical Changes at Both Ends of the Proterozoic:
  Insights from Phosphorites}.
\newblock {\em Astrobiology}, 10(2):165--181.

\bibitem[\protect\astroncite{{Parkinson} et~al.}{2008}]{PLY08}
{Parkinson}, C.~D., {Liang}, M.-C., {Yung}, Y.~L., and {Kirschivnk}, J.~L.
  (2008).
\newblock {Habitability of Enceladus: Planetary Conditions for Life}.
\newblock {\em Orig. Life Evol. Biosph.}, 38(4):355--369.

\bibitem[\protect\astroncite{{Parnell}}{2004}]{Par04}
{Parnell}, J. (2004).
\newblock {Mineral Radioactivity in Sands as a Mechanism for Fixation of
  Organic Carbon on the Early Earth}.
\newblock {\em Orig. Life Evol. Biosph.}, 34(6):533--547.

\bibitem[\protect\astroncite{{Parnell} and {McMahon}}{2016}]{PM16}
{Parnell}, J. and {McMahon}, S. (2016).
\newblock {Physical and chemical controls on habitats for life in the deep
  subsurface beneath continents and ice}.
\newblock {\em Phil. Trans. R. Soc. A}, 374(2059):20140293.

\bibitem[\protect\astroncite{{Pascal}}{2016}]{Pas16}
{Pascal}, R. (2016).
\newblock {Physicochemical Requirements Inferred for Chemical Self-Organization
  Hardly Support an Emergence of Life in the Deep Oceans of Icy Moons}.
\newblock {\em Astrobiology}, 16(5):328--334.

\bibitem[\protect\astroncite{{Pascal} et~al.}{2013}]{PPS13}
{Pascal}, R., {Pross}, A., and {Sutherland}, J.~D. (2013).
\newblock {Towards an evolutionary theory of the origin of life based on
  kinetics and thermodynamics}.
\newblock {\em Open Biol.}, 3(11):130156.

\bibitem[\protect\astroncite{{Pasek} and {Lauretta}}{2008}]{PL08}
{Pasek}, M. and {Lauretta}, D. (2008).
\newblock {Extraterrestrial Flux of Potentially Prebiotic C, N, and P to the
  Early Earth}.
\newblock {\em Orig. Life Evol. Biosph.}, 38(1):5--21.

\bibitem[\protect\astroncite{{Pasek} and {Greenberg}}{2012}]{PG12}
{Pasek}, M.~A. and {Greenberg}, R. (2012).
\newblock {Acidification of Europa's Subsurface Ocean as a Consequence of
  Oxidant Delivery}.
\newblock {\em Astrobiology}, 12(2):151--159.

\bibitem[\protect\astroncite{{Patel} et~al.}{2015}]{Pat15}
{Patel}, B.~H., {Percivalle}, C., {Ritson}, D.~J., {Duffy}, C.~D., and
  {Sutherland}, J.~D. (2015).
\newblock {Common origins of RNA, protein and lipid precursors in a
  cyanosulfidic protometabolism}.
\newblock {\em Nat. Chem.}, 7(4):301--307.

\bibitem[\protect\astroncite{{Paytan} and {McLaughlin}}{2007}]{PM07}
{Paytan}, A. and {McLaughlin}, K. (2007).
\newblock {The Oceanic Phosphorus Cycle}.
\newblock {\em Chem. Rev.}, 107(2):563--576.

\bibitem[\protect\astroncite{{Pedersen}}{1997}]{Ped97}
{Pedersen}, K. (1997).
\newblock {Microbial life in deep granitic rock}.
\newblock {\em FEMS Microbiol. Rev.}, 20(3-4):399--414.

\bibitem[\protect\astroncite{{Penn} et~al.}{2008}]{PHP08}
{Penn}, D.~C., {Holyoak}, K.~J., and {Povinelli}, D.~J. (2008).
\newblock {Darwin's mistake: Explaining the discontinuity between human and
  nonhuman minds}.
\newblock {\em Behav. Brain Sci.}, 31(2):109--130.

\bibitem[\protect\astroncite{{Penny} et~al.}{2016}]{PHC16}
{Penny}, M.~T., {Henderson}, C.~B., and {Clanton}, C. (2016).
\newblock {Is the Galactic Bulge Devoid of Planets?}
\newblock {\em Astrophys. J.}, 830(2):150.

\bibitem[\protect\astroncite{{Petrenko} and {Whitworth}}{1999}]{PW99}
{Petrenko}, V.~F. and {Whitworth}, R.~W. (1999).
\newblock {\em {Physics of Ice}}.
\newblock Oxford Univ. Press.

\bibitem[\protect\astroncite{{Picard} and {Daniel}}{2013}]{PD13}
{Picard}, A. and {Daniel}, I. (2013).
\newblock {Pressure as an environmental parameter for microbial life--A
  review}.
\newblock {\em Biophy. Chem.}, 183:30--41.

\bibitem[\protect\astroncite{{Pierazzo} and {Chyba}}{2002}]{PC02}
{Pierazzo}, E. and {Chyba}, C.~F. (2002).
\newblock {Cometary Delivery of Biogenic Elements to Europa}.
\newblock {\em Icarus}, 157(1):120--127.

\bibitem[\protect\astroncite{{Pizzarello}}{2006}]{Pizz06}
{Pizzarello}, S. (2006).
\newblock {The Chemistry of Life's Origin:  A Carbonaceous Meteorite
  Perspective}.
\newblock {\em Acc. Chem. Res.}, 39(4):231--237.

\bibitem[\protect\astroncite{{Pizzarello} and {Shock}}{2017}]{PS17}
{Pizzarello}, S. and {Shock}, E. (2017).
\newblock {Carbonaceous Chondrite Meteorites: the Chronicle of a Potential
  Evolutionary Path between Stars and Life}.
\newblock {\em Orig. Life Evol. Biosph.}, 47(3):249--260.

\bibitem[\protect\astroncite{{Planavsky} et~al.}{2014}]{PRW14}
{Planavsky}, N.~J., {Reinhard}, C.~T., {Wang}, X., {Thomson}, D., {McGoldrick},
  P., {Rainbird}, R.~H., {Johnson}, T., {Fischer}, W.~W., and {Lyons}, T.~W.
  (2014).
\newblock {Low Mid-Proterozoic atmospheric oxygen levels and the delayed rise
  of animals}.
\newblock {\em Science}, 346(6209):635--638.

\bibitem[\protect\astroncite{{Planavsky} et~al.}{2010}]{PR10}
{Planavsky}, N.~J., {Rouxel}, O.~J., {Bekker}, A., {Lalonde}, S.~V.,
  {Konhauser}, K.~O., {Reinhard}, C.~T., and {Lyons}, T.~W. (2010).
\newblock {The evolution of the marine phosphate reservoir}.
\newblock {\em Nature}, 467(7319):1088--1090.

\bibitem[\protect\astroncite{{Pl{\"u}mper} et~al.}{2017}]{PKG17}
{Pl{\"u}mper}, O., {King}, H.~E., {Geisler}, T., {Liu}, Y., {Pabst}, S.,
  {Savov}, I.~P., {Rost}, D., and {Zack}, T. (2017).
\newblock {Subduction zone forearc serpentinites as incubators for deep
  microbial life}.
\newblock {\em Proc. Natl. Acad. Sci. USA}, 114(17):4324--4329.

\bibitem[\protect\astroncite{{Poglitsch} et~al.}{2010}]{Pog10}
{Poglitsch}, A., {Waelkens}, C., {Geis}, N., {Feuchtgruber}, H.,
  {Vandenbussche}, B., {Rodriguez}, L., {Krause}, O., {Renotte}, E., {van
  Hoof}, C., {Saraceno}, P., {Cepa}, J., {Kerschbaum}, F., {Agn{\`e}se}, P.,
  {Ali}, B., {Altieri}, B., {Andreani}, P., {Augueres}, J.-L., {Balog}, Z.,
  {Barl}, L., {Bauer}, O.~H., {Belbachir}, N., {Benedettini}, M., {Billot}, N.,
  {Boulade}, O., {Bischof}, H., {Blommaert}, J., {Callut}, E., {Cara}, C.,
  {Cerulli}, R., {Cesarsky}, D., {Contursi}, A., {Creten}, Y., {De Meester},
  W., {Doublier}, V., {Doumayrou}, E., {Duband}, L., {Exter}, K., {Genzel}, R.,
  {Gillis}, J.-M., {Gr{\"o}zinger}, U., {Henning}, T., {Herreros}, J.,
  {Huygen}, R., {Inguscio}, M., {Jakob}, G., {Jamar}, C., {Jean}, C., {de
  Jong}, J., {Katterloher}, R., {Kiss}, C., {Klaas}, U., {Lemke}, D., {Lutz},
  D., {Madden}, S., {Marquet}, B., {Martignac}, J., {Mazy}, A., {Merken}, P.,
  {Montfort}, F., {Morbidelli}, L., {M{\"u}ller}, T., {Nielbock}, M.,
  {Okumura}, K., {Orfei}, R., {Ottensamer}, R., {Pezzuto}, S., {Popesso}, P.,
  {Putzeys}, J., {Regibo}, S., {Reveret}, V., {Royer}, P., {Sauvage}, M.,
  {Schreiber}, J., {Stegmaier}, J., {Schmitt}, D., {Schubert}, J., {Sturm}, E.,
  {Thiel}, M., {Tofani}, G., {Vavrek}, R., {Wetzstein}, M., {Wieprecht}, E.,
  and {Wiezorrek}, E. (2010).
\newblock {The Photodetector Array Camera and Spectrometer (PACS) on the
  Herschel Space Observatory}.
\newblock {\em Astron. Astrophys.}, 518:L2.

\bibitem[\protect\astroncite{{Portegies Zwart} et~al.}{2017}]{PZ17}
{Portegies Zwart}, S., {Pelupessy}, I., {Bedorf}, J., {Cai}, M., and {Torres},
  S. (2017).
\newblock {The origin of interstellar asteroidal objects like 1I/2017 U1}.
\newblock {\em submitted to Mon. Not. R. Astron. Soc. (arXiv:1711.03558)}.

\bibitem[\protect\astroncite{{Post} and {Palkovacs}}{2009}]{PP09}
{Post}, D.-M. and {Palkovacs}, E.-P. (2009).
\newblock {Eco-evolutionary feedbacks in community and ecosystem ecology:
  interactions between the ecological theatre and the evolutionary play}.
\newblock {\em Phil. Trans. R. Soc. B}, 364(1523):1629--1640.

\bibitem[\protect\astroncite{{Postberg} et~al.}{2011}]{PSH11}
{Postberg}, F., {Schmidt}, J., {Hillier}, J., {Kempf}, S., and {Srama}, R.
  (2011).
\newblock {A salt-water reservoir as the source of a compositionally stratified
  plume on Enceladus}.
\newblock {\em Nature}, 474(7353):620--622.

\bibitem[\protect\astroncite{{Prantzos}}{2008}]{Pran08}
{Prantzos}, N. (2008).
\newblock {On the ``Galactic Habitable Zone''}.
\newblock {\em Space Sci. Rev.}, 135(1-4):313--322.

\bibitem[\protect\astroncite{{Price}}{2000}]{Pri00}
{Price}, P.~B. (2000).
\newblock {A habitat for psychrophiles in deep Antarctic ice}.
\newblock {\em Proc. Natl. Acad. Sci. USA}, 97(3):1247--1251.

\bibitem[\protect\astroncite{{Price}}{2007}]{Pri07}
{Price}, P.~B. (2007).
\newblock {Microbial life in glacial ice and implications for a cold origin of
  life}.
\newblock {\em FEMS Microbiol. Ecol.}, 59(2):217--231.

\bibitem[\protect\astroncite{{Price} and {Sowers}}{2004}]{PS04}
{Price}, P.~B. and {Sowers}, T. (2004).
\newblock {Temperature dependence of metabolic rates for microbial growth,
  maintenance, and survival}.
\newblock {\em Proc. Natl. Acad. Sci. USA}, 101(13):4631--4636.

\bibitem[\protect\astroncite{{Priscu} et~al.}{1999}]{PAL99}
{Priscu}, J.~C., {Adams}, E.~E., {Lyons}, W.~B., {Voytek}, M.~A., {Mogk},
  D.~W., {Brown}, R.~L., {McKay}, C.~P., {Takacs}, C.~D., {Welch}, K.~A.,
  {Wolf}, C.~F., {Kirshtein}, J.~D., and {Avci}, R. (1999).
\newblock {Geomicrobiology of Subglacial Ice Above Lake Vostok, Antarctica}.
\newblock {\em Science}, 286(5447):2141--2144.

\bibitem[\protect\astroncite{{Raiswell} et~al.}{2006}]{RTB06}
{Raiswell}, R., {Tranter}, M., {Benning}, L.~G., {Siegert}, M., {De'ath}, R.,
  {Huybrechts}, P., and {Payne}, T. (2006).
\newblock {Contributions from glacially derived sediment to the global iron
  (oxyhydr)oxide cycle: Implications for iron delivery to the oceans}.
\newblock {\em Geochim. Cosmochim. Acta}, 70(11):2765--2780.

\bibitem[\protect\astroncite{{Ramirez} and {Kaltenegger}}{2014}]{RK14}
{Ramirez}, R.~M. and {Kaltenegger}, L. (2014).
\newblock {The Habitable Zones of Pre-main-sequence Stars}.
\newblock {\em Astrophys. J. Lett.}, 797(2):L25.

\bibitem[\protect\astroncite{{Ramirez} and {Kaltenegger}}{2016}]{RK16}
{Ramirez}, R.~M. and {Kaltenegger}, L. (2016).
\newblock {Habitable Zones of Post-Main Sequence Stars}.
\newblock {\em Astrophys. J.}, 823(1):6.

\bibitem[\protect\astroncite{{Rapf} and {Vaida}}{2016}]{RV16}
{Rapf}, R.~J. and {Vaida}, V. (2016).
\newblock {Sunlight as an energetic driver in the synthesis of molecules
  necessary for life}.
\newblock {\em Phys. Chem. Chem. Phys.}, 18(30):20067--20084.

\bibitem[\protect\astroncite{{Rasio} and {Ford}}{1996}]{RF96}
{Rasio}, F.~A. and {Ford}, E.~B. (1996).
\newblock {Dynamical instabilities and the formation of extrasolar planetary
  systems}.
\newblock {\em Science}, 274(5289):954--956.

\bibitem[\protect\astroncite{{Raymond} et~al.}{2018}]{RA17}
{Raymond}, S.~N., {Armitage}, P.~J., {Veras}, D., {Quintana}, E.~V., and
  {Barclay}, T. (2018).
\newblock {Implications of the interstellar object 1I/'Oumuamua for planetary
  dynamics and planetesimal formation}.
\newblock {\em Mon. Not. R. Astron. Soc.}, 476(3):3031--3038.

\bibitem[\protect\astroncite{{Raymond} et~al.}{2007}]{RSM}
{Raymond}, S.~N., {Scalo}, J., and {Meadows}, V.~S. (2007).
\newblock {A Decreased Probability of Habitable Planet Formation around
  Low-Mass Stars}.
\newblock {\em Astrophys. J.}, 669(1):606--614.

\bibitem[\protect\astroncite{{Reames}}{2013}]{Ream13}
{Reames}, D.~V. (2013).
\newblock {The Two Sources of Solar Energetic Particles}.
\newblock {\em Space Sci. Rev.}, 175(1-4):53--92.

\bibitem[\protect\astroncite{{Reinhard} et~al.}{2017}]{RPG17}
{Reinhard}, C.~T., {Planavsky}, N.~J., {Gill}, B.~C., {Ozaki}, K., {Robbins},
  L.~J., {Lyons}, T.~W., {Fischer}, W.~W., {Wang}, C., {Cole}, D.~B., and
  {Konhauser}, K.~O. (2017).
\newblock {Evolution of the global phosphorus cycle}.
\newblock {\em Nature}, 541(7637):386--389.

\bibitem[\protect\astroncite{{Reinhard} et~al.}{2016}]{RPO16}
{Reinhard}, C.~T., {Planavsky}, N.~J., {Olson}, S.~L., {Lyons}, T.~W., and
  {Erwin}, D.~H. (2016).
\newblock {Earth's oxygen cycle and the evolution of animal life}.
\newblock {\em Proc. Natl. Acad. Sci. USA}, 113(32):8933--8938.

\bibitem[\protect\astroncite{{Rendell} and {Whitehead}}{2001}]{RW01}
{Rendell}, L. and {Whitehead}, H. (2001).
\newblock {Culture in whales and dolphins}.
\newblock {\em Behav. Brain Sci.}, 24(2):309--324.

\bibitem[\protect\astroncite{{Requena-Torres} et~al.}{2008}]{RT08}
{Requena-Torres}, M.~A., {Mart{\'{\i}}n-Pintado}, J., {Mart{\'{\i}}n}, S., and
  {Morris}, M.~R. (2008).
\newblock {The Galactic Center: The Largest Oxygen-bearing Organic Molecule
  Repository}.
\newblock {\em Astrophys. J.}, 672(1):352--360.

\bibitem[\protect\astroncite{{Reynolds} et~al.}{1983}]{RS83}
{Reynolds}, R.~T., {Squyres}, S.~W., {Colburn}, D.~S., and {McKay}, C.~P.
  (1983).
\newblock {On the habitability of Europa}.
\newblock {\em Icarus}, 56(2):246--254.

\bibitem[\protect\astroncite{{Richerson} and {Boyd}}{2008}]{RB08}
{Richerson}, P.~J. and {Boyd}, R. (2008).
\newblock {\em {Not By Genes Alone: How Culture Transformed Human Evolution}}.
\newblock The Univ. of Chicago Press.

\bibitem[\protect\astroncite{{Robertson} and {Joyce}}{2012}]{RJ12}
{Robertson}, M.~P. and {Joyce}, G.~F. (2012).
\newblock {The Origins of the RNA World}.
\newblock {\em Cold Spring Harb. Perspect. Biol.}, 4(5):a003608.

\bibitem[\protect\astroncite{{Rogers}}{2015}]{Rog15}
{Rogers}, L.~A. (2015).
\newblock {Most 1.6 Earth-radius Planets are Not Rocky}.
\newblock {\em Astrophys. J.}, 801(1):41.

\bibitem[\protect\astroncite{{Rosenblum} et~al.}{2014}]{RPB14}
{Rosenblum}, E.~B., {Parent}, C.~E., and {Brandt}, E.~E. (2014).
\newblock {The Molecular Basis of Phenotypic Convergence}.
\newblock {\em Annu. Rev. Ecol. Evol. Syst.}, 45:203--226.

\bibitem[\protect\astroncite{{Rospars}}{2013}]{Ros13}
{Rospars}, J.-P. (2013).
\newblock {Trends in the evolution of life, brains and intelligence}.
\newblock {\em Int. J. Astrobiol.}, 12(3):186--207.

\bibitem[\protect\astroncite{{Ross} and {Deamer}}{2016}]{RD16}
{Ross}, D.~S. and {Deamer}, D. (2016).
\newblock {Dry/Wet Cycling and the Thermodynamics and Kinetics of Prebiotic
  Polymer Synthesis}.
\newblock {\em Life}, 6(3):28.

\bibitem[\protect\astroncite{{Rotelli} et~al.}{2016}]{RTR16}
{Rotelli}, L., {Trigo-Rodr{\'{\i}}guez}, J.~M., {Moyano-Cambero}, C.~E.,
  {Carota}, E., {Botta}, L., {di Mauro}, E., and {Saladino}, R. (2016).
\newblock {The key role of meteorites in the formation of relevant prebiotic
  molecules in a formamide/water environment}.
\newblock {\em Sci. Rep.}, 6:38888.

\bibitem[\protect\astroncite{{Roth} and {Dicke}}{2005}]{RD05}
{Roth}, G. and {Dicke}, U. (2005).
\newblock {Evolution of the brain and intelligence}.
\newblock {\em Trends Cogn. Sci.}, 9(5):250--257.

\bibitem[\protect\astroncite{{Rothschild} and {Mancinelli}}{2001}]{RoMa}
{Rothschild}, L.~J. and {Mancinelli}, R.~L. (2001).
\newblock {Life in extreme environments}.
\newblock {\em Nature}, 409(6823):1092--1101.

\bibitem[\protect\astroncite{{R{\o}y} et~al.}{2012}]{RKA12}
{R{\o}y}, H., {Kallmeyer}, J., {Adhikari}, R.~R., {Pockalny}, R.,
  {J{\o}rgensen}, B.~B., and {D'Hondt}, S. (2012).
\newblock {Aerobic Microbial Respiration in 86-Million-Year-Old Deep-Sea Red
  Clay}.
\newblock {\em Science}, 336(6083):922--925.

\bibitem[\protect\astroncite{{Ruiz-Mirazo} et~al.}{2014}]{RBD14}
{Ruiz-Mirazo}, K., {Briones}, C., and {de la Escosura}, A. (2014).
\newblock {Prebiotic Systems Chemistry: New Perspectives for the Origins of
  Life}.
\newblock {\em Chem. Rev.}, 114(1):285--366.

\bibitem[\protect\astroncite{{Rushby} et~al.}{2013}]{RCO13}
{Rushby}, A.~J., {Claire}, M.~W., {Osborn}, H., and {Watson}, A.~J. (2013).
\newblock {Habitable Zone Lifetimes of Exoplanets around Main Sequence Stars}.
\newblock {\em Astrobiology}, 13(9):833--849.

\bibitem[\protect\astroncite{{Russell} et~al.}{2016}]{RRA16}
{Russell}, C.~T., {Raymond}, C.~A., {Ammannito}, E., {Buczkowski}, D.~L., {De
  Sanctis}, M.~C., {Hiesinger}, H., {Jaumann}, R., {Konopliv}, A.~S.,
  {McSween}, H.~Y., {Nathues}, A., {Park}, R.~S., {Pieters}, C.~M.,
  {Prettyman}, T.~H., {McCord}, T.~B., {McFadden}, L.~A., {Mottola}, S.,
  {Zuber}, M.~T., {Joy}, S.~P., {Polanskey}, C., {Rayman}, M.~D.,
  {Castillo-Rogez}, J.~C., {Chi}, P.~J., {Combe}, J.~P., {Ermakov}, A., {Fu},
  R.~R., {Hoffmann}, M., {Jia}, Y.~D., {King}, S.~D., {Lawrence}, D.~J., {Li},
  J.-Y., {Marchi}, S., {Preusker}, F., {Roatsch}, T., {Ruesch}, O., {Schenk},
  P., {Villarreal}, M.~N., and {Yamashita}, N. (2016).
\newblock {Dawn arrives at Ceres: Exploration of a small, volatile-rich world}.
\newblock {\em Science}, 353(6303):1008--1010.

\bibitem[\protect\astroncite{{Russell} et~al.}{2014}]{RBB14}
{Russell}, M.~J., {Barge}, L.~M., {Bhartia}, R., {Bocanegra}, D., {Bracher},
  P.~J., {Branscomb}, E., {Kidd}, R., {McGlynn}, S., {Meier}, D.~H.,
  {Nitschke}, W., {Shibuya}, T., {Vance}, S., {White}, L., and {Kanik}, I.
  (2014).
\newblock {The Drive to Life on Wet and Icy Worlds}.
\newblock {\em Astrobiology}, 14(4):308--343.

\bibitem[\protect\astroncite{{Russell} et~al.}{2010}]{RHM}
{Russell}, M.~J., {Hall}, A.~J., and {Martin}, W. (2010).
\newblock {Serpentinization as a source of energy at the origin of life}.
\newblock {\em Geobiology}, 8(5):355--371.

\bibitem[\protect\astroncite{{Russell} et~al.}{2017}]{RMH17}
{Russell}, M.~J., {Murray}, A.~E., and {Hand}, K.~P. (2017).
\newblock {The Possible Emergence of Life and Differentiation of a Shallow
  Biosphere on Irradiated Icy Worlds: The Example of Europa}.
\newblock {\em Astrobiology}, 17(12):1265--1273.

\bibitem[\protect\astroncite{{Russell} and {Nitschke}}{2017}]{RN17}
{Russell}, M.~J. and {Nitschke}, W. (2017).
\newblock {Methane: Fuel or Exhaust at the Emergence of Life?}
\newblock {\em Astrobiology}, 17(10):1053--1066.

\bibitem[\protect\astroncite{{Russell} et~al.}{2013}]{RNB13}
{Russell}, M.~J., {Nitschke}, W., and {Branscomb}, E. (2013).
\newblock {The inevitable journey to being}.
\newblock {\em Phil. Trans. R. Soc. B}, 368(1622):20120254.

\bibitem[\protect\astroncite{{Ruxton} et~al.}{2014}]{RHM14}
{Ruxton}, G.~D., {Humphries}, S., {Morrell}, L.~J., and {Wilkinson}, D.~M.
  (2014).
\newblock {Why is eusociality an almost exclusively terrestrial phenomenon?}
\newblock {\em J. Animal Ecol.}, 83(6):1248--1255.

\bibitem[\protect\astroncite{{Sagan}}{1996}]{Sag96}
{Sagan}, C. (1996).
\newblock {Circumstellar Habitable Zones: An Introduction}.
\newblock In {Doyle}, L.~R., editor, {\em Circumstellar Habitable Zones}, pages
  3--16. Travis House Publications.

\bibitem[\protect\astroncite{{Sagan}}{1967}]{Sag67}
{Sagan}, L. (1967).
\newblock {On the origin of mitosing cells}.
\newblock {\em J. Theor. Biol.}, 14(3):225--274.

\bibitem[\protect\astroncite{{Saladino} et~al.}{2015}]{SC15}
{Saladino}, R., {Carota}, E., {Botta}, G., {Kapralov}, M., {Timoshenko}, G.~N.,
  {Rozanov}, A.~Y., {Krasavin}, E., and {Di Mauro}, E. (2015).
\newblock {Meteorite-catalyzed syntheses of nucleosides and of other prebiotic
  compounds from formamide under proton irradiation}.
\newblock {\em Proc. Natl. Acad. Sci. USA}, 112(21):E2746--E2755.

\bibitem[\protect\astroncite{{Saladino} et~al.}{2004}]{SCCD}
{Saladino}, R., {Crestini}, C., {Costanzo}, G., and {DiMauro}, E. (2004).
\newblock {Advances in the Prebiotic Synthesis of Nucleic Acids Bases:
  Implications for the Origin of Life}.
\newblock {\em Curr. Org. Chem.}, 8(15):1425--1443.

\bibitem[\protect\astroncite{{Saladino} et~al.}{2012}]{SC12}
{Saladino}, R., {Crestini}, C., {Pino}, S., {Costanzo}, G., and {Di Mauro}, E.
  (2012).
\newblock {Formamide and the origin of life}.
\newblock {\em Phys. Life Rev.}, 9(1):84--104.

\bibitem[\protect\astroncite{{Sanchez} et~al.}{1967}]{SFO67}
{Sanchez}, R.~A., {Ferris}, J.~P., and {Orgel}, L.~E. (1967).
\newblock {Studies in Prebiotic Synthesis: II. Synthesis of purine precursors
  and amino acids from aqueous hydrogen cyanide}.
\newblock {\em J. Mol. Biol.}, 30(2):223--253.

\bibitem[\protect\astroncite{{Schlesinger} and {Bernhardt}}{2013}]{SB13}
{Schlesinger}, W.~H. and {Bernhardt}, E.~S. (2013).
\newblock {\em {Biogeochemistry: An Analysis of Global Change}}.
\newblock Academic Press.

\bibitem[\protect\astroncite{{Schr{\"o}dinger}}{1944}]{Sch44}
{Schr{\"o}dinger}, E. (1944).
\newblock {\em {What Is Life?}}
\newblock Cambridge Univ. Press.

\bibitem[\protect\astroncite{{Schrum} et~al.}{2010}]{SZS}
{Schrum}, J.~P., {Zhu}, T.~F., and {Szostak}, J.~W. (2010).
\newblock {The Origins of Cellular Life}.
\newblock {\em Cold Spring Harb. Perspect. Biol.}, 2(9):a002212.

\bibitem[\protect\astroncite{{Schubert} et~al.}{2004}]{SASM}
{Schubert}, G., {Anderson}, J.~D., {Spohn}, T., and {McKinnon}, W.~B. (2004).
\newblock {Interior composition, structure and dynamics of the Galilean
  satellites}.
\newblock In {Bagenal}, F., {Dowling}, T.~E., and {McKinnon}, W.~B., editors,
  {\em Jupiter.~The Planet, Satellites and Magnetosphere}, pages 281--306.
  Cambridge Univ. Press.

\bibitem[\protect\astroncite{{Schubert} et~al.}{2001}]{STO01}
{Schubert}, G., {Turcotte}, D.~L., and {Olson}, P. (2001).
\newblock {\em {Mantle Convection in the Earth and Planets}}.
\newblock Cambridge Univ. Press.

\bibitem[\protect\astroncite{{Schulze-Makuch} and {Bains}}{2017}]{SMB17}
{Schulze-Makuch}, D. and {Bains}, W. (2017).
\newblock {\em {The Cosmic Zoo: Complex Life on Many Worlds}}.
\newblock Springer.

\bibitem[\protect\astroncite{{Schulze-Makuch} and {Guinan}}{2016}]{SG16}
{Schulze-Makuch}, D. and {Guinan}, E. (2016).
\newblock {Another Earth 2.0? Not So Fast}.
\newblock {\em Astrobiology}, 16(11):817--821.

\bibitem[\protect\astroncite{{Schulze-Makuch} and {Irwin}}{2002}]{SI02}
{Schulze-Makuch}, D. and {Irwin}, L.~N. (2002).
\newblock {Energy Cycling and Hypothetical Organisms in Europa's Ocean}.
\newblock {\em Astrobiology}, 2(1):105--121.

\bibitem[\protect\astroncite{{Schulze-Makuch} and {Irwin}}{2006}]{SMI06}
{Schulze-Makuch}, D. and {Irwin}, L.~N. (2006).
\newblock {The prospect of alien life in exotic forms on other worlds}.
\newblock {\em Naturwissenschaften}, 93(4):155--172.

\bibitem[\protect\astroncite{{Schulze-Makuch} and {Irwin}}{2008}]{SI08}
{Schulze-Makuch}, D. and {Irwin}, L.~N. (2008).
\newblock {\em {Life in the Universe: Expectations and Constraints}}.
\newblock Springer.

\bibitem[\protect\astroncite{{Schwartzman}}{1999}]{Sch99}
{Schwartzman}, D.~W. (1999).
\newblock {\em {Life, Temperature, and the Earth: The Self-organizing
  Biosphere}}.
\newblock Columbia Univ. Press.

\bibitem[\protect\astroncite{{Scorei}}{2012}]{Sco12}
{Scorei}, R. (2012).
\newblock {Is Boron a Prebiotic Element? A Mini-review of the Essentiality of
  Boron for the Appearance of Life on Earth}.
\newblock {\em Orig. Life Evol. Biosph.}, 42(1):3--17.

\bibitem[\protect\astroncite{{Seager} et~al.}{2007}]{SK07}
{Seager}, S., {Kuchner}, M., {Hier-Majumder}, C.~A., and {Militzer}, B. (2007).
\newblock {Mass-Radius Relationships for Solid Exoplanets}.
\newblock {\em Astrophys. J.}, 669(2):1279--1297.

\bibitem[\protect\astroncite{{Sephton}}{2002}]{Sep02}
{Sephton}, M.~A. (2002).
\newblock {Organic compounds in carbonaceous meteorites}.
\newblock {\em Nat. Prod. Rep.}, 19(3):292--311.

\bibitem[\protect\astroncite{{Shapiro}}{1984}]{Sha84}
{Shapiro}, R. (1984).
\newblock {The Improbability of Prebiotic Nucleic Acid Synthesis}.
\newblock {\em Orig. Life Evol. Biosph.}, 14(1-4):565--570.

\bibitem[\protect\astroncite{{Shields-Zhou} and {Och}}{2011}]{SZO11}
{Shields-Zhou}, G. and {Och}, L. (2011).
\newblock {The case for a Neoproterozoic Oxygenation Event: Geochemical
  evidence and biological consequences}.
\newblock {\em GSA Today}, 21(3):4--11.

\bibitem[\protect\astroncite{{Shock} and {Helgeson}}{1988}]{SH88}
{Shock}, E.~L. and {Helgeson}, H.~C. (1988).
\newblock {Calculation of the thermodynamic and transport properties of aqueous
  species at high pressures and temperatures: Correlation algorithms for ionic
  species and equation of state predictions to 5 kb and 1000$^\circ$C}.
\newblock {\em Geochim. Cosmochim. Acta}, 52(8):2009--2036.

\bibitem[\protect\astroncite{{Shock} and {Holland}}{2007}]{SH07}
{Shock}, E.~L. and {Holland}, M.~E. (2007).
\newblock {Quantitative Habitability}.
\newblock {\em Astrobiology}, 7(6):839--851.

\bibitem[\protect\astroncite{{Simpson}}{1964}]{Gay64}
{Simpson}, G.~G. (1964).
\newblock {The Nonprevalence of Humanoids}.
\newblock {\em Science}, 143(3608):769--775.

\bibitem[\protect\astroncite{{Sleep}}{2012}]{Sleep}
{Sleep}, N.~H. (2012).
\newblock {Maintenance of permeable habitable subsurface environments by
  earthquakes and tidal stresses}.
\newblock {\em Int. J. Astrobiol.}, 11(4):257--268.

\bibitem[\protect\astroncite{{Sloan} et~al.}{2017}]{SAL}
{Sloan}, D., {Alves Batista}, R., and {Loeb}, A. (2017).
\newblock {The Resilience of Life to Astrophysical Events}.
\newblock {\em Sci. Rep.}, 7:5419.

\bibitem[\protect\astroncite{{Smith} and {Morowitz}}{2016}]{SM16}
{Smith}, E. and {Morowitz}, H.~J. (2016).
\newblock {\em {The Origin and Nature of Life on Earth}}.
\newblock Cambridge Univ. Press.

\bibitem[\protect\astroncite{{Smith} and {Szathm{\'a}ry}}{1995}]{JMS95}
{Smith}, J.~M. and {Szathm{\'a}ry}, E. (1995).
\newblock {\em {The Major Transitions in Evolution}}.
\newblock Oxford Univ. Press.

\bibitem[\protect\astroncite{{Snyder} et~al.}{2005}]{SL05}
{Snyder}, L.~E., {Lovas}, F.~J., {Hollis}, J.~M., {Friedel}, D.~N., {Jewell},
  P.~R., {Remijan}, A., {Ilyushin}, V.~V., {Alekseev}, E.~A., and {Dyubko},
  S.~F. (2005).
\newblock {A Rigorous Attempt to Verify Interstellar Glycine}.
\newblock {\em Astrophys. J.}, 619(2):914--930.

\bibitem[\protect\astroncite{{Sohl} et~al.}{2010}]{SCK10}
{Sohl}, F., {Choukroun}, M., {Kargel}, J., {Kimura}, J., {Pappalardo}, R.,
  {Vance}, S., and {Zolotov}, M. (2010).
\newblock {Subsurface Water Oceans on Icy Satellites: Chemical Composition and
  Exchange Processes}.
\newblock {\em Space Sci. Rev.}, 153(1-4):485--510.

\bibitem[\protect\astroncite{{Sojo} et~al.}{2016}]{SHW16}
{Sojo}, V., {Herschy}, B., {Whicher}, A., {Camprub{\'{\i}}}, E., and {Lane}, N.
  (2016).
\newblock {The Origin of Life in Alkaline Hydrothermal Vents}.
\newblock {\em Astrobiology}, 16(2):181--197.

\bibitem[\protect\astroncite{{Sotin} et~al.}{2007}]{SGM07}
{Sotin}, C., {Grasset}, O., and {Mocquet}, A. (2007).
\newblock {Mass radius curve for extrasolar Earth-like planets and ocean
  planets}.
\newblock {\em Icarus}, 191(1):337--351.

\bibitem[\protect\astroncite{{Sousa} et~al.}{2013}]{STL13}
{Sousa}, F.~L., {Thiergart}, T., {Landan}, G., {Nelson-Sathi}, S., {Pereira},
  I.~A.~C., {Allen}, J.~F., {Lane}, N., and {Martin}, W.~F. (2013).
\newblock {Early bioenergetic evolution}.
\newblock {\em Phil. Trans. R. Soc. B}, 368(1622):20130088.

\bibitem[\protect\astroncite{{Sparks} et~al.}{2016}]{SH16}
{Sparks}, W.~B., {Hand}, K.~P., {McGrath}, M.~A., {Bergeron}, E., {Cracraft},
  M., and {Deustua}, S.~E. (2016).
\newblock {Probing for Evidence of Plumes on Europa with HST/STIS}.
\newblock {\em Astrophys. J.}, 829(2):121.

\bibitem[\protect\astroncite{{Sparks} et~al.}{2017}]{SSM17}
{Sparks}, W.~B., {Schmidt}, B.~E., {McGrath}, M.~A., {Hand}, K.~P., {Spencer},
  J.~R., {Cracraft}, M., and {E Deustua}, S. (2017).
\newblock {Active Cryovolcanism on Europa?}
\newblock {\em Astrophys. J. Lett.}, 839(2):L18.

\bibitem[\protect\astroncite{{Spencer} et~al.}{2009}]{SBE09}
{Spencer}, J.~R., {Barr}, A.~C., {Esposito}, L.~W., {Helfenstein}, P.,
  {Ingersoll}, A.~P., {Jaumann}, R., {McKay}, C.~P., {Nimmo}, F., and {Waite},
  J.~H. (2009).
\newblock {Enceladus: An Active Cryovolcanic Satellite}.
\newblock In {Dougherty}, M.~K., {Esposito}, L.~W., and {Krimigis}, S.~M.,
  editors, {\em Saturn from Cassini-Huygens}, pages 683--724. Springer.

\bibitem[\protect\astroncite{{Spencer} and {Nimmo}}{2013}]{SN13}
{Spencer}, J.~R. and {Nimmo}, F. (2013).
\newblock {Enceladus: An Active Ice World in the Saturn System}.
\newblock {\em Annu. Rev. Earth Planet. Sci.}, 41:693--717.

\bibitem[\protect\astroncite{{Spiegel} and {Turner}}{2012}]{ST12}
{Spiegel}, D.~S. and {Turner}, E.~L. (2012).
\newblock {Bayesian analysis of the astrobiological implications of life's
  early emergence on Earth}.
\newblock {\em Proc. Natl. Acad. Sci. USA}, 109(2):395--400.

\bibitem[\protect\astroncite{{Spitzer}}{2017}]{Spit17}
{Spitzer}, J. (2017).
\newblock {Emergence of Life on Earth: A Physicochemical Jigsaw Puzzle}.
\newblock {\em J. Mol. Evol.}, 84(1):1--7.

\bibitem[\protect\astroncite{{Spohn} and {Schubert}}{2003}]{SS03}
{Spohn}, T. and {Schubert}, G. (2003).
\newblock {Oceans in the icy Galilean satellites of Jupiter?}
\newblock {\em Icarus}, 161(2):456--467.

\bibitem[\protect\astroncite{{Steel} et~al.}{2017}]{SDM17}
{Steel}, E.~L., {Davila}, A., and {McKay}, C.~P. (2017).
\newblock {Abiotic and Biotic Formation of Amino Acids in the Enceladus Ocean}.
\newblock {\em Astrobiology}, 17(9):862--875.

\bibitem[\protect\astroncite{{Stelmach} et~al.}{2018}]{SNVM}
{Stelmach}, K.~B., {Neveu}, M., {Vick-Majors}, T.~J., {Mickol}, R.~L., {Chou},
  L., {Webster}, K.~D., {Tilley}, M., {Zacchei}, F., {Escudero}, C., {Flores
  Martinez}, C.~L., {Labrado}, A., and {Fern{\'a}ndez}, E.~J.~G. (2018).
\newblock {Secondary Electrons as an Energy Source for Life}.
\newblock {\em Astrobiology}, 18(1):73--85.

\bibitem[\protect\astroncite{{Stevenson}}{1999}]{S99}
{Stevenson}, D.~J. (1999).
\newblock {Life-sustaining planets in interstellar space?}
\newblock {\em Nature}, 400(6739):32.

\bibitem[\protect\astroncite{{Stevenson} et~al.}{2015}]{SLC15}
{Stevenson}, J., {Lunine}, J., and {Clancy}, P. (2015).
\newblock {Membrane alternatives in worlds without oxygen: Creation of an
  azotosome}.
\newblock {\em Sci. Adv.}, 1(1):1400067.

\bibitem[\protect\astroncite{{Stribling} and {Miller}}{1987}]{SM87}
{Stribling}, R. and {Miller}, S.~L. (1987).
\newblock {Energy yields for hydrogen cyanide and formaldehyde syntheses: The
  hcn and amino acid concentrations in the primitive ocean}.
\newblock {\em Orig. Life Evol. Biosph.}, 17(3-4):261--273.

\bibitem[\protect\astroncite{{Strigari} et~al.}{2012}]{SBMB}
{Strigari}, L.~E., {Barnab{\`e}}, M., {Marshall}, P.~J., and {Blandford}, R.~D.
  (2012).
\newblock {Nomads of the Galaxy}.
\newblock {\em Mon. Not. R. Astron. Soc.}, 423(2):1856--1865.

\bibitem[\protect\astroncite{{St{\"u}eken} et~al.}{2013}]{SAB13}
{St{\"u}eken}, E.~E., {Anderson}, R.~E., {Bowman}, J.~S., {Brazelton}, W.~J.,
  {Colangelo-Lillis}, J., {Goldman}, A.~D., {Som}, S.~M., and {Baross}, J.~A.
  (2013).
\newblock {Did life originate from a global chemical reactor?}
\newblock {\em Geobiology}, 11(2):101--126.

\bibitem[\protect\astroncite{{Suddendorf}}{2013}]{Sud13}
{Suddendorf}, T. (2013).
\newblock {\em {The Gap: The Science of What Separates Us from Other Animals}}.
\newblock Basic Books.

\bibitem[\protect\astroncite{{Suddendorf} and {Corballis}}{2007}]{SC07}
{Suddendorf}, T. and {Corballis}, M.~C. (2007).
\newblock {The evolution of foresight: What is mental time travel, and is it
  unique to humans?}
\newblock {\em Behavioral and Brain Sciences}, 30(3):299--313.

\bibitem[\protect\astroncite{{Sutherland}}{2017}]{Suth17}
{Sutherland}, J.~D. (2017).
\newblock {Studies on the origin of life -- the end of the beginning}.
\newblock {\em Nat. Rev. Chem.}, 1:0012.

\bibitem[\protect\astroncite{{Szathm{\'a}ry}}{2015}]{Sza15}
{Szathm{\'a}ry}, E. (2015).
\newblock {Toward major evolutionary transitions theory 2.0}.
\newblock {\em Proc. Natl. Acad. Sci. USA}, 112(33):10104--10111.

\bibitem[\protect\astroncite{{Szathm{\'a}ry} and {Smith}}{1995}]{SS95}
{Szathm{\'a}ry}, E. and {Smith}, J.~M. (1995).
\newblock The major evolutionary transitions.
\newblock {\em Nature}, 374(6519):227--232.

\bibitem[\protect\astroncite{{Szathm{\'a}ry} and {Smith}}{1997}]{SS97}
{Szathm{\'a}ry}, E. and {Smith}, J.~M. (1997).
\newblock {From Replicators to Reproducers: the First Major Transitions Leading
  to Life}.
\newblock {\em J. Theor. Biol.}, 187(4):555--571.

\bibitem[\protect\astroncite{{Takano} et~al.}{2007}]{TT07}
{Takano}, Y., {Takahashi}, J.-i., {Kaneko}, T., {Marumo}, K., and {Kobayashi},
  K. (2007).
\newblock {Asymmetric synthesis of amino acid precursors in interstellar
  complex organics by circularly polarized light}.
\newblock {\em Earth Planet. Sci. Lett.}, 254(1-2):106--114.

\bibitem[\protect\astroncite{{Tasker} et~al.}{2017}]{TT17}
{Tasker}, E., {Tan}, J., {Heng}, K., {Kane}, S., {Spiegel}, D., {Brasser}, R.,
  {Casey}, A., {Desch}, S., {Dorn}, C., {Hernlund}, J., {Houser}, C.,
  {Laneuville}, M., {Lasbleis}, M., {Libert}, A.-S., {Noack}, L., {Unterborn},
  C., and {Wicks}, J. (2017).
\newblock {The language of exoplanet ranking metrics needs to change}.
\newblock {\em Nature Astron.}, 1:0042.

\bibitem[\protect\astroncite{{The KamLAND Collaboration}}{2011}]{KC11}
{The KamLAND Collaboration} (2011).
\newblock {Partial radiogenic heat model for Earth revealed by geoneutrino
  measurements}.
\newblock {\em Nat. Geosci.}, 4(9):647--651.

\bibitem[\protect\astroncite{{Thomas} and {Dieckmann}}{2002}]{TD02}
{Thomas}, D.~N. and {Dieckmann}, G.~S. (2002).
\newblock {Antarctic Sea Ice-a Habitat for Extremophiles}.
\newblock {\em Science}, 295(5555):641--644.

\bibitem[\protect\astroncite{{Thomas} et~al.}{2016}]{TT16}
{Thomas}, P.~C., {Tajeddine}, R., {Tiscareno}, M.~S., {Burns}, J.~A., {Joseph},
  J., {Loredo}, T.~J., {Helfenstein}, P., and {Porco}, C. (2016).
\newblock {Enceladus's measured physical libration requires a global subsurface
  ocean}.
\newblock {\em Icarus}, 264:37--47.

\bibitem[\protect\astroncite{{Throop}}{2011}]{Roop}
{Throop}, H.~B. (2011).
\newblock {UV photolysis, organic molecules in young disks, and the origin of
  meteoritic amino acids}.
\newblock {\em Icarus}, 212(2):885--895.

\bibitem[\protect\astroncite{{Tian} and {Ida}}{2015}]{TI15}
{Tian}, F. and {Ida}, S. (2015).
\newblock {Water contents of Earth-mass planets around M dwarfs}.
\newblock {\em Nat. Geosci.}, 8(3):177--180.

\bibitem[\protect\astroncite{{Travis} et~al.}{2012}]{TPS12}
{Travis}, B.~J., {Palguta}, J., and {Schubert}, G. (2012).
\newblock {A whole-moon thermal history model of Europa: Impact of hydrothermal
  circulation and salt transport}.
\newblock {\em Icarus}, 218(2):1006--1019.

\bibitem[\protect\astroncite{{Trilling} et~al.}{2017}]{TR17}
{Trilling}, D.~E., {Robinson}, T., {Roegge}, A., {Chandler}, C.~O., {Smith},
  N., {Loeffler}, M., {Trujillo}, C., {Navarro-Meza}, S., and {Glaspie}, L.~M.
  (2017).
\newblock {Implications for Planetary System Formation from Interstellar Object
  1I/2017 U1 (`Oumuamua)}.
\newblock {\em Astrophys. J. Lett.}, 850(2):L38.

\bibitem[\protect\astroncite{{Trinks} et~al.}{2005}]{TSB05}
{Trinks}, H., {Schr{\"o}der}, W., and {Biebricher}, C.~K. (2005).
\newblock {Ice And The Origin Of Life}.
\newblock {\em Orig. Life Evol. Biosph.}, 35(5):429--445.

\bibitem[\protect\astroncite{{Turcotte} and {Schubert}}{2002}]{TS02}
{Turcotte}, D.~L. and {Schubert}, G. (2002).
\newblock {\em {Geodynamics}}.
\newblock Cambridge Univ. Press.

\bibitem[\protect\astroncite{{Tyack}}{2001}]{Ty01}
{Tyack}, P.~L. (2001).
\newblock {Cetacean culture: Humans of the sea?}
\newblock {\em Behav. Brain Sci.}, 24(2):358--359.

\bibitem[\protect\astroncite{{Tyrrell}}{1999}]{Ty99}
{Tyrrell}, T. (1999).
\newblock {The relative influences of nitrogen and phosphorus on oceanic
  primary production}.
\newblock {\em Nature}, 400(6744):525--531.

\bibitem[\protect\astroncite{{Tyrrell}}{2013}]{Ty13}
{Tyrrell}, T. (2013).
\newblock {\em {On Gaia: A Critical Investigation of the Relationship between
  Life and Earth}}.
\newblock Princeton Univ. Press.

\bibitem[\protect\astroncite{{Valencia} and {O'Connell}}{2009}]{VC09}
{Valencia}, D. and {O'Connell}, R.~J. (2009).
\newblock {Convection scaling and subduction on Earth and super-Earths}.
\newblock {\em Earth Planet. Sci. Lett.}, 286(3-4):492--502.

\bibitem[\protect\astroncite{{Valencia} et~al.}{2006}]{VOCS06}
{Valencia}, D., {O'Connell}, R.~J., and {Sasselov}, D. (2006).
\newblock {Internal structure of massive terrestrial planets}.
\newblock {\em Icarus}, 181(2):545--554.

\bibitem[\protect\astroncite{{Valencia} et~al.}{2007}]{VS07}
{Valencia}, D., {Sasselov}, D.~D., and {O'Connell}, R.~J. (2007).
\newblock {Detailed Models of Super-Earths: How Well Can We Infer Bulk
  Properties?}
\newblock {\em Astrophys. J.}, 665(2):1413--1420.

\bibitem[\protect\astroncite{{van Dishoeck} et~al.}{2014}]{VDB14}
{van Dishoeck}, E.~F., {Bergin}, E.~A., {Lis}, D.~C., and {Lunine}, J.~I.
  (2014).
\newblock {Water: From Clouds to Planets}.
\newblock In {Beuther}, H., {Klessen}, R.~S., {Dullemond}, C.~P., and
  {Henning}, T., editors, {\em Protostars and Planets VI}, pages 835--858.
  Univ. of Arizona Press.

\bibitem[\protect\astroncite{{van Sluijs} and {Van Eylen}}{2018}]{VSVE}
{van Sluijs}, L. and {Van Eylen}, V. (2018).
\newblock {The occurrence of planets and other substellar bodies around white
  dwarfs using K2}.
\newblock {\em Mon. Not. R. Astron. Soc.}, 474(4):4603--4611.

\bibitem[\protect\astroncite{{Vance} et~al.}{2007}]{VHK07}
{Vance}, S., {Harnmeijer}, J., {Kimura}, J., {Hussmann}, H., {deMartin}, B.,
  and {Brown}, J.~M. (2007).
\newblock {Hydrothermal Systems in Small Ocean Planets}.
\newblock {\em Astrobiology}, 7(6):987--1005.

\bibitem[\protect\astroncite{{Vance} et~al.}{2016}]{VHP16}
{Vance}, S.~D., {Hand}, K.~P., and {Pappalardo}, R.~T. (2016).
\newblock {Geophysical controls of chemical disequilibria in Europa}.
\newblock {\em Geophys. Res. Lett.}, 43(10):4871--4879.

\bibitem[\protect\astroncite{{Vance} et~al.}{2018}]{VKP18}
{Vance}, S.~D., {Kedar}, S., {Panning}, M.~P., {St{\"a}hler}, S.~C., {Bills},
  B.~G., {Lorenz}, R.~D., {Huang}, H.-H., {Pike}, W.~T., {Castillo}, J.~C.,
  {Lognonn{\'e}}, P., {Tsai}, V.~C., and {Rhoden}, A.~R. (2018).
\newblock {Vital Signs: Seismology of Icy Ocean Worlds}.
\newblock {\em Astrobiology}, 18(1):37--53.

\bibitem[\protect\astroncite{{Vermeij}}{2006}]{Verm06}
{Vermeij}, G.~J. (2006).
\newblock {Historical contingency and the purported uniqueness of evolutionary
  innovations}.
\newblock {\em Proc. Natl. Acad. Sci. USA}, 103(6):1804--1809.

\bibitem[\protect\astroncite{{Vermeij}}{2015}]{Verm15}
{Vermeij}, G.~J. (2015).
\newblock {Forbidden phenotypes and the limits of evolution}.
\newblock {\em Interface Focus}, 5(6):20150028.

\bibitem[\protect\astroncite{{Vermeij}}{2017}]{Verm17}
{Vermeij}, G.~J. (2017).
\newblock {How the Land Became the Locus of Major Evolutionary Innovations}.
\newblock {\em Curr. Biol.}, 27(20):3178--3182.

\bibitem[\protect\astroncite{{Wachtershauser}}{1990}]{Wac90}
{Wachtershauser}, G. (1990).
\newblock {Evolution of the First Metabolic Cycles}.
\newblock {\em Proc. Natl. Acad. Sci. USA}, 87(1):200--204.

\bibitem[\protect\astroncite{{Waite} et~al.}{2017}]{WG17}
{Waite}, J.~H., {Glein}, C.~R., {Perryman}, R.~S., {Teolis}, B.~D., {Magee},
  B.~A., {Miller}, G., {Grimes}, J., {Perry}, M.~E., {Miller}, K.~E.,
  {Bouquet}, A., {Lunine}, J.~I., {Brockwell}, T., and {Bolton}, S.~J. (2017).
\newblock {Cassini finds molecular hydrogen in the Enceladus plume: Evidence
  for hydrothermal processes}.
\newblock {\em Science}, 356(6334):155--159.

\bibitem[\protect\astroncite{{Waite} et~al.}{2009}]{WLM09}
{Waite}, Jr., J.~H., {Lewis}, W.~S., {Magee}, B.~A., {Lunine}, J.~I.,
  {McKinnon}, W.~B., {Glein}, C.~R., {Mousis}, O., {Young}, D.~T., {Brockwell},
  T., {Westlake}, J., {Nguyen}, M.-J., {Teolis}, B.~D., {Niemann}, H.~B.,
  {McNutt}, R.~L., {Perry}, M., and {Ip}, W.-H. (2009).
\newblock {Liquid water on Enceladus from observations of ammonia and $^{40}$Ar
  in the plume}.
\newblock {\em Nature}, 460(7254):487--490.

\bibitem[\protect\astroncite{{Walker}}{2017}]{SW17}
{Walker}, S.~I. (2017).
\newblock {Origins of life: a problem for physics, a key issues review}.
\newblock {\em Rep. Prog. Phys.}, 80(9):092601.

\bibitem[\protect\astroncite{{Walker} et~al.}{2017}]{Walk17}
{Walker}, S.~I., {Bains}, W., {Cronin}, L., {DasSarma}, S., {Danielache}, S.,
  {Domagal-Goldman}, S., {Kacar}, B., {Kiang}, N.~Y., {Lenardic}, A.,
  {Reinhard}, C.~T., {Moore}, W., {Schwieterman}, E.~W., {Shkolnik}, E.~L., and
  {Smith}, H.~B. (2017).
\newblock {Exoplanet Biosignatures: Future Directions}.
\newblock {\em Astrobiology (arXiv: 1705.08071)}.

\bibitem[\protect\astroncite{{Wallmann}}{2010}]{Wall10}
{Wallmann}, K. (2010).
\newblock {Phosphorus imbalance in the global ocean?}
\newblock {\em Global Biogeochem. Cycles}, 24(4):GB4030.

\bibitem[\protect\astroncite{{Waltham}}{2011}]{Wal11}
{Waltham}, D. (2011).
\newblock {Anthropic selection and the habitability of planets orbiting M and K
  dwarfs}.
\newblock {\em Icarus}, 215(2):518--521.

\bibitem[\protect\astroncite{{Ward} and {Brownlee}}{2000}]{WaB00}
{Ward}, P. and {Brownlee}, D. (2000).
\newblock {\em {Rare Earth: Why Complex Life Is Uncommon in the Universe}}.
\newblock Copernicus.

\bibitem[\protect\astroncite{{Watson}}{2008}]{Wat08}
{Watson}, A.~J. (2008).
\newblock {Implications of an Anthropic Model of Evolution for Emergence of
  Complex Life and Intelligence}.
\newblock {\em Astrobiology}, 8(1):175--185.

\bibitem[\protect\astroncite{{Weinberg}}{2008}]{Wein08}
{Weinberg}, S. (2008).
\newblock {\em {Cosmology}}.
\newblock Oxford Univ. Press.

\bibitem[\protect\astroncite{{Weiss} et~al.}{2016}]{Weiss16}
{Weiss}, M.~C., {Sousa}, F.~L., {Mrnjavac}, N., {Neukirchen}, S., {Roettger},
  M., {Nelson-Sathi}, S., and {Martin}, W.~F. (2016).
\newblock {The physiology and habitat of the last universal common ancestor}.
\newblock {\em Nat. Microbiol.}, 1:16116.

\bibitem[\protect\astroncite{{Wesson}}{2010}]{Wess10}
{Wesson}, P.~S. (2010).
\newblock {Panspermia, Past and Present: Astrophysical and Biophysical
  Conditions for the Dissemination of Life in Space}.
\newblock {\em Space Sci. Rev.}, 156(1-4):239--252.

\bibitem[\protect\astroncite{{Westheimer}}{1987}]{West87}
{Westheimer}, F.~H. (1987).
\newblock {Why Nature Chose Phosphates}.
\newblock {\em Science}, 235(4793):1173--1178.

\bibitem[\protect\astroncite{{Wheat} et~al.}{2003}]{WM02}
{Wheat}, C.~G., {McManus}, J., {Mottl}, M.~J., and {Giambalvo}, E. (2003).
\newblock {Oceanic phosphorus imbalance: Magnitude of the mid-ocean ridge flank
  hydrothermal sink}.
\newblock {\em Geophys. Res. Lett.}, 30(17):1895.

\bibitem[\protect\astroncite{{Whitehead}}{2017}]{HW17}
{Whitehead}, H. (2017).
\newblock {Gene-culture coevolution in whales and dolphins}.
\newblock {\em Proc. Natl. Acad. Sci. USA}, 114(30):7814--7821.

\bibitem[\protect\astroncite{{Whitehead} and {Rendell}}{2015}]{WR15}
{Whitehead}, H. and {Rendell}, L. (2015).
\newblock {\em {The Cultural Lives of Whales and Dolphins}}.
\newblock The Univ. of Chicago Press.

\bibitem[\protect\astroncite{{Whiten} and {van Schaik}}{2007}]{WVS07}
{Whiten}, A. and {van Schaik}, C.~P. (2007).
\newblock {The evolution of animal ‘cultures’ and social intelligence}.
\newblock {\em Phil. Trans. R. Soc. B}, 362(1480):603--620.

\bibitem[\protect\astroncite{{Wickramasinghe}}{2010}]{Wick10}
{Wickramasinghe}, C. (2010).
\newblock {The astrobiological case for our cosmic ancestry}.
\newblock {\em Int. J. Astrobiol.}, 9(2):119--129.

\bibitem[\protect\astroncite{{Wickramasinghe} et~al.}{2012}]{WW12}
{Wickramasinghe}, N.~C., {Wallis}, J., {Wallis}, D.~H., {Schild}, R.~E., and
  {Gibson}, C.~H. (2012).
\newblock {Life-bearing primordial planets in the solar vicinity}.
\newblock {\em Astrophys. Space Sci.}, 341(2):295--299.

\bibitem[\protect\astroncite{{Williams} and {Fra\'usto Da Silva}}{2003}]{WFDS}
{Williams}, R.~J.~P. and {Fra\'usto Da Silva}, J.~J.~R. (2003).
\newblock {Evolution was Chemically Constrained}.
\newblock {\em J. Theor. Biol.}, 220(3):323--343.

\bibitem[\protect\astroncite{{Woese}}{2004}]{Woe04}
{Woese}, C.~R. (2004).
\newblock {A New Biology for a New Century}.
\newblock {\em Microbiol. Mol. Biol. Rev.}, 68(2):173--186.

\bibitem[\protect\astroncite{{Worth} et~al.}{2013}]{WSH13}
{Worth}, R.~J., {Sigurdsson}, S., and {House}, C.~H. (2013).
\newblock {Seeding Life on the Moons of the Outer Planets via Lithopanspermia}.
\newblock {\em Astrobiology}, 13(12):1155--1165.

\bibitem[\protect\astroncite{{Ye} et~al.}{2017}]{YZ17}
{Ye}, Q.-Z., {Zhang}, Q., {Kelley}, M.~S.~P., and {Brown}, P.~G. (2017).
\newblock {1I/2017 U1 (`Oumuamua) is Hot: Imaging, Spectroscopy, and Search of
  Meteor Activity}.
\newblock {\em Astrophys. J. Lett.}, 851(1):L5.

\bibitem[\protect\astroncite{{Youngblood} et~al.}{2017}]{YF17}
{Youngblood}, A., {France}, K., {Parke Loyd}, R.~O., {Brown}, A., {Mason},
  J.~P., {Schneider}, P.~C., {Tilley}, M.~A., {Berta-Thompson}, Z.~K.,
  {Buccino}, A., {Froning}, C.~S., {Hawley}, S.~L., {Linsky}, J., {Mauas},
  P.~J.~D., {Redfield}, S., {Kowalski}, A., {Miguel}, Y., {Newton}, E.~R.,
  {Rugheimer}, S., {Segura}, A., {Roberge}, A., and {Vieytes}, M. (2017).
\newblock {The MUSCLES Treasury Survey. IV. Scaling Relations for Ultraviolet,
  Ca ii K, and Energetic Particle Fluxes from M Dwarfs}.
\newblock {\em Astrophys. J.}, 843(1):31.

\bibitem[\protect\astroncite{{Zackrisson} et~al.}{2016}]{ZCG16}
{Zackrisson}, E., {Calissendorff}, P., {Gonz{\'a}lez}, J., {Benson}, A.,
  {Johansen}, A., and {Janson}, M. (2016).
\newblock {Terrestrial Planets across Space and Time}.
\newblock {\em Astrophys. J.}, 833(2):214.

\bibitem[\protect\astroncite{{Zag{\'o}rski}}{2003}]{Zag03}
{Zag{\'o}rski}, Z.~P. (2003).
\newblock {Radiation chemistry and origins of life on earth}.
\newblock {\em Rad. Phys. Chem.}, 66:329--334.

\bibitem[\protect\astroncite{{Zhang} et~al.}{2016}]{ZWW16}
{Zhang}, S., {Wang}, X., {Wang}, H., {Bjerrum}, C.~J., {Hammarlund}, E.~U.,
  {Mafalda Costa}, M., {Connelly}, J.~N., {Zhang}, B., {Su}, J., and
  {Canfield}, D.~E. (2016).
\newblock {Sufficient oxygen for animal respiration 1,400 million years ago}.
\newblock {\em Proc. Natl. Acad. Sci. USA}, 113(7):1731--1736.

\bibitem[\protect\astroncite{{Zolotov} and {Shock}}{2003}]{ZS03}
{Zolotov}, M.~Y. and {Shock}, E.~L. (2003).
\newblock {Energy for biologic sulfate reduction in a hydrothermally formed
  ocean on Europa}.
\newblock {\em J. Geophys. Res. E}, 108(E4):3.1--3.8.

\bibitem[\protect\astroncite{{Zolotov} and {Shock}}{2004}]{ZS04}
{Zolotov}, M.~Y. and {Shock}, E.~L. (2004).
\newblock {A model for low-temperature biogeochemistry of sulfur, carbon, and
  iron on Europa}.
\newblock {\em J. Geophys. Res. E}, 109(E6):E06003.

\end{thebibliography}

\end{document}